\documentclass[11pt]{article}
\pdfoutput=1 

\usepackage{jheppub}
                     
\usepackage{amsmath}
\usepackage{bm}
\usepackage[T1]{fontenc}
\usepackage{caption}
\usepackage{subcaption}
\usepackage{color}
\usepackage[export]{adjustbox}
\usepackage[normalem]{ulem}
\usepackage{xcolor}
\usepackage{tocvsec2}

\newcommand{\Mp}{M_{\phi}}
\newcommand{\f}{_{\phi}}
\newcommand{\dd}[2]{\frac{\textrm{d} #1}{\textrm{d} #2}}
\newcommand{\vp}{|\vec{p}|}
\newcommand{\p}{\tilde{p}}
\newcommand{\s}{\tilde{s}}
\newcommand{\tl}{\tilde{t}}
\newcommand{\Mf}{\tilde{M}_{\phi}}
\newcommand{\br}[1]{#1}
\newcommand{\rh}{_\textrm{rh}}
\newcommand{\pushright}[1]{\ifmeasuring@#1\else\omit\hfill$\displaystyle#1$\fi\ignorespaces}
\newcommand{\pushleft}[1]{\ifmeasuring@#1\else\omit$\displaystyle#1$\hfill\fi\ignorespaces}

\title{\boldmath Reheating in two-sector cosmology}

\makeatother

\author{Peter Adshead,}
\author{Pranjal Ralegankar,}
\author{and Jessie Shelton}

\affiliation{Department of Physics, University of Illinois at Urbana-Champaign, Urbana, IL 61801, USA }

\emailAdd{adshead@illinois.edu}
\emailAdd{pranjal6@illinois.edu}
\emailAdd{sheltonj@illinois.edu}

\abstract{We analyze reheating scenarios where a hidden sector is populated during reheating along with the sector containing the Standard Model.  We numerically solve the Boltzmann equations describing perturbative reheating of the two sectors, including the full dependence on quantum statistics, and study how quantum statistical effects during reheating as well as the non-equilibrium inflaton-mediated energy transfer between the two sectors affects the temperature evolution of the two radiation baths.  We obtain new power laws describing the temperature evolution of fermions and bosons when quantum statistics are important during reheating.  We show that inflaton-mediated scattering is generically most important at radiation temperatures $T\sim M_\phi/4$, and build on this observation to obtain analytic estimates for the temperature asymmetry produced by asymmetric reheating.    We find that for reheating temperatures $T_{\textrm{rh}} \ll M_{\phi}/4$, classical perturbative reheating provides an excellent approximation to the final temperature asymmetry, while for $T_{\textrm{rh}}\gg M_{\phi}/4$, inflaton-mediated scattering dominates the population of the colder sector and thus the final temperature asymmetry.  We additionally present new techniques to calculate energy transfer rates between two relativistic species at different temperatures.}

\begin{document}
\maketitle
\flushbottom

\section{Introduction}

A host of observational data from galactic rotation curves to the cosmic microwave background (CMB) points towards an unknown form of matter comprising more than 85\% of the matter quota in our Universe \cite{Ade:2015xua,Aghanim:2018eyx}.  While this dark matter (DM) could be a single particle, or a family of particles missing from the standard model of particle physics (SM), to date no effects have been detected that require dark matter interact non-gravitationally with the SM.  Further, the traditional weakly interacting massive particle (WIMP) paradigm is under ever-increasing pressure from the dearth of observational signatures in collider, direct, and indirect detection experiments \cite{Escudero:2016gzx,Alexander:2016aln}.

A straightforward alternative to the WIMP scenario is that the dark matter is in a `hidden sector' containing new particles and forces that may or may not interact non-gravitationally with the standard model and its extensions.  Hidden sector DM refers to a class of models where the DM resides in an internally thermalized hidden sector, and has a relic abundance that is determined by physics unrelated to its couplings to the visible sector.  
When the hidden sector is out of equilibrium with the SM in the early universe, a wide variety of possible cosmic histories for dark matter are newly possible,  together with avenues for discovery distinct from standard WIMP search strategies.  Exotic thermal evolution in a decoupled dark sector can lead to substantial changes in DM properties relative to a traditional WIMP  (e.g., \cite{Carlson:1992fn,Feng:2008mu,Sigurdson:2009uz,Cheung:2010gj,Pappadopulo:2016pkp,Dror:2016rxc,Berlin:2016gtr,Dror:2017gjq,Georg:2019jld,Faraggi:2000pv,Forestell:2016qhc,Dienes:2016vei,Forestell:2018dnu,Kang:2019izi,Blanco:2019eij}).   Decoupled dark sectors can also induce early departures from the standard radiation-dominated evolution of the universe \cite{Zhang:2015era,Berlin:2016gtr}, or admit substantial amounts of relic dark radiation without violating the stringent bounds from Planck, $\Delta N_{\rm eff} < 0.36$ at 95\% confidence \cite{Aghanim:2018eyx}.  

There is now strong evidence from observations of the fluctuations in the CMB that the thermal era was preceded by an epoch of early accelerated expansion---inflation. Inflation exponentially dilutes any pre-existing matter and radiation leaving the Universe cold and empty. The population of otherwise decoupled sectors cannot therefore be put in `by hand' as an initial condition. Instead it must be generated dynamically in the post-inflationary evolution of the Universe. In the simplest scenarios, the accelerated expansion is driven by a single fundamental scalar degree of freedom, whose weak couplings to matter reheat the Universe via perturbative decays. One of the simplest mechanisms for populating hidden sectors is to couple them to the inflaton so that they are populated at reheating along with the visible sector. By arranging the couplings so that the hidden sector couples differently to the inflaton than the SM, reheating can be asymmetric, whereby the SM and the hidden sector are heated to different temperatures \cite{Hodges:1993yb,Berezhiani:1995am,Adshead:2016xxj,Halverson:2019kna}.  However, coupling both the SM and a hidden sector to the inflaton in the UV necessarily results in inflaton-mediated interactions between the two sectors.  As demonstrated in reference \cite{Adshead:2016xxj}, this irreducible inflaton-mediated scattering can thermalize the two sectors under fairly generic conditions.

In this work, we extend the analysis of reference \cite{Adshead:2016xxj} to explore the effects of out-of-equilibrium inflaton-mediated interactions on asymmetric reheating. Along the way, we develop and implement methods to numerically solve the Boltzmann equations describing the reheating of two otherwise-decoupled sectors from the perturbative decay of the inflaton. In particular, we develop accurate approximations (including the effects of quantum statistics) for the collision terms that describe the inflaton-mediated scattering between thermalized gases of fermionic and bosonic particles.  We take an effective field theory approach and consider combinations of trilinear scalar, Yukawa, and pseudo-scalar couplings between fermions, bosons, and the inflaton.  
When inflaton couplings to matter become sufficiently large, both non-perturbative effects such as preheating and collective effects in the radiation baths such as Landau damping and thermal masses can provide important corrections to the inflaton decay rate and hence the evolution of the temperature asymmetry, particularly at very high radiation temperatures \cite{Kolb:2003ke,Drewes:2013iaa,Hardy:2017wkr}.   However, as we show here,  both inflaton decays and inflaton-mediated scattering furnish cosmological attractor solutions during the perturbative phase of reheating, making the final temperature asymmetry largely sensitive to the dynamics of the system at and below the perturbative reheat temperature $T_{\rm rh}$.  Thus  the perturbative reheating process that we analyze in the present paper will often serve as a good guide to the final temperature asymmetry despite the presence of richer dynamics at early times.

This paper is organized as follows. In section \ref{sec:singquant}, we review standard  perturbative reheating  and extend the well-known single-sector results to include the effects of quantum statistics on the decay width of the inflaton. In section \ref{sec:twosecquant}, we begin our study of reheating into two sectors by considering the effects of quantum statistics on reheating into combinations of otherwise-decoupled fermionic and bosonic sectors. In section \ref{sec:twosecreheat}, we reintroduce the inflaton-mediated interactions between the two sectors (required by self-consistency) and show how inter-sector scattering dominates over any features from quantum statistics in most of the parameter space.
We conclude in section \ref{sec:conclusions}. Appendix~\ref{appendix:attractor} elaborates on the construction of cosmological attractor solutions, while appendix~\ref{appendix:preheating} demonstrates that the novel behavior found for bosonic radiation baths in section~\ref{sec:twosecquant} can indeed be realized during perturbative reheating.  In appendices~\ref{appendix:s-channel} and \ref{appendix:t-channel}, we lay out the procedure for analytically evaluating energy transfer rates between two relativistic particles at different temperatures mediated by a massive scalar field.

We work in units where $\hbar = c = k_B = 1$, and denote by $M_{\rm Pl} = 2.435\times  10^{18}$ GeV the reduced Planck mass.
%
\section{Quantum statistics in single-sector reheating}\label{sec:singquant}
%

In this section we revisit the perturbative reheating of a radiation bath.  After reviewing the classic treatment, we demonstrate that at temperatures $T>M_{\phi}/4$, where $M_{\phi}$ is the inflaton mass, quantum effects such as Bose enhancement and Pauli blocking can significantly affect the evolution of the temperature of the radiation bath during reheating. We show that the effects of quantum statistics disappear once $T$ drops below $M_{\phi}/4$, and thus alter the outcome of reheating only when $T_{\rm rh} \gtrsim M_{\phi}/4$. 
 While we refer to the decaying particle as an inflaton and have post-inflationary reheating primarily in mind, our results  apply also to other ``reheatons'' such as curvatons or moduli (see also \cite{Reece:2015lch}). 

\subsection{Perturbative reheating}

A generic scenario of inflation \cite{Guth:1980zm,Linde:1981mu,Albrecht:1982wi} consists of one or more scalar fields $\phi_i$ slowly rolling on a sufficiently flat potential, $V(\phi_i)$ (see, for example, \cite{Martin:2013tda} and references within). Inflation ends when the slow-roll conditions are violated, and the fields $\phi_i$ roll quickly to the potential minima and start oscillating. For this work, we assume that only one field $\phi$ is relevant during the reheating process, and that its potential is analytic and can be expanded in a Taylor series about its minimum. We further assume that only the leading quadratic term in this Taylor series is needed.\footnote{We are explicitly ignoring anharmonic corrections to the inflaton potential that may be relevant during reheating. These anharmonic terms can be important for non-perturbative effects during reheating, such as the formation of oscillons, as  recently reviewed in \cite{Amin:2014eta}. Conversely, the absence of a quadratic minima generically leads to a radiation equation-of-state very quickly following inflation \cite{Lozanov:2016hid}.} The time-averaged equation of state of a field oscillating in a quadratic potential is that of a stationary massive particle, and thus the Universe undergoes a period of matter domination while the inflaton energy density dominates \cite{Turner:1983he}. During this oscillating phase, the inflaton condensate starts to decay through its couplings to matter, initiating reheating. If these couplings are large enough, the first stage of reheating can proceed through a period of parametric resonance known as preheating \cite{Traschen:1990sw, Kofman:1994rk}. In the preheating regime, particle production is non-perturbative and typically requires numerical treatment (however, see \cite{Emond:2018ybc}). As the amplitude of inflaton oscillations decreases, due to both Hubble friction and inflaton decay, preheating ceases and particle production can be treated perturbatively.  Unless preheating is violent enough to drain an $\mathcal{O}(1)$ fraction of energy out of the inflaton condensate, this final epoch of perturbative reheating  typically dominates the properties of the radiation bath produced by inflaton decays.

For this work, we thus consider perturbative reheating in  a quadratic potential \cite{Abbott:1982hn, Albrecht:1982mp}. We consider the generic case where all particle masses besides the inflaton mass are negligible at the energies we consider, and therefore treat all matter species as radiation. We further neglect inverse decays from radiation into inflaton quanta; this is a good approximation provided the number of species in the radiation bath is large, $g_* \gg 1$.  With these approximations, the Boltzmann equations describing reheating read (see, for example, \cite{Chung:1998rq})
\begin{eqnarray}\label{eq:boltz_rho}
\dd{{\rho}\f}{t}+3{H}{\rho}\f &=&-{\Gamma}{\rho}\f \\ \label{eq:boltz_rhorad}
\dd{{\rho}}{t}+4{H}{\rho} &=&{\Gamma}{\rho}\f,
\end{eqnarray}
where the Hubble rate  ${H}$ is given by the Friedmann equation
\begin{eqnarray}
{H} =  \frac{1}{\sqrt{3}{M}_{\textrm{Pl}}}\sqrt{{\rho}+{\rho}\f }\; .
\end{eqnarray}
The inflaton width is denoted by $\Gamma$, and $\rho_{\phi}$ and $\rho$ are the inflaton and radiation energy densities, respectively.  These equations are (approximately) valid from the end of inflation at some scale factor $a=a_I$, which we take as our initial point. The radiation sector is initially empty,\footnote{This is generally a good approximation for models with tri-linear scalar couplings and Yukawa interactions with fermions, as in these cases the daughter fields get a large mass during inflation, shutting off inflaton decays.  However, for a pseudo-scalar inflaton coupling to either fermions or gauge bosons, there can be significant energy density already in the radiation sector as inflation ends (see, for example, \cite{Adshead:2015pva, Adshead:2015kza}). However, as we demonstrate below, the specific initial conditions are largely irrelevant for the detailed outcome of reheating.} $\rho_{,I}=0$, whereas the initial energy density of the inflaton is given in terms of  the mean-square value of the inflaton field just after the end of inflation, $\langle\phi^2_I\rangle$, as $\rho_{\phi,I}=M^2\f\langle\phi^2_I\rangle/2$.

In figure~\ref{fig:single_sector_power_law}, we plot the inflaton and radiation energy densities obtained by numerically solving eqns.\ \eqref{eq:boltz_rho} and \eqref{eq:boltz_rhorad} with a constant inflaton decay width, $\Gamma =\Gamma_0$. Initially,
$\Gamma\ll H$ and therefore inflaton decays are inefficient. Thus the inflaton energy density during this phase can be well approximated as diluting only through redshifting, $\rho\f \approx \rho_{\phi,I}(a/a_I)^{-3}$. The evolution of the radiation sector, however, is dominated by the energy injection from inflaton decays. Initially, the radiation energy density grows rapidly until the rate at which energy is injected into the radiation bath by inflaton decays, governed by ${\Gamma}{\rho}_\phi$, matches the rate at which the radiation bath loses energy due to the expansion of the universe, governed by $4H\rho$. After this point, the evolution of the radiation sector follows an attractor solution, which realizes a quasi-static equilibrium between energy injection and dilution (see appendix \ref{appendix:attractor}),
\begin{align}\label{eq:attractor_propto}
4H\rho= \frac{4}{4+\frac{q}{1-p}}\Gamma\rho\f,
\end{align}
where
\begin{align}
q(a)=\frac{\partial \ln(\Gamma\rho\f/H)}{\partial \ln a}, \qquad \text{and}  &\qquad p(a)=\frac{\partial \ln(\Gamma\rho\f/H)}{\partial \ln \rho}.
\end{align}
We call the above evolution imposed on the radiation sector by inflaton decays the \textit{reheating attractor curve}.
For a temperature-independent decay width, $\Gamma = \Gamma_0$, the factors $q$ and $p$ are readily determined to be constants, $q = -3/2$ and $p = 0$, yielding the relation $4 H \rho = (8/5) \Gamma_0\rho\f$.
On this attractor solution the radiation bath evolves as \cite{Chung:1998rq}
\begin{align}\label{eq:rho_rh_classical}
\rho= \Big(\frac{2}{5}\Gamma_0 M_{\textrm{Pl}}\sqrt{\rho_{\phi,I}}\Big) \left(\frac{a}{a_I}\right)^{-3/2},
\end{align}
 when $a \gg a_I$.
The attractor nature of this solution means that the evolution of the energy density of the radiation bath during reheating is relatively insensitive to its initial conditions.  Radiation baths with energy density initially  below the attractor solution rise rapidly to meet the attractor.  Meanwhile radiation densities initially above the attractor curve redshift as $\rho \propto a^{-4}$ until they meet the attractor, as can occur in (e.g.) scenarios where a modulus comes to dominate the post-reheating universe \cite{Reece:2015lch, Kane:2015jia,Co:2015pka}.

The attractor solution can also be obtained by solving the Boltzmann equations during reheating. Well before reheating, the  inflaton condensate dominates the energy budget of the Universe; its comoving energy density is approximately constant and  the Boltzmann equation describing the radiation bath can be simplified to yield \cite{Chung:1998rq}
\begin{eqnarray}\label{eq:Rstart}
\dd{}{a} \left(\rho a^4\right)&=& \sqrt{3}{M}_{\textrm{Pl}}a^{3}\sqrt{\rho_\phi}\,{\Gamma_0}.
\end{eqnarray}
Solving eq.~\eqref{eq:Rstart} with the initial condition $\rho(a_I)=0$ also allows us to determine the maximum energy density attained by the radiation sector \cite{Chung:1998rq},
\begin{align}\label{eq:classical_tmax}
{\rho}_{\textrm{max}}=0.24{M}_{\textrm{Pl}}{\Gamma}_0\sqrt{{\rho}_{\phi,I}}.
\end{align}

\subsection{Quantum statistics during single-sector reheating}

The preceding discussion neglected the possible effects of quantum statistics during reheating. Typically, the inflaton decays at rest, producing pairs of particles at a fixed energy $M_{\phi}/2$.
To quantify the possible effects of Pauli blocking or Bose enhancement of the inflaton decay, we need to specify the  phase space distribution in the radiation sector.  For simplicity, we assume the radiation is in thermal equilibrium,
\begin{align}\label{eqn:tempdef}
\rho=\frac{\pi^2g_{*}}{30} T^4\equiv\alpha T^4,
\end{align}
 which amounts to assuming that the thermalization time scale for the radiation sector is much faster than any other time scale in the problem. This is in some sense a conservative assumption for the purpose of analyzing the scattering and reheating processes discussed in this paper: a less equilibrated sector has a greater fraction of particles with energies concentrated near $M_{\phi}/2$, making both inflaton-mediated scattering and quantum statistics more important.  However, as we demonstrate below, the post-reheating properties of the radiation baths are typically determined by the late-time behavior of the system, making the detailed approach to thermal equilibrium within each radiation bath largely immaterial for the final outcome of reheating.  This separation of timescales generally makes prompt thermalization a robust assumption.

For temperatures $T \lesssim M_{\phi}/4$, a constant (zero-temperature) inflaton decay width is a good approximation. At these temperatures, the phase space where particles are injected by inflaton decays, $E\sim M\f/2$ , is sparsely populated due to the fast thermalization of the injected particles, and thus the effects of Pauli blocking or Bose enhancement are negligible. However, at higher temperatures, $T \gtrsim M_{\phi}/4$, the equilibrium thermal distributions have significant support at $E\sim M\f/2$, and the inflaton decay rate can be significantly altered. 
The partial decay width of a parent scalar to pairs of particles in equilibrium at finite temperature is given by 
\begin{align}\label{eq:decay_widths}
\Gamma(T)=\Gamma_0\frac{\exp(\frac{\Mp}{2T})\pm 1}{\exp(\frac{\Mp}{2T})\mp 1} ,
\end{align}
where $\Gamma_0$ is the zero temperature decay width, and the  upper (lower) sign holds for bosons (fermions) in the final state.
At high temperatures, the decay width is enhanced (suppressed) for bosons (fermions) due to Bose enhancement (Pauli blocking).  We now consider reheating to boson and fermionic radiation separately.

\begin{figure}
\begin{subfigure}{.5\textwidth}
\includegraphics[width=1.00\textwidth]{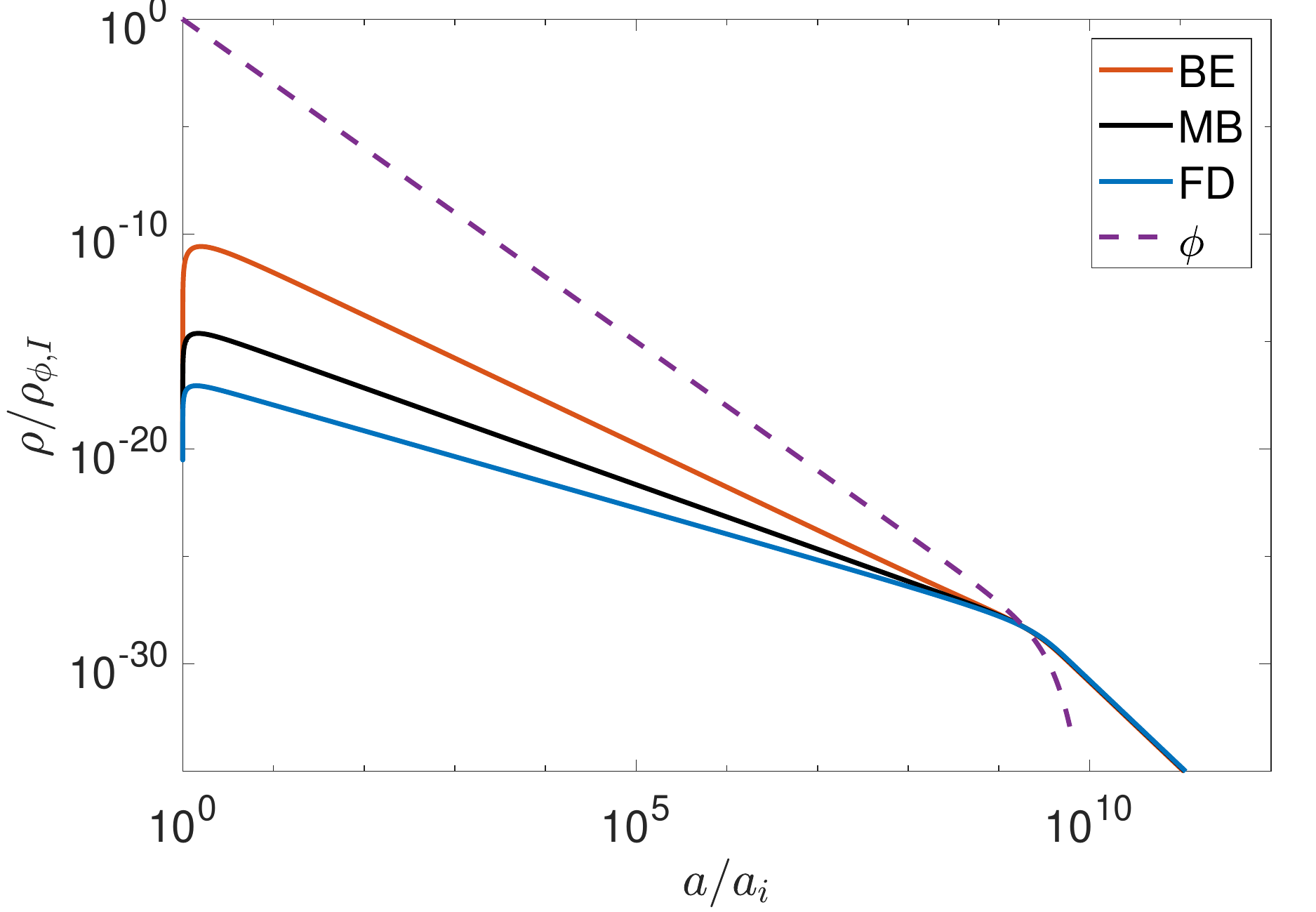}
\end{subfigure}
\begin{subfigure}{.5\textwidth}
\includegraphics[width=1.00\textwidth]{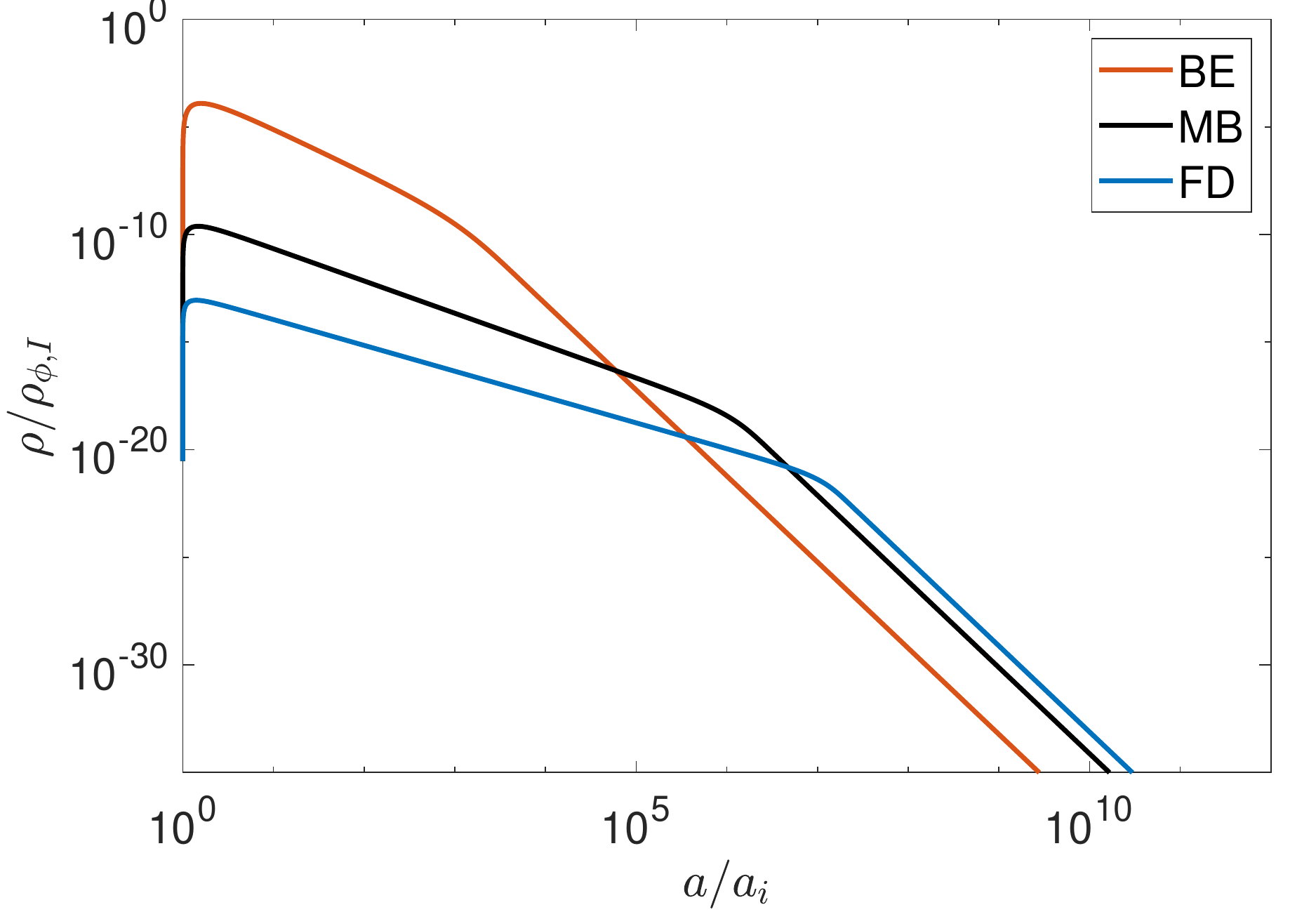}
\end{subfigure}
\caption{\emph{Left panel:} Energy density as a function of scale factor during reheating for $\Gamma_0=10^{-14}M\f$, $\rho_{\phi,I}=M_{\textrm{Pl}}^2M_{\phi}^2$, and $M\f=2.4\times 10^5$~GeV.  Solid lines show the energy density of a thermal radiation bath with different colors indicating different statistics: orange for Bose-Einstein (BE), black for Maxwell-Boltzman (MB) and blue for Fermi-Dirac (FD).
The energy density in the inflaton field is shown by the purple dashed line. \emph{Right panel:} Same as the left panel, with inflaton width given by $\Gamma_0=10^{-9}M\f$. For these parameters the reheat temperature is larger than the inflaton mass and hence different quantum statistics lead to different reheat temperatures.}
\label{fig:single_sector_power_law}
\end{figure}

\paragraph{Bosonic reheating:} In the case of  decays to bosons, for $T\gg M_{\phi}/4$ the inflaton decay width is approximately given by $\Gamma\approx 4T\Gamma_0/M\f$. Using this decay width in eq.~\eqref{eq:attractor_propto} immediately yields a new quasi-static equilibrium solution for the radiation bath ($q=-3/2$ and $p=1/4$), with power law evolution 
\begin{align}\label{eq:boson_power_law}
T= M\f\Big(\frac{2\sqrt{3}}{\alpha}\frac{{M}_{\textrm{Pl}}}{M\f^4}{\Gamma}_0\sqrt{{\rho}_{\phi,I}}\Big)^{1/3}\left(\frac{a}{a_I}\right)^{-1/2}.
\end{align}
This reheating attractor curve can again be found analytically by solving the approximate Boltzmann equation describing the radiation bath during reheating, analogous to the Maxwell-Boltzmann result.  With initial condition $\rho (a_I)=0$, the full temperature evolution is
\begin{align}
T(a)=M\f\Bigg(\frac{2\sqrt{3}}{\alpha}\frac{{M}_{\textrm{Pl}}}{M\f^4}{\Gamma}_0\sqrt{{\rho}_{\phi,I}}\bigg[\bigg(\frac{a}{a_I}\bigg)^{-3/2}-\bigg(\frac{a}{a_I}\bigg)^{-3}\bigg]\Bigg)^{1/3}.
\end{align}
The radiation bath attains its maximum temperature,
\begin{align}\label{eq:boson_tmax}
T_{\textrm{max}}=0.95 M\f\Big(\frac{1}{\alpha}\frac{{M}_{\textrm{Pl}}}{M\f^4}{\Gamma}_0\sqrt{{\rho}_{\phi,I}}\Big)^{1/3},
\end{align}
at $a=1.6 a_I$.
For bosons, the inflaton decay width decreases with temperature, making energy injection into the radiation sector less efficient as the temperature decreases. This results in the temperature dropping as $T \propto a^{-1/2}$, faster than the classical result $T \propto a^{-3/8}$ (eq.~\eqref{eq:rho_rh_classical}). 

Obtaining radiation temperatures high enough to realize this power law requires a relatively large inflaton width.  One might be concerned that such large couplings to matter place the inflaton in the regime where  preheating dominates over perturbative reheating.  However, we demonstrate in appendix \ref{appendix:preheating}
that, for an inflaton with trilinear couplings to scalars, there is indeed a region of parameter space where the novel Bose power law $T\propto a^{-1/2}$ is realized during perturbative reheating.

\paragraph{Fermionic reheating:} For an inflaton decaying to fermions at $T\gg M\f$,  the decay width can be well approximated by $\Gamma\approx\Gamma_0 M\f/(4T)$, which gives $q=-3/2$ and $p=-1/4$.   In this regime, the radiation sector evolves as
\begin{align}
T= M\f\Big(\frac{5\sqrt{3}}{56\alpha}\frac{{M}_{\textrm{Pl}}}{M\f^4}{\Gamma}_0\sqrt{{\rho}_{\phi,I}}\Big)^{1/5}\left(\frac{a}{a_I}\right)^{-3/10}.
\end{align}
The full solution to the Boltzmann equations with initial condition $\rho (a_I)=0$ is
\begin{align}
T(a)=M\f\Bigg(\frac{5\sqrt{3}}{56\alpha}\frac{{M}_{\textrm{Pl}}}{M\f^4}{\Gamma}_0\sqrt{{\rho}_{\phi,I}}\bigg[\bigg(\frac{a}{a_I}\bigg)^{-3/2}-\bigg(\frac{a}{a_I}\bigg)^{-5}\bigg]\Bigg)^{1/5}.
\end{align}
The maximum temperature attained by the radiation bath is
\begin{align}\label{eq:fermion_tmax}
T_{\textrm{max}}=0.58M\f\Big(\frac{1}{\alpha}\frac{{M}_{\textrm{Pl}}}{M\f^4}{\Gamma}_0\sqrt{{\rho}_{\phi,I}}\Big)^{1/5},
\end{align}
attained at $a=1.4 a_I$.

\medskip
In  the analytic treatment of the Boltzmann equations for reheating in the fermionic and bosonic cases above, we have taken $T(a_I)=0$ as our initial condition. Strictly this is inconsistent with the high temperature expansion used for the inflaton width. A more complete analytic treatment would use the zero-temperature inflaton width to describe the early evolution of the radiation bath until its temperature rises to $M\f$ before implementing the high temperature expansion. However, such a procedure only alters the scale factor at which the maximum temperature is attained and not its value. Moreover, since the maximum temperature is attained very quickly compared to other timescales in our problem, the error due to this simplifying assumption is negligible.   Perhaps the more consequential assumption in this region is that we have taken the radiation bath to attain internal thermal equilibrium nearly instantaneously. In the very early periods of reheating, the thermalization rate is likely to be smaller than the very rapid  rate at which the energy density of the radiation bath grows. The simple solutions presented here for the decay width and the initial evolution of the energy densities  are thus probably incorrect for describing these very early regions.  

Once the temperature falls below the inflaton mass scale, the  temperature dependence of the inflaton decay width in eq.~\eqref{eq:decay_widths} becomes unimportant as inflaton decays now populate sparsely occupied regions of phase space.   Subsequently the radiation sector evolves as $T\propto a^{-3/8}$.

Reheating completes when the inflaton decays become efficient, $\Gamma\sim H$, and the inflaton energy density decreases exponentially.  During this epoch the Universe transitions from the matter-dominated era of reheating to a radiation-dominated expansion, where  the temperature of the radiation sector redshifts adiabatically as $T\propto a^{-1}$. If $\Gamma\sim H$ occurs while  $T\gtrsim M_{\phi}/4$, then the temperature of the radiation sector directly transitions to $T\propto a^{-1}$ without going through the classical $T\propto a^{-3/8}$ regime. In this scenario, the resulting reheat temperature depends on the quantum statistics of the inflaton decay products.  Estimating the reheat temperature by setting $H=\Gamma(T)$ and taking $H$ to be dominated by the radiation bath, we find for $T_{\textrm{rh}}\gg M\f $
\begin{align}\label{eq:reheat_temperatures}
T_{\textrm{rh}}=\begin{cases}
\dfrac{4\sqrt{3}}{\sqrt{\alpha}}\dfrac{{M}_{\textrm{Pl}}\Gamma_0}{M\f} & \textrm{boson }\\ 
\bigg(\dfrac{\sqrt{3}}{4\sqrt{\alpha}} M_{\textrm{Pl}}\Gamma_0 M\f\bigg)^{1/3}& \textrm{fermion }
\end{cases} ,
\end{align}
in contrast to the classical result  
\begin{align}
T_{\textrm{rh}}= \bigg(\frac{\sqrt{3}}{\sqrt{\alpha}}{M}_{\textrm{Pl}}\Gamma_0\bigg)^{1/2}  ,
\end{align}
which holds for $T_{\textrm{rh}}\ll M\f$.
We summarize these different power law behaviors of the radiation temperature in figure~\ref{fig:single_sector_power_law}. In the left panel we show a case where the reheat temperature $T_{\textrm{rh}}$ is below the inflaton mass. In this case, quantum statistics are unimportant for determining $T_{\textrm{rh}}$, as all three scenarios converge onto the
attractor solution governing classical perturbative reheating, eq.~\eqref{eq:attractor_propto}.
In the right panel of figure~\ref{fig:single_sector_power_law} we show a case where $T_{\textrm{rh}}$ is above the inflaton mass.  As the inflaton decay width gets significant corrections from quantum statistics at these high temperatures, we observe the different reheat temperatures of eq.~\eqref{eq:reheat_temperatures} expected for different quantum statistics at fixed zero-temperature decay width. 

In the above scenario we have assumed that all particles coupled to the inflaton have the same quantum statistics (bosons or fermions).  If the inflaton couples to both bosons and fermions then the energy density of the radiation sector as a whole evolves depending on the total inflaton decay width. In this scenario,  the inflaton width is dominated by the Bose-enhanced partial widths at very high temperatures, and hence the radiation sector evolves according to the bosonic power law ($T\propto a^{-1/2}$). If the zero-temperature partial-width into fermions is larger than that to bosons,  $\Gamma^{\rm fermion}_{0}>\Gamma^{\rm boson}_{0}$, then (assuming $T_{\textrm{rh}}<M_{\phi}/4$) there is a temperature, $T_*$, for which $\Gamma^{\rm boson}(T_*) = \Gamma^{\rm fermion}(T_*)$, while $\Gamma^{\rm boson}(T) < \Gamma^{\rm fermion}(T)$ for $T < T_*$. For $M_{\phi}/4< T < T_*$, the radiation bath transitions to the power law $T\propto a^{-3/10}$ (characteristic of high-temperature fermionic reheating) before ultimately transitioning to the classical $T\propto a^{-3/8}$ for  $T < M_{\phi}/4$.

Despite model-dependent uncertainties associated with the initial evolution of the radiation baths, the attractor nature of these perturbative reheating solutions renders the later temperature evolution, and the resulting reheat temperatures, insensitive to variations to the initial conditions and early evolution provided that the attractor solution is obtained.  Reaching the attractor solution requires that 1) the energy density of the oscillating inflaton dominates the Hubble rate for some time, during which inflaton decays become a cosmologically important source for the radiation bath, and 2) that the thermalization timescale is short compared to the duration of inflaton domination.  As we demonstrate in the remainder of this paper, similarly general results can be obtained for the more complicated scenarios that arise in two-sector reheating as well. 

\section{Two-sector reheating and quantum statistics}\label{sec:twosecquant}

In section \ref{sec:singquant}, we demonstrated how quantum statistics alter the temperature evolution of the radiation sector prior to reheating. In this section we explore the extent to which these quantum statistical effects can impact the final temperature asymmetry in two-sector reheating.   To isolate the effects of quantum statistics in the inflaton decay width on the resulting temperature asymmetries, in this section we temporarily ignore inflaton-mediated scattering.
We begin by writing down the Boltzmann equations for two-sector reheating, and then consider the case where the inflaton couples to fields with the same statistics in both sectors and the case where the  inflaton couples to fields with different statistics in each sector.

\subsection{Two-sector reheating}

We restrict ourselves to the regime of perturbative reheating and neglect inflaton-mediated scattering.  For simplicity we continue to assume that the inflaton is only coupled to one species of particle (boson or fermion) in each sector.  Ignoring any inflaton quanta, the Boltzmann equations in this limit read
\begin{eqnarray}\label{eq:boltz_2sector}
\dd{{\rho}\f}{t}+3{H}{\rho}\f &=&-({\Gamma}_a+{\Gamma}_b){\rho}\f,\nonumber\\
\dd{{\rho}_a}{t}+4{H}{\rho}_a &=&{\Gamma}_a{\rho}\f,\nonumber\\
\dd{{\rho}_b}{t}+4{H}{\rho}_b &=&{\Gamma}_b{\rho}\f.
\end{eqnarray}
Here ${H}$ is the Hubble rate, which is given by the Friedmann equation
\begin{eqnarray}\label{eq:friedmann2sector}
{H}=\frac{1}{\sqrt{3}{M}_{\textrm{Pl}}}\sqrt{{\rho}_a+{\rho}_b+{\rho}\f },
\end{eqnarray}
and $\Gamma_{a,b}$ are the (temperature-dependent) decay rates of $\phi$ to the respective sectors. The subscript `$a$' denotes the sector that attains the larger temperature at the end of reheating. Generically, this corresponds to the sector with the largest zero-temperature decay width. However, as we demonstrate below, this is not the case when the inflaton couples to bosons in one sector and fermions in the other  and the resulting reheat temperature is large, $T_{\rm rh} \gtrsim M_{\phi}/4$.  `Reheat temperature' in this context refers to the temperature of the hotter sector when the universe transitions from matter to radiation domination. We define the transition from matter domination to radiation domination at the point where energy density in the radiation becomes equal to the energy density in the inflaton, $\rho_a(a\rh)+$~$\rho_b(a\rh)=\rho\f(a\rh)$.

Although no scattering terms appear in eq.\ \eqref{eq:boltz_2sector}, the radiation baths are still coupled gravitationally through the Friedmann equation, eq.\ \eqref{eq:friedmann2sector}.
Prior to reheating, when the Hubble rate is dominated by the inflaton, the two sectors are effectively decoupled and evolve independently of each other. During this phase, the temperature evolution in both radiation sectors is determined by their respective reheating attractor curves, as discussed in section~\ref{sec:singquant}. The reheat temperature is eventually determined by the hotter sector, and following reheating both sectors evolve adiabatically.

For the numerical results in the rest of the paper we adopt a common reference set of  numerical values for the inflaton mass and initial energy density as well as the number of degrees of freedom in each radiation bath, 
\begin{align}\label{eq:constants}
\alpha_a =  \frac{\pi^2g_{*, a}}{30} =\alpha_b = \frac{\pi^2g_{*, b}}{30}=30, & \quad 
M\f=10^{-13}M_{\rm Pl} = 2.4\times10^5  \textrm{\, GeV}, 
\\ \nonumber\rho_{\phi,I} =  \frac{1}{2}M^2\f\langle\phi_I^2\rangle &=M_{\rm Pl}^2M_{\phi}^2 .
\end{align}
We  assume for simplicity that $\alpha_a$ and $\alpha_b$ are constant over the range of temperatures we consider. While in what follows we have fixed the value of $M_{\phi}$, our results are broadly independent of its precise value.  As we demonstrate below, our results for the final temperature asymmetry depend on $M_\phi$ only through $T_\mathrm{rh}$ and the ratio $T_\mathrm {rh}/M_\phi$. The specific value of $M_\phi$ is generally only important insofar as smaller values of $M_\phi$ make it easier to obtain larger $T_\mathrm {rh}/M_\phi$.  

When solving the Boltzmann equations numerically, we start the computation immediately after the end of inflation with $\rho_{\phi}(a_I)=\rho_{\phi,I}$. However, instead of assuming the radiation sectors to be initially empty, we take them to be at their maximum temperatures,  $\left.T_{a,b}(a_I)=T_{a,b}\right|_{\textrm{max}}$ (given by eqs.\ \eqref{eq:boson_tmax}, \eqref{eq:fermion_tmax}, and  \eqref{eq:classical_tmax}), to improve computational speed. Since the maximum temperatures are obtained very quickly around $a\approx 1.5a_I$, the above approximation deviates only marginally from the exact result. Moreover, the attractor nature of the reheating process ensures that such small deviations in the early evolution of the radiation baths leave their late time evolution entirely unaffected.

%
\subsection{Temperature asymmetries for sectors with the same quantum statistics}
%

\begin{figure}[t!]
\includegraphics[width=1.0\textwidth]{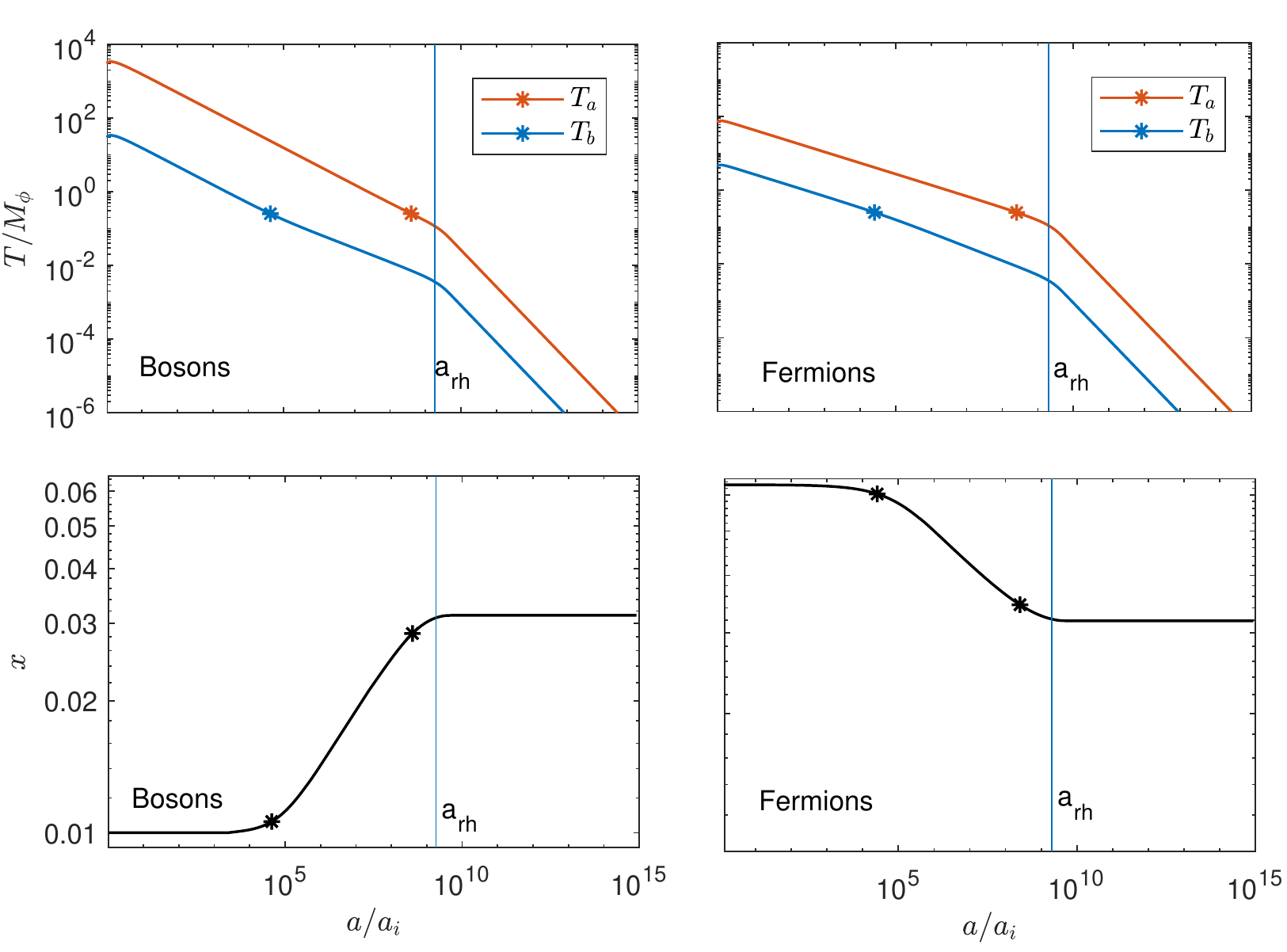}
\caption{\textit{Upper panels:} Radiation temperature as a function of scale factor for two-sector reheating without scattering,  when the inflaton decays to particles with the same statistics in both sectors.  The upper left panel shows results for an inflaton coupled to bosons in both sectors with $\Gamma_{0,a}=10^{-14}M\f$ (red) and $\Gamma_{0,b}=10^{-20}M\f$  (blue); the upper right panel shows the same for inflaton decays to fermions. Other parameters are specified in eq.~\eqref{eq:constants}. Reheating is indicated by the vertical blue line at $a_{\textrm{rh}}$. \textit{Lower panels:} Temperature ratio of the two sectors, $x=T_b/T_a$, as a function of the scale factor. In each panel, asterisks indicate   $T_b= M_{\phi}/4$  (left) and   $T_a= M_{\phi}/4$ (right), which is approximately where quantum statistics become unimportant in each sector. The left panel corresponds to bosons in both sectors, while the right panel corresponds to fermions in both sectors.} 
\label{fig:scalar_temp_evolve}
\end{figure}

We first consider the case where the inflaton couples to fields with the same statistics in both sectors.  Figure \ref{fig:scalar_temp_evolve} 
shows the evolution of the temperatures of the two sectors as a function of the scale factor when the inflaton is coupled to either bosons (left panels) or fermions (right panels) in both sectors. The temperature evolution changes as the temperatures of the sectors drop below $M_{\phi}/4$ (indicated in the figure by asterisks), and quantum statistics become unimportant. As the two sectors reach this critical temperature $M_{\phi}/4$ at different times, the temperature ratio
\begin{align}
x=T_b/T_a
\end{align}
changes during the period where quantum statistics are important in one sector but not the other. This is displayed in the right panels of figure \ref{fig:scalar_temp_evolve}. Note that  $x$ increases in the case of bosons (lower left panel) while it decreases for fermions (lower right panel). This is because Bose enhancement causes the hotter sector to redshift more quickly, $T\propto a^{-1/2}$, compared to the classical behavior $T\propto a^{-3/8}$ at $T\lesssim M_{\phi}/4$. On the other hand, fermions redshift more slowly due to Pauli blocking, $T\propto a^{-3/10}$.  After reheating, both sectors evolve adiabatically with $T\propto 1/a$, and the temperature asymmetry between the two sectors is frozen in.

\begin{figure}
\begin{subfigure}{.5\textwidth}
\includegraphics[width=1.00\textwidth]{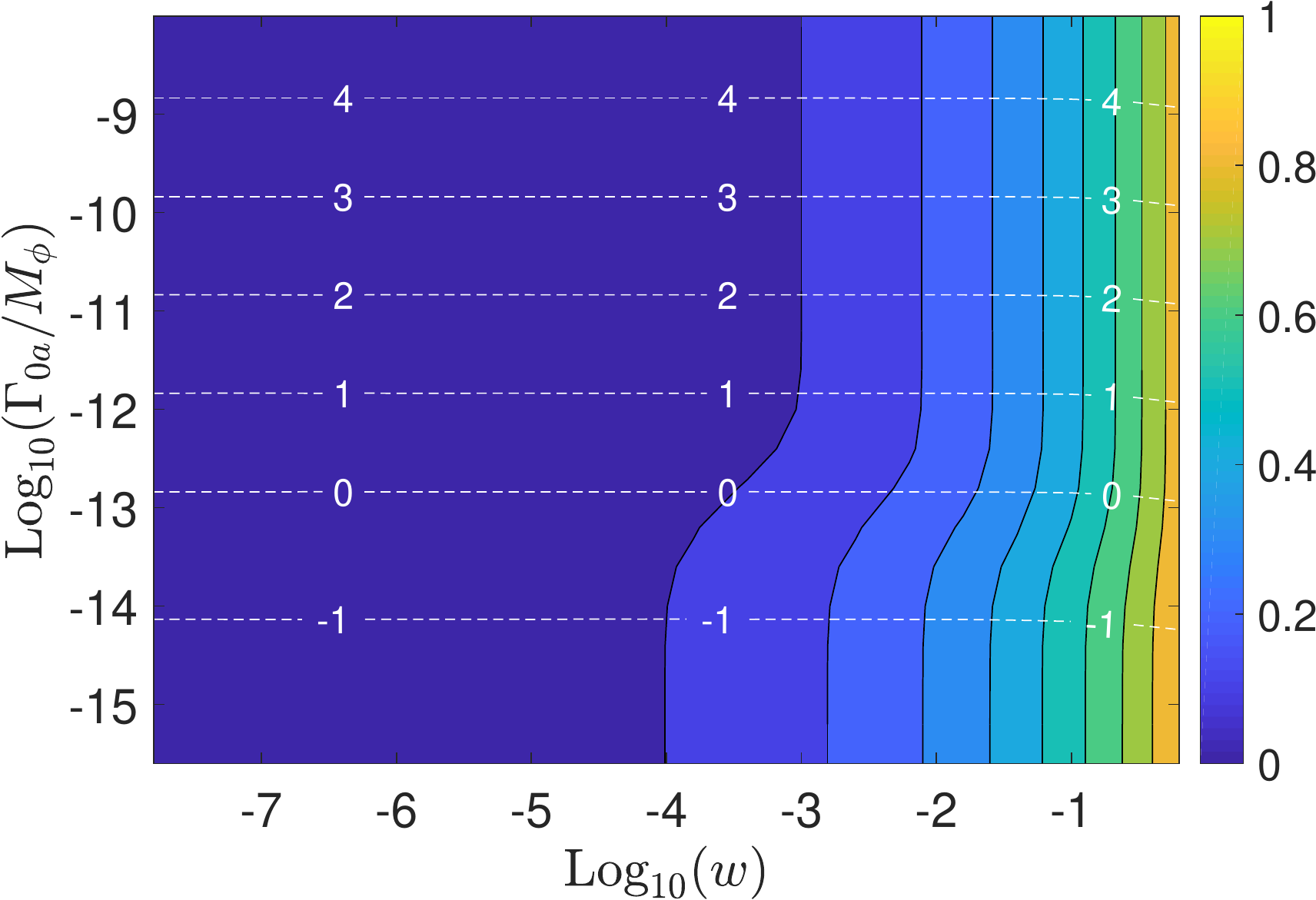}
\end{subfigure}
\begin{subfigure}{.5\textwidth}
\includegraphics[width=1.00\textwidth]{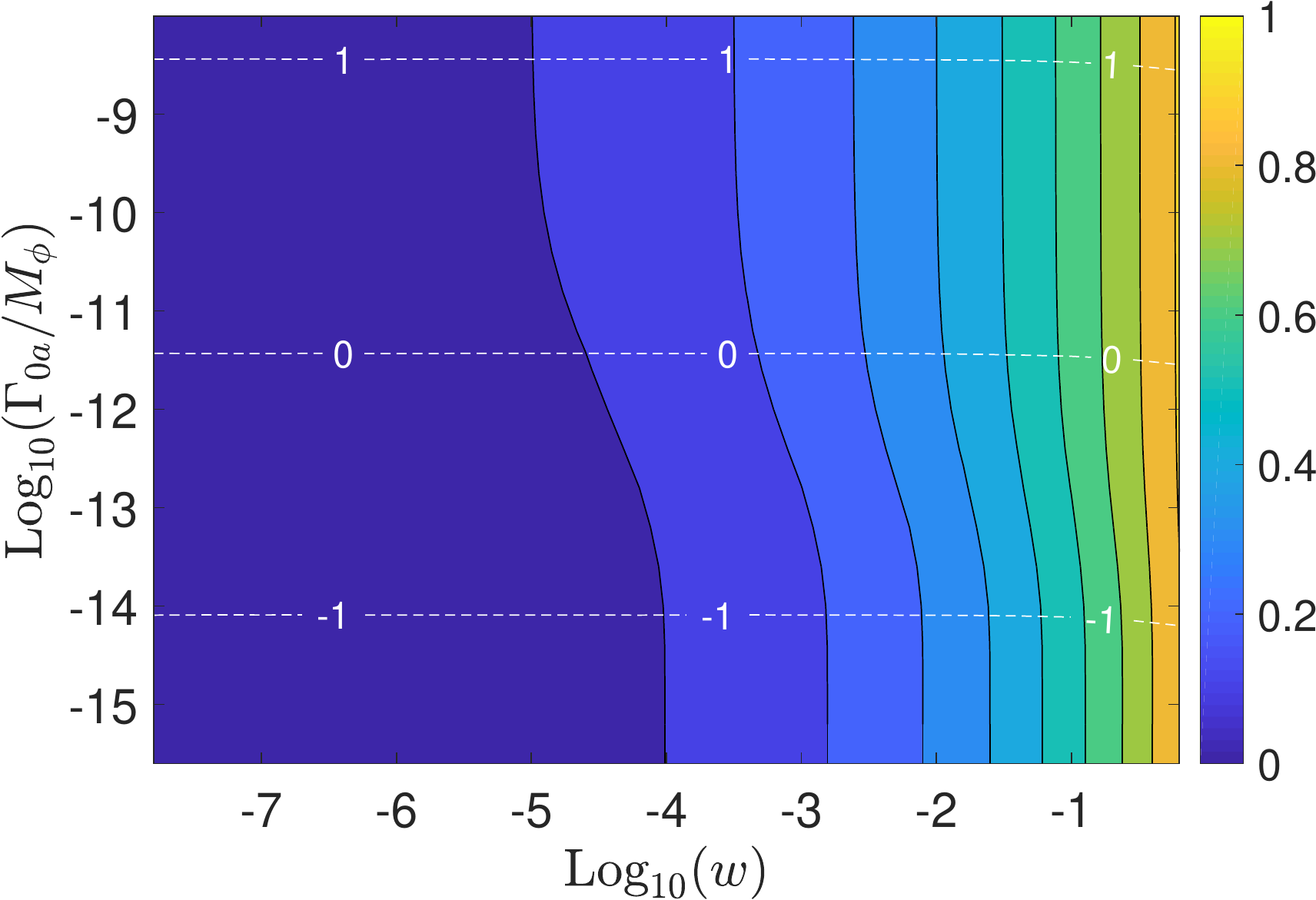}
\end{subfigure}
\caption{Contours of the post-reheating temperature ratio in the absence of inflaton-mediated scattering $x\rh$,  as a function of ${\Gamma}_{0,a}$ and $w={\Gamma}_{0,b}/{\Gamma}_{0,a}$ for the case when inflaton is coupled to bosons in both sectors (left) and when it is coupled to fermions in both sectors (right). Other parameters are as specified in eq.~\eqref{eq:constants}. The yellow region corresponds to near-equal temperatures, $x\geq 0.9$, whereas the blue region corresponds to a final temperature ratio $x\leq 0.1$. The white dashed lines are contours of $\log_{10}(T_{\textrm{rh}}/M\f)$. }
\label{fig:xfnoscat}
\end{figure}

We define the final post-reheating temperature ratio as
\begin{align}
x\rh=\frac{T_b(a')}{T_a(a')},
\end{align}
where $a'>a\rh$ is some value of the scale factor at which both radiation sectors are evolving adiabatically, i.e.\ $\rho \propto a^{-4}$. In figure~\ref{fig:xfnoscat}, we display the contours of $x\rh$ as a function of the ratio of the zero-temperature decay widths into each sector, 
\begin{align}
w  = \Gamma_{0,b}/\Gamma_{0,a}
\end{align} 
for inflaton decays to bosons (left) or fermions (right) in both sectors. When $T_{\textrm{rh}}\ll M\f$ and $T_{\textrm{rh}}\gg M\f$, the final temperature ratio  depends only on the ratio of the zero-temperature decay widths. In the intermediate region,  $T_{\textrm{rh}}\sim M\f$, the contours tilt to the right (left) for the boson (fermion) case, since here reheating occurs when $T_a\gtrsim M_{\phi}/4$ but $T_b \lesssim M_{\phi}/4$. 

In the next section, we show that these results, obtained in the absence of scattering, are a good guide to the temperature asymmetries obtained in the full theory in the regime when  $T\rh\lesssim M\f$ (see figure \ref{fig:xfscat_scalar} and \ref{fig:xfscat_fermion}, below).  Above  $T\rh\gtrsim M\f$, however,
inflaton-mediated scattering between the two sectors is significant in determining the final temperature asymmetries.

%
\subsection{Temperature asymmetries for sectors with differing quantum statistics}
%

Finally, we consider the case where the inflaton is coupled to fermions in one sector and bosons in the other. For this scenario, we  label the sector with bosons `$a$', and the sector with fermions `$b$'.
\begin{figure}
\begin{subfigure}{.5\textwidth}
\includegraphics[width=1.00\textwidth]{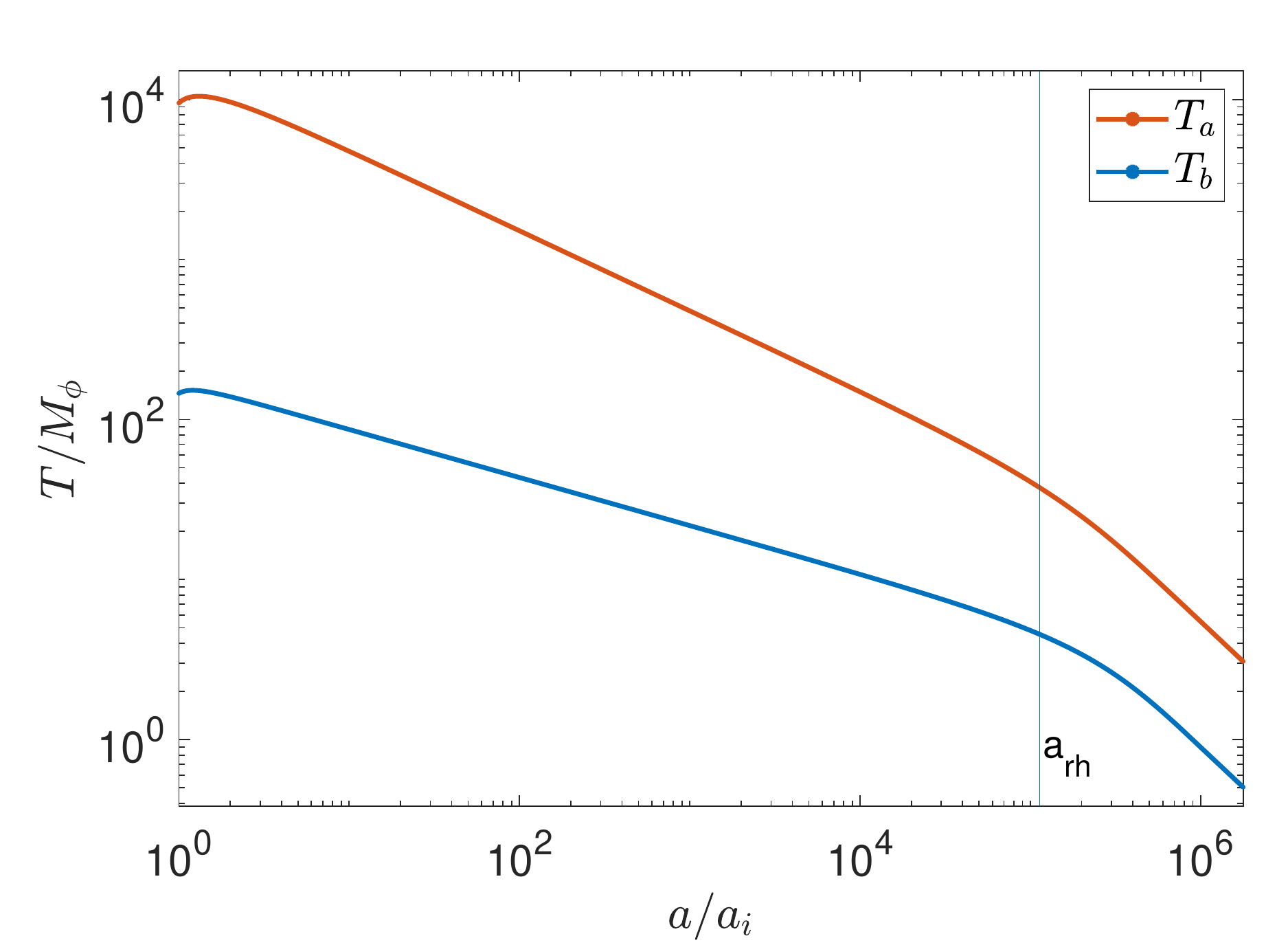}
\end{subfigure}
\begin{subfigure}{.5\textwidth}
\includegraphics[width=1.00\textwidth]{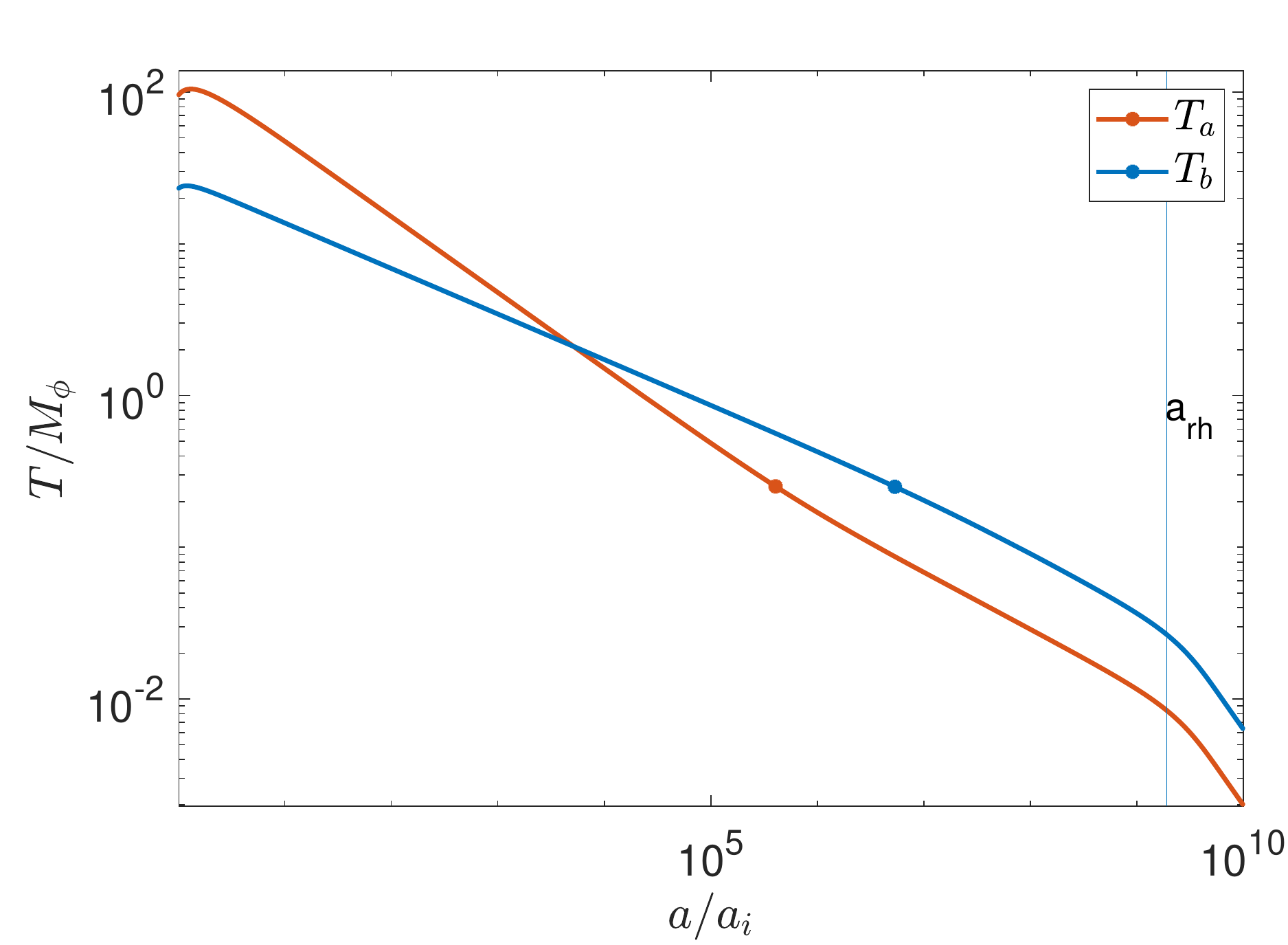}
\end{subfigure}
\caption{\textit{Left panel:} Temperature of the radiation baths as a function of scale factor in the case where the inflaton decays to bosons in one sector (orange line) and fermions in the other (blue line). The inflaton decay widths are  $\Gamma_{0,a}=\Gamma_{0,b}=10^{-9}M\f$; other constants are specified in eq.~\eqref{eq:constants}. \textit{Right panel:} Same as left panel, but with decay widths set at $\Gamma_{0,a}=10^{-16}$ and $\Gamma_{0,b}=10^{-14}M\f$. The dots on the curves denote the point $T=M_{\phi}/4$, below which the effects from quantum statistics become negligible.}
\label{fig:fb_temp_evolve}
\end{figure}

As in the cases discussed above, prior to reheating the two radiation baths evolve independently.
When $T\gg \Mp$ the different quantum statistics results in different temperature evolution in the two sectors. As the temperature drops below the inflaton mass scale, $T\lesssim \Mp/4$, the inflaton decay width becomes, to an excellent approximation, temperature-independent, and the temperature of both sectors evolves as $T\propto a^{-3/8}$. For reheating at temperatures $T \lesssim M_{\phi}/4$, the attractor nature of the temperature evolution in both sectors ensures that, regardless of the earlier history  of both sectors, the final temperature asymmetry is set by the ratio of the zero-temperature decay widths. However, when $T_{\rm rh} \gtrsim M_{\phi}/4$, the differing temperature evolution in the two sectors is frozen into the resulting temperature asymmetries, as we now discuss. 

In the case where the zero-temperature decay width is the same to both sectors, ${\Gamma}_{0b}={\Gamma}_{0a}$, if reheating occurs after quantum statistics become unimportant, the two sectors subsequently attain the same temperature. This is evident in the left panel of figure~\ref{fig:single_sector_power_law}. However, for reheating  at $T_{\textrm{rh}}>M_{\phi}/4$, the details of the quantum statistics become important. This is demonstrated in the left panel of figure~\ref{fig:fb_temp_evolve}, which shows how early reheating can freeze in a temperature asymmetry due to quantum statistics in an otherwise symmetric reheating scenario. 

If the zero-temperature partial width to fermions is larger than that to bosons, ${\Gamma}_{0b}\geq {\Gamma}_{0a}$,
the bosonic sector could still initially be hotter (due to the Bose-enhanced larger initial $T_{\textrm{max}}$) than the fermionic sector, and then evolve to become cooler. An example of this behavior is illustrated in the right panel of figure~\ref{fig:fb_temp_evolve}. In this case, the identity of the `hotter' sector depends on the temperature at which reheating happens. In particular, for $T_{\textrm{rh}}\gtrsim M_{\phi}/4$, the fermionic sector can have a smaller temperature than the bosonic sector even when the inflaton has a larger zero-temperature partial width into fermions.

\begin{figure}
\centering
\includegraphics[width=0.7\textwidth]{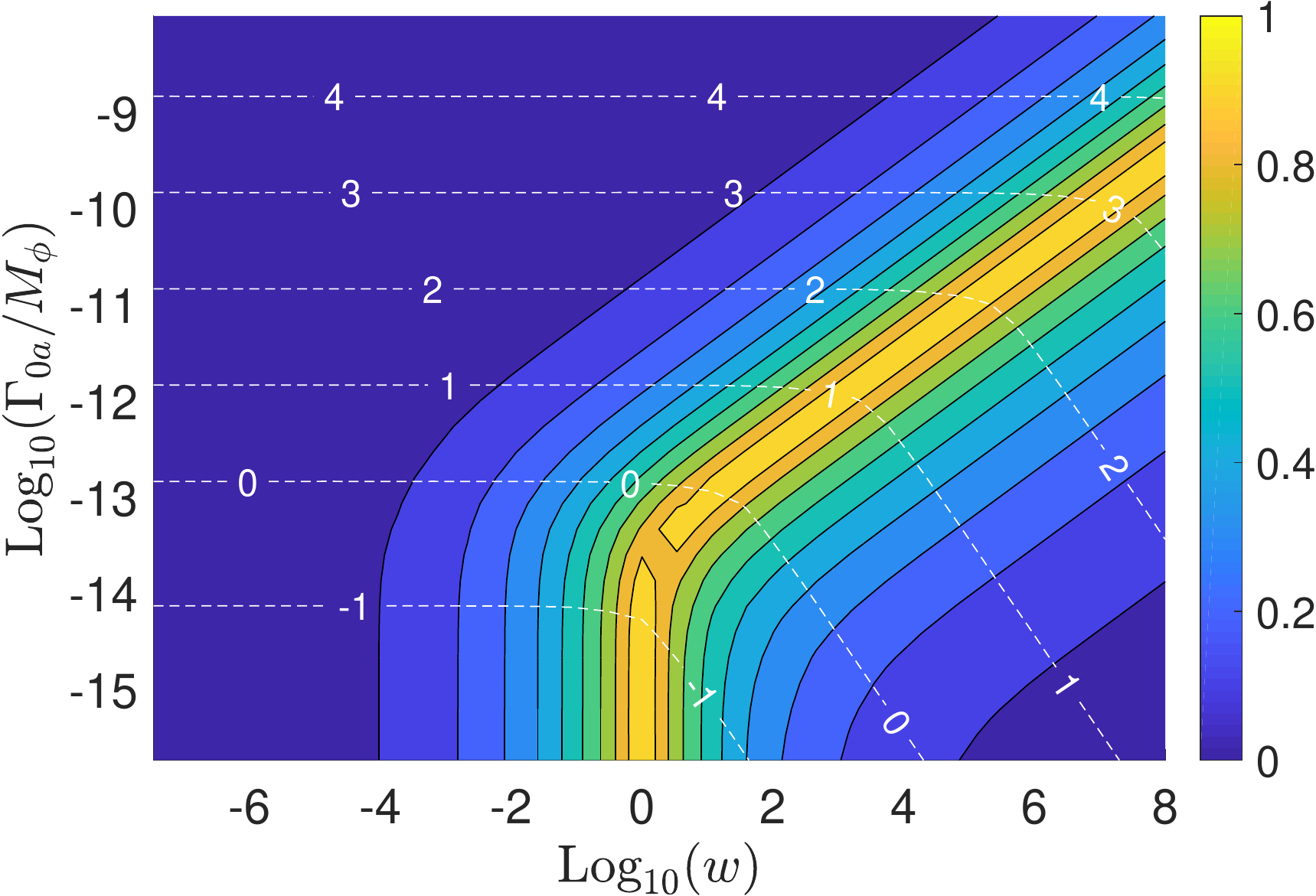}
\caption{Contours of final temperature ratio, $x\rh$, for the case when inflaton is coupled to bosons in one sector (sector $a$) and fermion in the other (sector $b$). The parameters are as specified in eq.~\eqref{eq:constants}. White dashed lines show contours of $\log_{10}(T_{\textrm{rh}}/M\f)$.}
\label{fig:xfnoscat_fb}
\end{figure}

In figure~\ref{fig:xfnoscat_fb}, we plot the post-reheating temperature ratio, $x\rh =$ $\textrm{min}(T_{b}/T_{a},T_{a}/T_{b})$, as a function of ${\Gamma}_a$ and $w={\Gamma}_b/{\Gamma}_a$. When reheating happens below the scale of the inflaton mass, $T\rh \ll \Mp$, the final temperature asymmetry is insensitive to the effects of quantum statistics and the final contours depend only on the ratio of the widths, $w$. For $T_{\textrm{rh}}\gtrsim \Mp/4$, on the other hand, $x_{\rm rh}$ depends on both ${\Gamma}_a$ and $w$ due to the asymmetric effects from quantum statistics. In the blue region on the left, bosons are the hotter sector, while in the blue region on the right, fermions are the hotter sector.  The transition between these situations occurs in the yellow region.  As in the cases above with identical statistics,  these no-scattering results provide an accurate guide to the final temperature asymmetries for $T\rh\lesssim M\f$ (see figure~\ref{fig:xfscat_fb}, below). 

Unless there are additional non-adiabatic processes at late times, the final temperature asymmetry between the sectors is determined by their temperatures at reheating. Since the radiation temperatures are governed by the reheating attractor curve $4H\rho\propto \Gamma\rho\f$, the temperature asymmetry depends only on the ratio of decay widths at reheating. Hence, if reheating happens at temperatures $T_{\textrm{rh}}\lesssim M_{\phi}/4$ the final temperature asymmetry is  determined by the ratio of the zero-temperature decay widths, $x=(\alpha_a{\Gamma}_{0,b}/\alpha_b{\Gamma}_{0,a})^{1/4}$. However, larger reheat temperatures probe the quantum statistics of the inflaton decay products, which can leave imprints in enhanced or suppressed temperature asymmetries.

\section{Two-sector reheating with inflaton-mediated interactions}\label{sec:twosecreheat}

We now incorporate inflaton-mediated scattering between the two sectors and study its effect on the temperature evolution in both sectors and the final temperature asymmetry.  As we demonstrate in this section, inflaton-mediated energy transfer between sectors also yields an attractor solution for the temperature of the colder radiation bath, which allows us to make analytic predictions for the final temperature asymmetries in the regime where inflaton-mediated scattering is important.

 We begin by establishing our notation. Introducing the scattering terms in the Boltzmann equations, we have \cite{Adshead:2016xxj}
\begin{eqnarray}\label{eqn:boltzwints}
\frac{d{\rho}\f}{dt}+3{H}{\rho}\f &=&-({\Gamma}_a+{\Gamma}_b){\rho}\f\\
\dd{{\rho}_a}{t}+4{H}{\rho}_a &=&{\Gamma}_a{\rho}\f-{\mathcal{C}}_E\\
\dd{{\rho}_b}{t}+4{H}{\rho}_b &=&{\Gamma}_b{\rho}\f+{\mathcal{C}}_E,
\end{eqnarray}
where $\mathcal{C}_E$ is the collision term describing the energy transfer from the hotter radiation sector to the colder radiation sector via two-to-two scattering processes of the form $1+2\rightarrow 3+4$, 
\begin{align}
\label{eq:scattering_demo}
\mathcal{C}_{E}=&\int \frac{d^3p_1}{2E_1(2\pi)^3}\frac{d^3p_2}{2E_2(2\pi)^3}\frac{d^3p_3}{2E_3(2\pi)^3}\frac{d^3p_4}{2E_4(2\pi)^3} 
(2\pi)^4\delta^4(p_1+p_2-p_3-p_4)|\overline{\mathcal{M}}|^2 S(E_1+E_2) \nonumber\\ &\times[f_1(p_1)f_2(p_2)(1\pm f_3(p_3))(1\pm f_4(p_4))-f_3(p_3)f_4(p_4)(1\pm f_1(p_1))(1\pm f_2(p_2))] \nonumber\\
\equiv& \mathcal{C}_{E}^f-\mathcal{C}_{E}^b.
\end{align}
Here $\mathcal{C}_{E}^f$ and $\mathcal{C}_{E}^b$ are the collision terms for forward and backward reactions respectively, $|{\overline{\mathcal{M}}}|^2$ is the spin-summed scattering amplitude as determined by the particular inflaton-radiation interaction, and $S$ is a symmetry factor accounting for possible identical particles in the initial and/or final state.  
We retain the full dependence on quantum statistics to accurately describe the energy transfer between two relativistic radiation baths \cite{Adshead:2016xxj}, which makes the evaluation of the collision term more challenging. 
In appendix~\ref{appendix:s-channel}, we analytically evaluate eq.\ \eqref{eq:scattering_demo}  for $s$-channel processes between two relativistic species at different temperatures for the specific interactions we analyze below. In appendix~\ref{appendix:t-channel}, we describe a novel procedure to reduce the integral in eq.~\eqref{eq:scattering_demo} for a $t$-channel scattering amplitude and verify explicitly that the energy transfer rate through $t$-channel processes is always orders of magnitude smaller than the energy transfer via $s$-channel processes.   

We next demonstrate that the inflaton-mediated energy transfer yields a cosmological attractor solution for the colder radiation bath, using the model where the inflaton has trilinear couplings to scalar fields in both sectors as an illustrative example.  We then analyze in detail how the interplay between this scattering attractor solution and the reheating attractor curve of the previous section determines the final temperature asymmetry. We then extend this analysis to other forms of the inflaton couplings to matter. In particular, we consider theories where the inflaton has: Yukawa couplings to fermions in both sectors; axion-like couplings to gauge bosons in both sectors; and a mixed scenario with a trilinear coupling to scalars in one sector and Yukawa coupling to fermions in the other. 

The numerical analysis  in the following sections uses the same parameter values specified in eq.~\eqref{eq:constants}.  For the collision term, $\mathcal{C}_E$, we use the analytic approximations derived in appendix~\ref{appendix:s-channel}.
For simplicity we continue to assume that the inflaton only couples to a single species in each sector.

\subsection{Scalar trilinear couplings}\label{sec:scalarbothsectors}

We begin by considering a theory where the inflaton is coupled to scalar fields in both sectors, $\chi_a$ and $\chi_b$, via trilinear couplings
\begin{eqnarray}
\mathcal{L}_{\textrm{int}}=\frac{1}{2}\mu_a\phi\chi_a^2+\frac{1}{2}\mu_b\phi\chi_b^2.
\end{eqnarray}
This interaction results in zero-temperature decay widths given by
\begin{eqnarray}
\Gamma_{0a,b}=\frac{1}{32\pi}\frac{\mu_{a,b}^2}{M\f}\sqrt{1-\frac{4m_{a,b}^2}{M\f^2}}\approx \frac{{\mu}_{a,b}^2}{32\pi M\f},
\end{eqnarray}
where $m_{a,b}$ denotes the mass of the fields $\chi_{a,b}$, which we have assumed to be much smaller than the inflaton mass, $m_{a,b}\ll \Mp$. Our convention is that sector $a$ is the hotter sector, and accordingly we take  $\mu_a \geq \mu_b$ in what follows. 

\subsubsection{The collision term}

The $s$-channel amplitude for $\chi_a \chi_a \leftrightarrow \chi_b \chi_b$ scattering mediated by inflaton exchange is given by
\begin{align}
|\mathcal{M}(s)|^2=\frac{\mu_a^2\mu_b^2}{(s-M^2_{\phi})^2+(\Gamma_{0a}+\Gamma_{0b})^2}.
\end{align}
In appendix~\ref{appendix:s_scalar}, we compute the collision term, $\mathcal{C}_E$, following from this amplitude. 

The collision term in general is a function of both $T_a$ and $T_b$. However, for large asymmetries $T_b\ll T_a$, the forward energy transfer term governing energy injection into the colder sector dwarfs the backward energy transfer term.  Moreover, in this regime we can also ignore the final state Bose enhancement of $C_E^f$: while  $\chi_b$  particles produced in the forward reaction typically have energies of order $\sim T_a$, for $T_b\ll T_a$ those energy levels are mostly unpopulated.  In this simplified regime, the collision term thus depends only on $T_a$
as
\begin{align}\label{eqn:CEscalars}
\mathcal{C}_E= \frac{1}{16\pi^3}\times\begin{cases}
\dfrac{\mu^2_a\mu^2_b}{\mu_a^2+\mu_b^2}T_a^3\Big[1.6 \log\Big(\frac{T_a}{M\f}\Big)+1.3\Big] & T_a\gtrsim M_{\phi}
\\ \\
\dfrac{\mu^2_a\mu^2_b}{\mu_a^2+\mu_b^2}M\f^2\dfrac{T_a}{4} K_2\Big(\frac{M_{\phi}}{T_a}\Big) & T_a\lesssim M_{\phi}
\\ \\
\dfrac{7.9}{32\pi^2}\dfrac{\mu_a^2\mu_b^2}{M\f^4} T_a^5 & m_{a,b}\ll T_a\ll M_{\phi},
\end{cases}
\end{align}
as derived in appendix~\ref{appendix:s_scalar}; see figure~\ref{fig:scalar_boson_analytical_compare}.
For $T_a\gtrsim M\f$, the collision term is substantially enhanced by the resonant exchange of inflaton particles.  The divergence in the Bose-Einstein distributions at $E\to 0$ combined with the resonant peak in the scattering amplitude results in a logarithmic dependence on $T_a/M\f$ for $T_a\gg M_\phi$. As $T_a$ drops below the inflaton mass, the scattering goes off resonance and $\mathcal{C}_E$ drops rapidly.  Because the scattering is dominated by the energetic tail of the phase space distribution, this fall-off of the energy transfer rate can be accurately described assuming Maxwell-Boltzmann statistics, thus yielding a Bessel function $K_{2}(M_\phi/T_a)\sim (M_\phi/T_a)^{3/2}\exp(-M_\phi/T_a)$. Note that, in the resonant regime, the energy transfer rate depends more strongly on  the smaller coupling $\mu_b$ than the larger coupling $\mu_a$, and in particular, when $\mu_b \ll \mu_a$, the rate is almost independent of $\mu_a$. Below the resonance, the energy transfer rate drops rapidly until it reaches the low-temperature regime $T_a \ll M_{\phi}$. In this regime, the inflaton can be integrated out of the theory, leaving a constant scattering amplitude, $|\mathcal{M}(s)|^2\approx\mu_a^2\mu_b^2/M\f^4$.  Thus we obtain the $\mathcal{C}_E \propto T_a^5$ behavior in the last line of eq.~\eqref{eqn:CEscalars}. Finally, at temperatures low enough that one or both of the scattering species becomes non-relativistic, the energy transfer rate becomes Boltzmann-suppressed;  we do not include this effect, as we find that  generically the behavior of $\mathcal{C}_E$ below $T_a<M_{\phi}/4$ is inconsequential to determining the final temperature asymmetry.

Finally, we stress that the expression for $\mathcal{C}_E$ given in eq.~\eqref{eqn:CEscalars} is a limiting version that neglects its dependence on $T_b$.  Dependence on $T_b$ can enter in two ways: first, via the backward energy transfer term, and second, from Bose enhancement of $\mathcal{C}_E^f$.  The backward energy transfer term becomes important when $T_b\gtrsim 0.9 T_a$, and as the two sectors approach equilibration the net energy transfer rate rapidly drops.  
The Bose enhancement of the forward energy transfer term is more involved to model.  This Bose enhancement largely serves to increase $\mathcal{C}_E^f$ in the high and low temperature regimes in eq.~\eqref{eqn:CEscalars} with increasing $T_b$.  The middle regime in eq.~\eqref{eqn:CEscalars}, however, is insensitive to the possible Bose enhancement terms, as that regime is effectively described by Maxwell-Boltzmann statistics. As we show below, this last property enables us to obtain analytic predictions of the final temperature asymmetry without needing to keep track of the full behavior of the Bose enhancements.

\begin{figure}
\begin{subfigure}{.5\textwidth}
\includegraphics[width=1.00\textwidth]{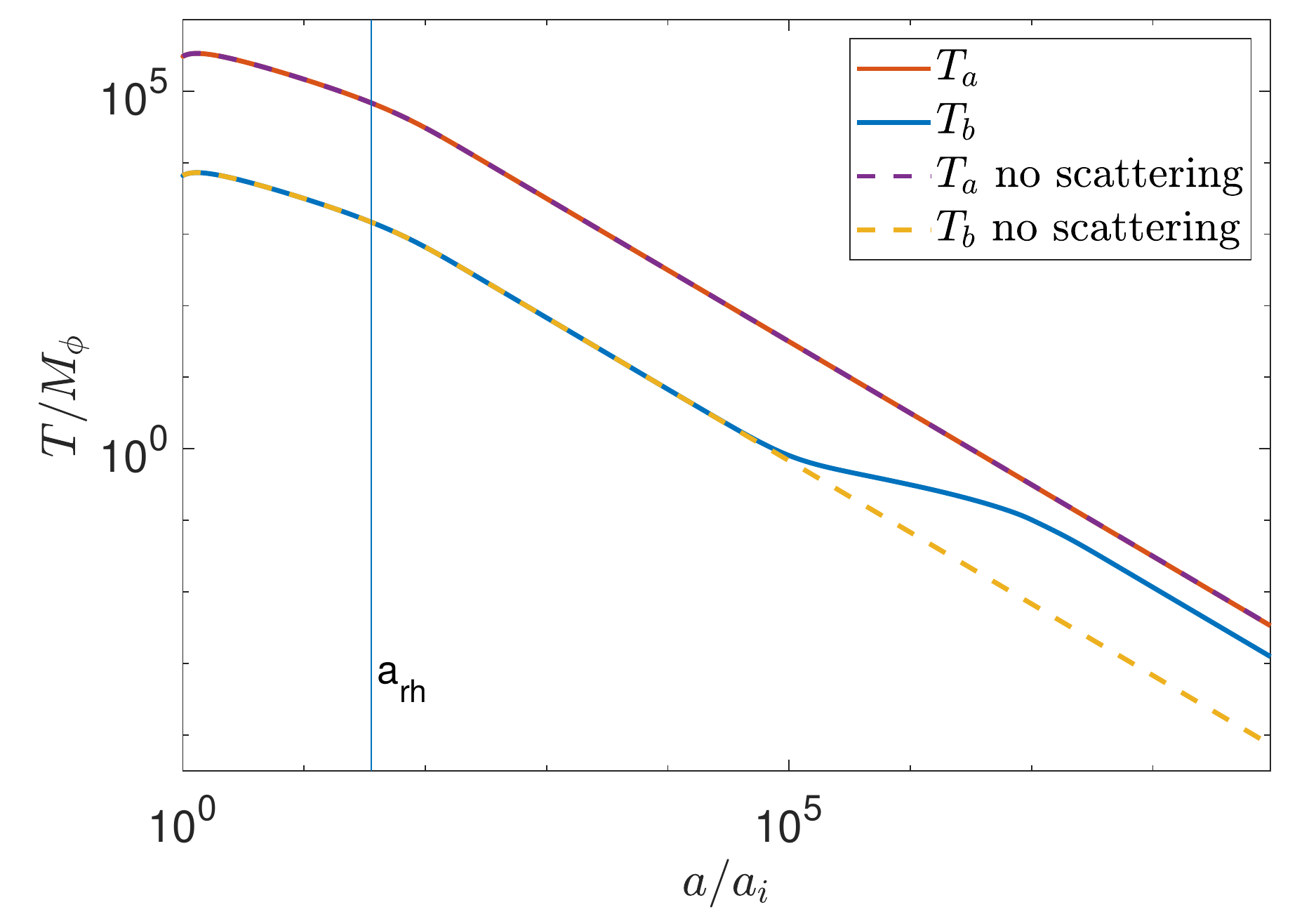}
\end{subfigure}
\begin{subfigure}{.5\textwidth}
\includegraphics[width=1.00\textwidth]{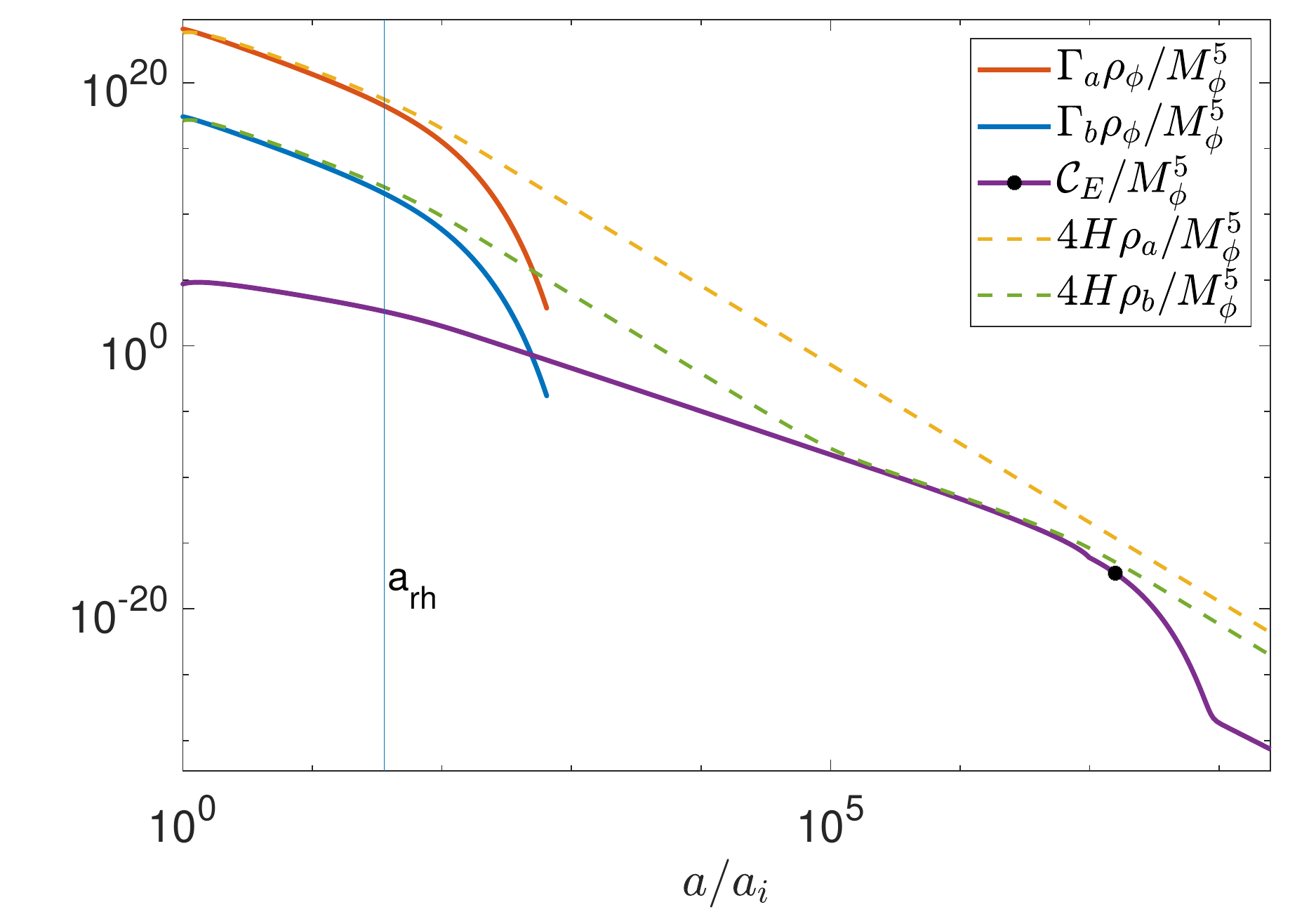}
\end{subfigure}
\caption{\textit{Left panel:} Example temperature evolution of radiation sectors during and after reheating via scalar trilinear couplings. The solid (dashed) lines denote the temperatures obtained from numerically solving the Boltzmann equations including (neglecting) the inflaton-mediated scatterings between the sectors. Reheating is denoted by the vertical blue line at $a_{\textrm{rh}}$. The figure has been plotted for $\mu_a=10^{-3}M\f$, $\mu_b=10^{-5.5}M\f$, $m_{a,b}=10^{-3}M\f$ with other parameters as specified in eq.~\eqref{eq:constants}. \textit{Right panel:} Comparison of the collision term to the redshifting of both sectors as well as inflaton decays for the same parameters as that in left panel. The black dot on the curve for the collision term indicates $T_a= M_{\phi}/4$, which approximately indicates the temperature below which the $s$-channel energy transfer rate becomes exponentially suppressed.  }
\label{fig:scalar_int_temp_evolve_3}
\end{figure}

\subsubsection{The scattering attractor solution}\label{sec:scattering_attractor}

Now we discuss the impact of the collision term on the temperature evolution of both sectors, and derive the corresponding scattering attractor curve for the temperature of the colder sector.  We begin by considering scenarios where $T\rh\gg M\f$, where, as we demonstrate, scattering becomes important post-reheating. One such parameter point is shown in figure~\ref{fig:scalar_int_temp_evolve_3}, which plots numerical solutions for the radiation temperatures obtained using the collision term as given in eq.~\eqref{eq:scalar_fit}. To highlight the importance of scattering, we also show the temperature evolution when the scattering has been turned off. In the right panel, we show the evolution of the collision term, $\mathcal{C}_E$, in comparison to the combinations $4H\rho_{a,b}$ and $\Gamma_{a,b}\rho\f$. We mark the point $T_a=M_{\phi}/4$ around which $\mathcal{C}_E$ begins to exhibit substantial Boltzmann suppression.  As the scattering process affects the temperature evolution substantially post-reheating in this example, we can cleanly separate the effects of scattering  from the contributions of reheating; in this discussion, reheating itself is only important insofar as it provides initial conditions for the subsequent post-reheating evolution of $T_a$ and $T_b$.

As figure~\ref{fig:scalar_int_temp_evolve_3} shows, $T_b$ begins to deviate from the no-scattering solution as soon as the fractional energy transfer rate  into the colder sector becomes comparable to the Hubble rate, $\Gamma_{E,b}=\mathcal{C}_E/\rho_b \sim H$.   When this happens we say that inflaton-mediated scattering becomes \textit{effective}. In contrast, when the fractional energy transfer rate out of the hotter sector becomes comparable to the Hubble rate, $\Gamma_{E,a}=\mathcal{C}_E/\rho_a \sim H$, inflaton-mediated scattering becomes \textit{efficient} and the two sectors attain thermal equilibrium. In the scenario shown in figure~\ref{fig:scalar_int_temp_evolve_3}, inflaton-mediated scattering becomes effective but never efficient. 
The solution to the Boltzmann equation for $T_b$ when scattering becomes effective is approximated by the quasi-static attractor solution (see appendix \ref{appendix:attractor})
\begin{align}\label{eq:twosectorattractor}
\rho_b (T_b) =\frac{1}{4+\frac{q}{1-p}}\frac{\mathcal{C}_E(T_a,T_b)}{H(T_a)},
\end{align}
where
 \begin{align}
p(a)=\dfrac{\partial \ln (\mathcal{C}_E/H)}{\partial \ln\rho_b}, \qquad \text{and}\qquad q(a)=\dfrac{\partial \ln (\mathcal{C}_E/H)}{\partial \ln a}.
 \end{align}
We call this evolution of $\rho_b$ the \textit{scattering attractor curve}. 
In evaluating $q(a)$, the scale factor dependence in $\mathcal{C}_E/H$ comes through $T_a$, which in the present scenario is evolving adiabatically.  For a given value of $T_a$, there is a single corresponding value of $T_b$ that satisfies  eq.~\eqref{eq:twosectorattractor}.  In general, solving eq.~\eqref{eq:twosectorattractor} for $T_b$ is non-trivial given the dependence of $\mathcal{C}_E$ on $T_b$ through Bose enhancement. 
The attractor curve exists as long as $4+q/(1-p)>0$, which translates to the condition that $\mathcal{C}_E$ falls off more slowly with scale factor than $Ha^{-4}$.  At temperatures below $T\sim M_{\phi}/4$ the collision term falls off exponentially (eq.~\eqref{eqn:CEscalars}), marking the end of the attractor evolution. Beyond that point, $\rho_b$ evolves adiabatically as seen in figure~\ref{fig:scalar_int_temp_evolve_3}.  Thus the scattering attractor curve yields a final temperature asymmetry simply given by the asymmetry at  $T_a\approx M_{\phi}/4$.

\begin{figure}
\begin{subfigure}{.5\textwidth}
\includegraphics[width=1.00\textwidth]{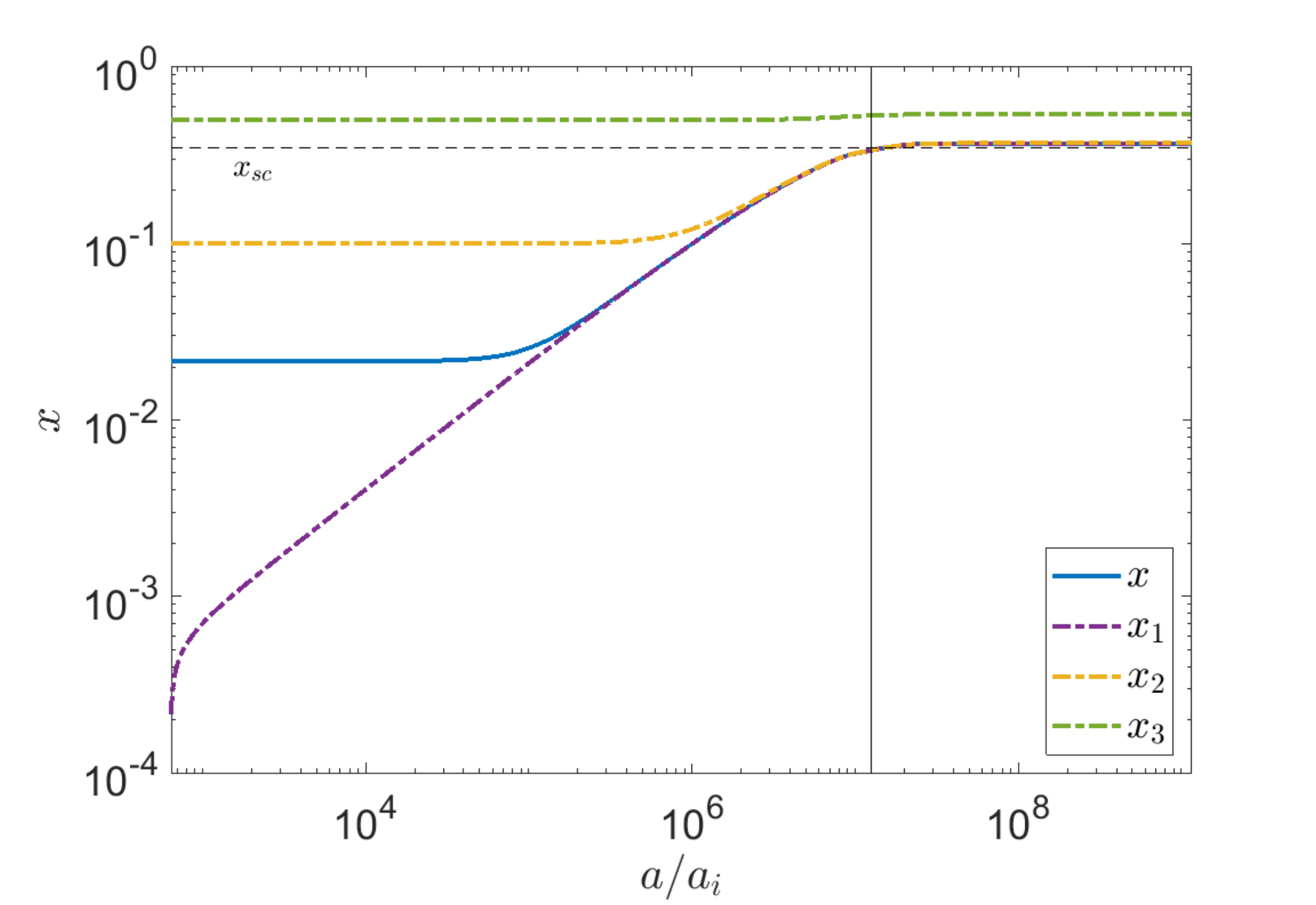}
\end{subfigure}
\begin{subfigure}{.5\textwidth}
\includegraphics[width=1.00\textwidth]{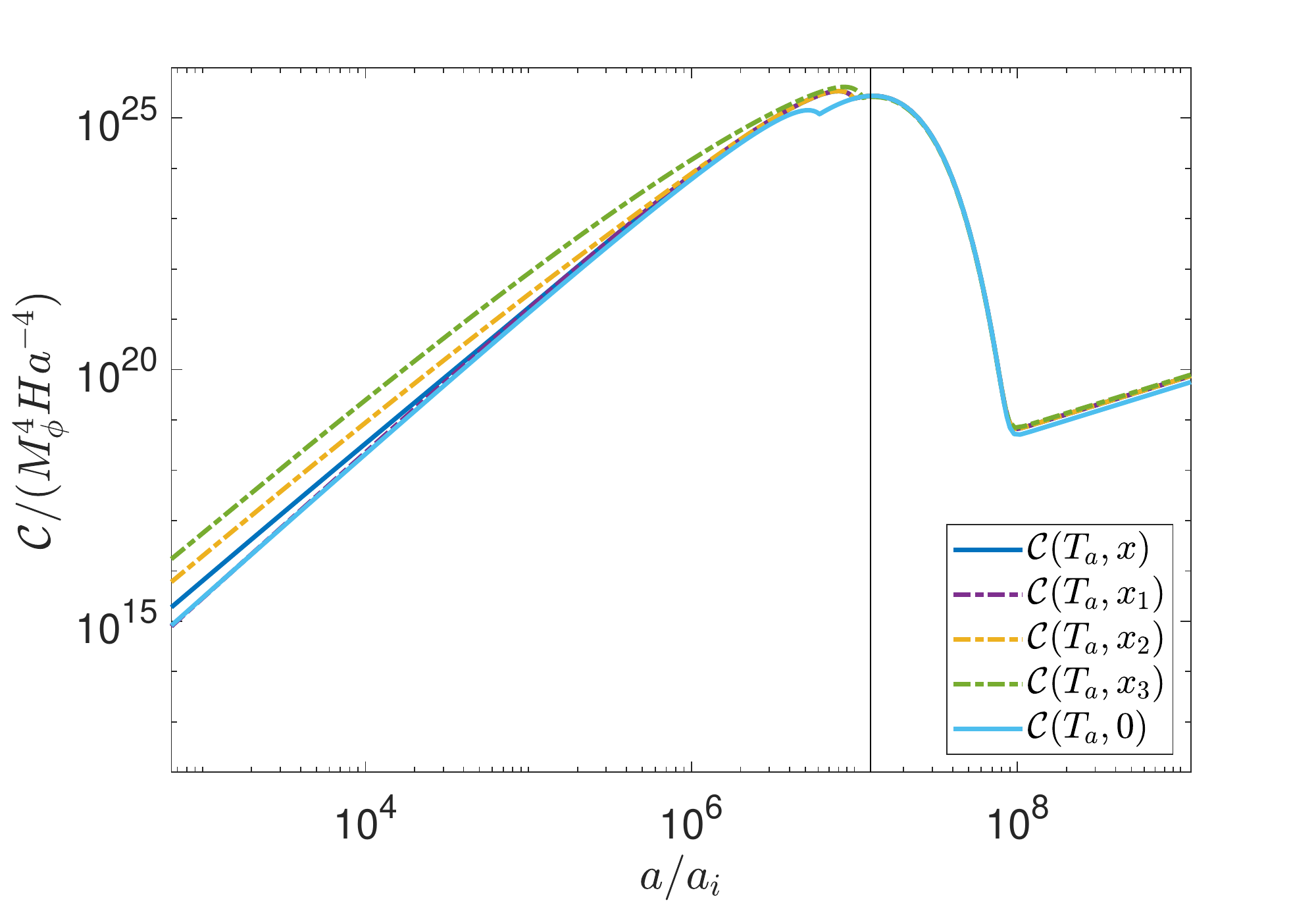}
\end{subfigure}
\caption{\textit{Left panel:} Evolution of post-reheating temperature ratios for several different initial conditions. The solid blue line shows the evolution $x=T_b/T_a$ from figure~\ref{fig:scalar_int_temp_evolve_3}. The dot-dashed lines ($x_1,\ x_2$ and $x_3$) indicate the temperature ratio evolution after artificially varying the initial condition for $T_b$ ($x_{1,i}=0$, $x_{2,i}=0.1$ and $x_{3,i}=0.5$). The horizontal black dashed line corresponds to the analytic estimation of the final temperature ratio derived in eq.~\eqref{eq:scalar_xfscat}. \textit{Right panel:} The collision terms (normalized by $M\f^4 Ha^{-4}$) experienced by the colder sectors plotted on the left panel. The solid light blue line represents the evolution of the collision term when the $T_b$-dependence of $\mathcal{C}_E$ is neglected. The vertical black line in both panels indicates the scale where $T_a=M_{\phi}/4$.}
\label{fig:collision_attractor}
\end{figure}

To further highlight the attractor nature of the collision term, figure~\ref{fig:collision_attractor}  shows the post-reheating evolution of the temperature ratio,  $x=(\alpha_a\rho_b/(\alpha_b\rho_a))^{1/4}$, for the parameter point of figure~\ref{fig:scalar_int_temp_evolve_3}, but now considering a range of (post-reheating) initial conditions for $\rho_b$ (or equivalently $x$).   In the left panel, the solid blue line tracks the evolution of $T_b/T_a$ following from figure~\ref{fig:scalar_int_temp_evolve_3}, where the initial conditions are determined self-consistently from inflaton decays, $x_i=x\rh=0.02$. The purple dot-dashed line shows the evolution when the initial temperature ratio is instead zero, $x_{1,i}=0$; again, initial densities below the attractor solution rise rapidly to attain the attractor. The yellow dot-dashed line shows the evolution with an initial temperature ratio $x_{2,i}=0.1> x\rh$; this solution still attains the scattering attractor curve, eq.~\eqref{eq:twosectorattractor}.   The green dot-dashed line denotes evolution with an initial temperature ratio, $x_{3,i}=0.5$, much above the final temperature ratio determined by the scattering attractor solution. In this case $T_b$ remains mostly unaffected by inflaton-mediated interactions. Thus we see that the inflaton-mediated interactions impose a minimum value for the final temperature ratio: any initial temperature ratio  below this minimum value is increased to that value by the scattering attractor solution while initial temperature ratios above this minimum remain largely unaffected.  This minimum final temperature ratio is simply determined by the behavior of $\mathcal{C}_E$ near $T_a\sim M_{\phi}/4$, as we elaborate below.

The right panel of figure~\ref{fig:collision_attractor} shows the ratio $\mathcal{C}_E(T_a, x)/(H a^{-4})$ for all scenarios.  The different behavior of the curves at early times shows the differing importance of Bose enhancement on the final state  in the different scenarios.  All curves converge onto a single common solution for $T_a\sim M\f$, when scattering is well described by Maxwell-Boltzmann statistics.   As $H\rho_a$ evolves with scale factor as $Ha^{-4}$, the value of $a$ where $\mathcal{C}_E/Ha^{-4}$ is maximized\footnote{The  visible hiccup near the peak in $\mathcal{C}_E(T_a, 0)$ is a feature of the imperfect fitting used in our analytic approximation of $\mathcal{C}_E$, eq. \eqref{eq:scalar_fit}.} coincides with the value of $a$ where the scattering attractor solution for $\rho_b$ ends.

To accurately determine the final temperature asymmetry predicted by the scattering attractor-curve, we need to integrate the Boltzmann equations around the region where $T_a\lesssim M\f$.  Assuming no $T_b$ dependence in $\mathcal{C}_E$ or $H$, the Boltzmann equation for $\rho_b$ can be directly integrated,
\begin{align}\label{eq:integrating_near_off_resonance}
{\rho}_b a^4-({\rho}_b a^4)_{a=a_1}&=\int_{a_1}^a\frac{a'^3{\mathcal{C}}_E({T}_a)}{{H}}da'
=a_1^4\int_{1}^{a/a_1}\frac{z^5\sqrt{3}{M}_{\textrm{Pl}}{\mathcal{C}}_E(M\f/z)}{\alpha_a^{1/2}M\f^2}dz,
\end{align}
where we have defined $a_1$ as the scale factor at which $T_a(a_1)=M\f$, and $z=a'/a_1$. We evaluate $\mathcal{C}_E$ by taking the limit $x\rightarrow 0$ around $T_a\sim M\f$ in eq.\ \eqref{eq:scalar_fit} to obtain
\begin{align}
{\rho}_b a^4-({\rho}_b a^4)_{a=a_1}  & =a_1^4\bigg(\frac{2}{\pi^2}M\f^2\dfrac{\Gamma_{0,a}\Gamma_{0,b}}{\Gamma_{0,a}+\Gamma_{0,b}}\frac{\sqrt{3}{M}_{\textrm{Pl}}}{\alpha_a^{1/2}}\bigg)\nonumber\\&\qquad\times\int_{1}^{a/a_1}\!\!\!\textrm{max}\Big\{z^2\Big(1.3-1.6\log(z)\Big),\frac{1}{4}z^4K_2(z)\Big\}dz\nonumber\\
&\xrightarrow{a/a_1 \textrm{large}}6.3 a_1^4  \bigg(\frac{2}{\pi^2} M\f^2\dfrac{\Gamma_{0,a}\Gamma_{0,b}}{\Gamma_{0,a}+\Gamma_{0,b}}\frac{\sqrt{3}{M}_{\textrm{Pl}}}{\alpha_a^{1/2}}\bigg).
\end{align}
We determine the initial energy density of the colder sector, $({\rho}_b a^4)_{a=a_1}$,  by assuming that the colder sector is already on the scattering attractor curve defined by evaluating eq.~\eqref{eq:twosectorattractor} using $\mathcal{C}_E= \mathcal{C}_E(T_a, 0)$. However, the Maxwell-Boltzmann behavior of the collision term in this temperature range helps to ensure that the final result is insensitive to the specific choice of $x=0$. We find
\begin{align}
{\rho}_{b}(a_1)=0.71  \bigg(\frac{2}{\pi^2} M\f^2\dfrac{\Gamma_{0,a}\Gamma_{0,b}}{\Gamma_{0,a}+\Gamma_{0,b}}\frac{\sqrt{3}{M}_{\textrm{Pl}}}{\alpha_a^{1/2}}\bigg).
\end{align}
The final energy density of the colder sector is then given by
\begin{align}
\br{\rho}_b(a)=7.0\Big(\frac{a_1}{a}\Big)^4 \bigg(\frac{2}{\pi^2} M\f^2\dfrac{\Gamma_{0,a}\Gamma_{0,b}}{\Gamma_{0,a}+\Gamma_{0,b}}\frac{\sqrt{3}{M}_{\textrm{Pl}}}{\alpha_a^{1/2}}\bigg).
\end{align}
The final temperature ratio between the two sectors predicted by inflaton-mediated scattering is then 
\begin{align}\label{eq:scalar_xfscat}
x_{sc} \equiv \left(\frac{\alpha_a\rho_b}{\alpha_b\rho_a}\right)^{1/4}_{a>a_1}= 1.25 \bigg(\frac{1}{M\f^2}\dfrac{\Gamma_{0,a}\Gamma_{0,b}}{\Gamma_{0,a}+\Gamma_{0,b}}\frac{{M}_{\textrm{Pl}}}{\sqrt{\alpha_a}\alpha_b}\bigg)^{1/4},
\end{align}
where we have used $\rho_a(a)=\alpha_a M\f^4(a_1/a)^4$.
This is the value that the temperature ratios $x, x_1$ and $x_2$ asymptote to as shown by the horizontal black dashed line in figure~\ref{fig:collision_attractor}. 
Eq.\ \eqref{eq:scalar_xfscat} holds as long as $x_{sc}\lesssim 0.9$. Once the temperature ratio approaches unity, backward energy transfer and the contribution of $\rho_b$ to the Hubble parameter become important, and the attractor solution no longer captures the full behavior of the system.  In these cases, where the two sectors approach thermalization, a more detailed numerical study is required.

Finally, it is worth emphasizing that the scattering attractor curve discussed here is dominated by the resonant behavior of the energy transfer rate, and depends on the properties of the radiation baths at $T_a \sim M\f$.  In the trilinear scalar model, a second attractor phase appears  at temperatures well below the resonance ($T_a\ll M\f$).
This is evident from the late-time increase in $\mathcal{C}_E/(H a^{-4})$ in the right panel of figure~\ref{fig:collision_attractor}, after the resonant enhancement ends.  This possibility of IR thermalization is a special feature of the trilinear scalar model, where integrating out the inflaton introduces a renormalizeable quartic interaction between the two sectors. In all other cases $\mathcal{C}_E$ falls off much faster at lower temperatures due to the higher ($\geq 4$) mass dimensions of the operators that couple the inflaton to the radiation baths.  
 Once $T_{a,b}\sim m_{a,b}$,  $\mathcal{C}_E$ becomes exponentially suppressed and scattering is cut off. Thus, thermalization in the IR depends on the mass scales in the matter sectors coupled to the inflaton, as well as the inflaton mass and $T_{\rm rh}$.  Late-time equilibration through scalar portal interactions is studied in detail in \cite{Krnjaic:2015mbs,Adshead:2016xxj,Evans:2017kti}  and we do not discuss it further here.

\subsubsection{Final temperature ratios}

\begin{figure}
\begin{subfigure}{.5\textwidth}
\includegraphics[width=1.00\textwidth]{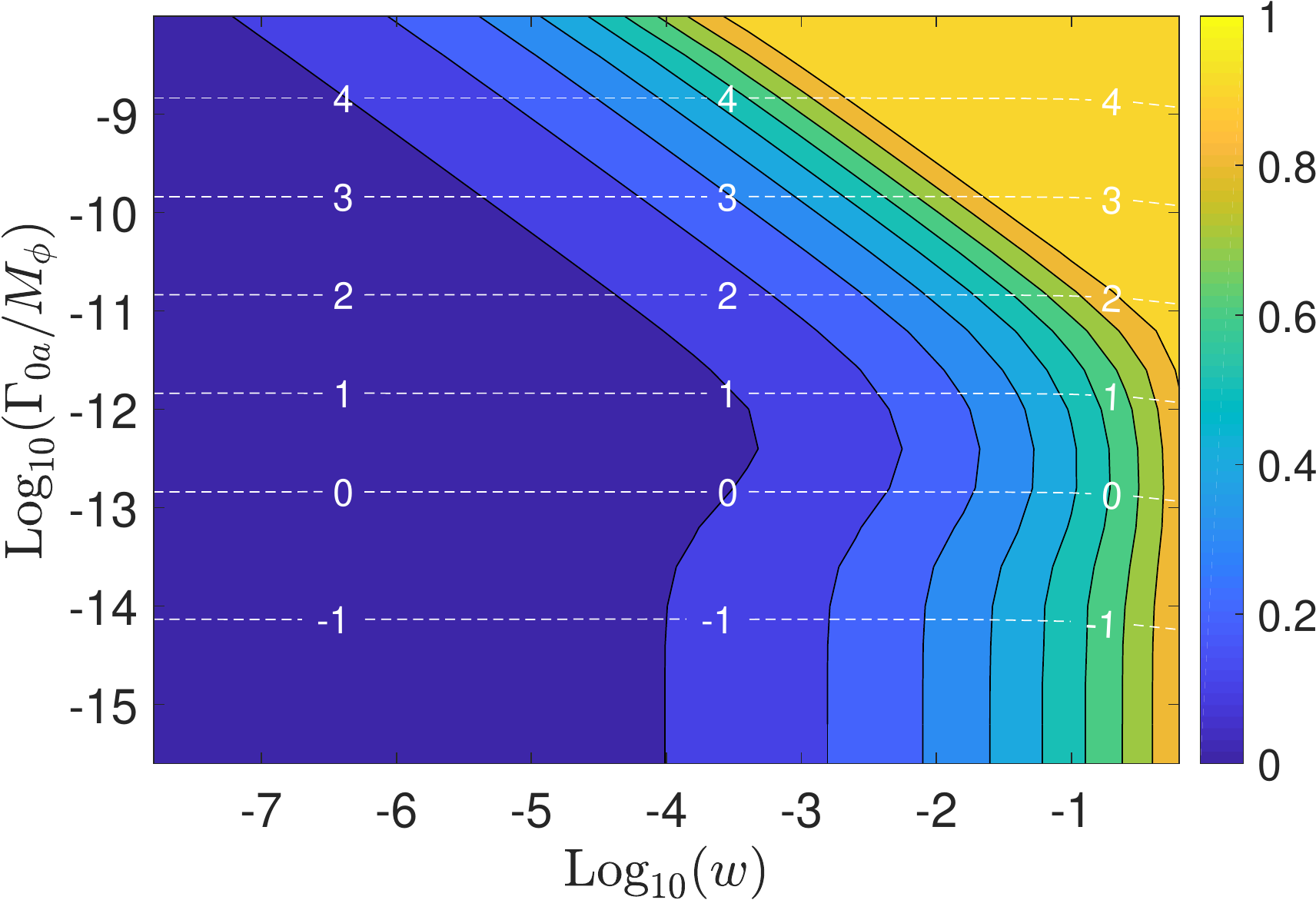}
\end{subfigure}
\begin{subfigure}{.5\textwidth}
\includegraphics[width=1.00\textwidth]{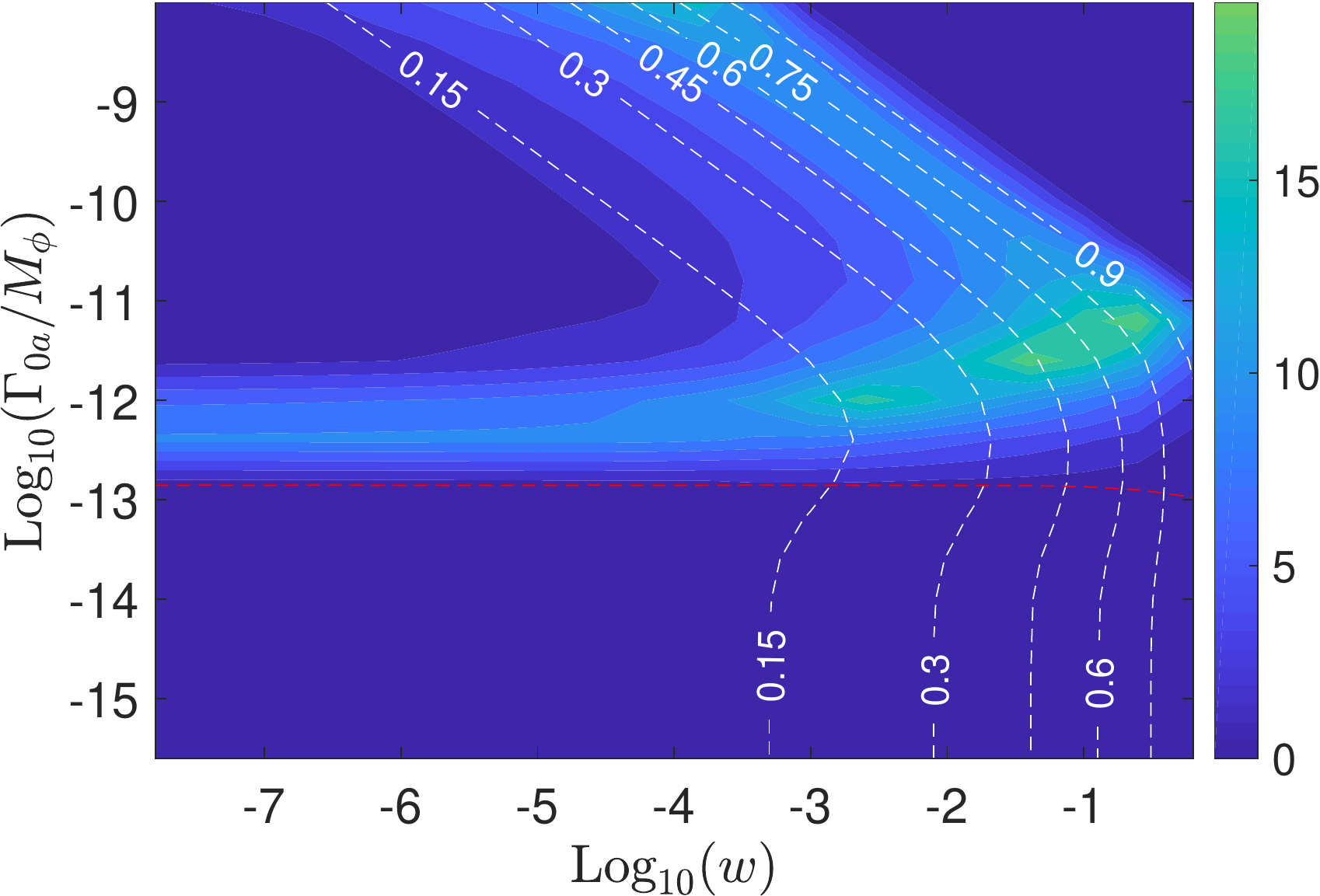}
\end{subfigure}
\caption{\emph{Left panel}: Contours of the final temperature ratio obtained numerically, $x_{f,n}=T_b/T_a$, shown in color, for the case when the inflaton is coupled to scalars in both  sectors. The white dashed lines show contours of $\log_{10}(T_{\textrm{rh}}/M\f)$. \emph{Right panel}: Contours of relative fractional discrepancy, $|x_{f,a}-x_{f,n}|/x_{f,n}$, where $x_{f,a}$ is the analytic estimate of eq.~\eqref{eq:x_predict} with   eq.~\eqref{eq:scalar_xfscat}, in percent. The white dashed contours depict $x_{f,n}$ and the red dashed line marks the region where $T_{\textrm{rh}}=M\f$.}
\label{fig:xfscat_scalar}
\end{figure}

Both reheating and scattering, considered independently, produce attractor solutions.  In most of parameter space, one attractor solution dominates over the other and thus is primarily responsible for determining the  final temperature asymmetry.  As demonstrated above in section~\ref{sec:scattering_attractor}, when $T_{\rm rh} > M\f$, a good semi-analytic approximation to the final temperature asymmetry is therefore
\begin{equation}
\label{eq:x_predict}
x_f = \max [ x\rh, x_{sc}, 1 ].
\end{equation}
Both $x\rh$ and $x_{sc}$ can be straightforwardly computed from the Lagrangian parameters without any need to solve the full Boltzmann equations.

In the case $T\rh\ll M\f/4$, the reheating attractor solution dominates, as we now show.  As $\mathcal{C}_E$ redshifts more slowly than $\Gamma_b \rho_\phi$  during reheating, it suffices to show that $\mathcal{C}_E/\Gamma_b\rho\f < 1$ at $T_a\approx M\f/4$ when $\mathcal{C}_E$ is maximized.  Using eq.~\eqref{eq:attractor_propto} for $\rho\f$ at $T_a=M\f/4$ and  eq.~\eqref{eq:scalar_MB} for $\mathcal{C}_{E}(T_a=M\f/4)$ (at $x=0$) we find
\begin{align}
\left(\frac{\mathcal{C}_{E}}{\Gamma_{0b}\rho\f}\right)_{T_a=M\f/4} \approx \left(\frac{25\pi^2\alpha_a}{3\times 2^{15}K_2(4)}\right)^{1/4}\frac{T\rh}{M\f} \approx  \left(\frac{\alpha_a}{10}\right)^{1/4}\frac{T\rh}{M\f}.
\end{align}
This ratio is small by assumption for SM-scale values of $\alpha_a$.
 Hence for low reheat temperatures inflaton-mediated scattering is unimportant during reheating.  After reheating, scattering cannot thermalize the two sectors as the resonant enhancement has already ended. Thus, for $T\rh \ll M\f$ the final temperature asymmetry is simply given by $x\rh=(\alpha_a\Gamma_{0b}/\alpha_b\Gamma_{0a})^{1/4}$. Although $x_{sc}$, as defined in eq.~\eqref{eq:scalar_xfscat}  as the temperature ratio obtained by the {\em post-reheating} scattering attractor curve, does not strictly pertain in this case, one can check that its value is always less than $x\rh$ when $T\rh\ll M\f/4$. Thereby we can extend eq.~\eqref{eq:x_predict} to hold for $T\rh \ll M\f/4$ as well.

We are now ready to consider the full numerical solution to the Boltzmann equations, eq.~\eqref{eqn:boltzwints}.  
The resulting numerical  temperature asymmetry, $x_{f,n}$, is shown in the left panel of figure~\ref{fig:xfscat_scalar} as a function of the ratio of the zero-temperature widths $w =\Gamma_{0,b}/ \Gamma_{0,a}$. 

As expected, for low reheat temperatures the reheating attractor curve dominates the evolution.  The contours below $T\rh\lesssim M\f$ show the same behavior as in the absence of inflaton-mediated scattering; compare the left panel of figure~\ref{fig:xfnoscat}. Inflaton-mediated scattering becomes important roughly for $T\rh\gtrsim M\f$, and dominates for $T\rh\gtrsim 10M\f$.  In this high-temperature regime the contours are almost diagonal, reflecting the fact that $x_{sc}$ is dominantly governed by the smaller decay width $\Gamma_{0,b}$.  In the right panel of figure~\ref{fig:xfscat_scalar} we compare our analytic estimate of eq.~\eqref{eq:x_predict}\footnote{The temperature ratio $x\rh$ resulting from the reheating attractor curve used in eq.~\eqref{eq:x_predict} is evaluated by solving eq.~\eqref{eq:boltz_2sector} numerically.} with the result obtained from numerically solving the Boltzmann equations.  The analytic estimate agrees with the numerical results within 20\%. The discrepancies are greatest exactly where the scattering and reheating attractor curves are no longer individually sufficient to capture the full behavior of the system: when both scattering and inflaton decays are important for determining the final asymmetry, around $T\rh\sim \,\mathrm{few}\,M\f$, and when the sectors are approaching (but not obtaining) thermalization, $x_f \sim 0.7-0.8$.  

Finally, let us note the  important point that our numerical results for $x_{f,n}$ are themselves based on analytic approximations to the collision term.  Our analytic fits to the collision term deviate from the exact numerical values by almost 50\% near $T\sim M\f$ (see figure~\ref{fig:scalar_boson_analytical_compare}). As the final temperature ratio is predominantly determined by the behavior of the collision term near $T\sim M\f$, this error is unfortunately not negligible for our final results. However, this error is made less numerically consequential once we take the fourth root to find the temperature (eq.~\eqref{eq:scalar_xfscat}), inducing uncertainties of up to $\sim 15\%$ in the numerical temperature ratio plots at high $T\rh$, figure~\ref{fig:xfscat_scalar}.

\subsection{Final temperature asymmetry for other theories}

The two key properties of $\mathcal{C}_E$---the exponential suppression at $T_a\lesssim M_{\phi}/4$ and the weak dependence on $T_b$ in this range---that allowed us to analytically determine the final temperature asymmetry for the scalar trilinear case are generic features of resonant $s$-channel interactions.  Much  of our analysis in the previous section can thus be applied directly to other interaction structures. As we demonstrate, in models where the inflaton has renormalizeable couplings to matter, scattering is only important for determining the final temperature asymmetry when the endpoint of the scattering attractor curve occurs post-reheating.   However, scattering during reheating can also be important when the inflaton is a pseudoscalar with dimension-five couplings to gauge fields in both sectors.

\subsubsection{Yukawa couplings}
\label{sec:fermionsbothsectors}

We begin with a model where the inflaton has Yukawa couplings to fermions in both sectors,
\begin{eqnarray}
\mathcal{L}_{\textrm{int}}=y_a\phi\bar{\psi}_a\psi_a+y_b\phi\bar{\psi}_b\psi_b.
\end{eqnarray}
This interaction results in zero-temperature inflaton decay widths given by
\begin{eqnarray}
\Gamma_{0a,b}=\frac{y_{a,b}^2}{8\pi}M\f\sqrt{1-\frac{4m_{a,b}^2}{M\f^2}}\approx \frac{y_{a,b}^2}{8\pi} M\f,
\end{eqnarray}
where $m_{a,b}\ll \Mp$ denotes the mass of fields $\psi_{a,b}$. The $s$-channel spin-summed scattering amplitude between the two species is
\begin{align}
|\overline{\mathcal{M}}(s)|^2=4y_a^2y_b^2\bigg(1-\frac{4m_a^2}{s}\bigg)\bigg(1-\frac{4m_b^2}{s}\bigg)\frac{s^2}{(s-M^2_{\phi})^2+(\Gamma_{0a}+\Gamma_{0b})^2}.
\end{align}
The total energy transfer collision term, $\mathcal{C}_E$, following from this amplitude is discussed in appendix~\ref{appendix:s_fermions} and shown in figure \ref{fig:fermion_analytical_compare}. Unlike the scalar case discussed in section \ref{sec:scalarbothsectors}, the collision term is almost insensitive to the temperature of the colder sector unless the two temperatures are very close and $\mathcal{C}_E^b$ becomes important. In the limit that the temperature ratio between the two sectors is very small, $x\ll 1$, $\mathcal{C}_E$ is approximately given by 
\begin{align}
\mathcal{C}_E=\frac{1}{4\pi^3}\times\begin{cases}
\dfrac{3.0}{2\pi^2}y_a^2y_b^2 T_a^5 & T_a\gg M\f\\
\\
0.29\dfrac{y^2_ay^2_b}{y_a^2+y_b^2}M\f^2 T_a^3& T_a\gtrsim M_{\phi}
\\
\\
\dfrac{y^2_ay^2_b}{y_a^2+y_b^2}M\f^4\dfrac{T_a}{4} K_2\Big(\frac{M_{\phi}}{T_a}\Big) & T_a\lesssim M_{\phi}
\\
\\
\dfrac{1.4\times 10^3}{2\pi^2}y_a^2y_b^2 \dfrac{T_a^9}{M\f^4} & m_{a,b}\ll T_a\ll M_{\phi}.
\end{cases}
\end{align}
At temperatures much larger than the inflaton mass, the inflaton mass can be neglected and the scattering amplitude is approximately constant, $|\overline{\mathcal{M}}(s)|^2\approx y_a^2y_b^2$, yielding the $\mathcal{C}_E \propto T^5$ behavior required from dimensional analysis. At temperatures closer to the inflaton mass, the energy transfer rate is resonantly enhanced, yielding $\mathcal{C}_E \propto T^3$ behavior. As the temperature drops below the inflaton mass, the energy transfer rate is dominated by resonant scattering in the Boltzmann-suppressed tails. Analogously to the scalar case, $\mathcal{C}_E$ can be well modeled in this region using Maxwell-Boltzmann statistics. In the low temperature regime the scattering amplitude can be approximated as $|\overline{\mathcal{M}}(s)|^2\approx y_a^2y_b^2s^2/M\f^4$, yielding the steep $\mathcal{C}_E \propto T^9$ behavior.  Note that, like the scalar trilinear case, the energy transfer rate  depends most strongly on the smaller coupling in the resonant regime.

\begin{figure}
\begin{subfigure}{.5\textwidth}
\includegraphics[width=1.00\textwidth]{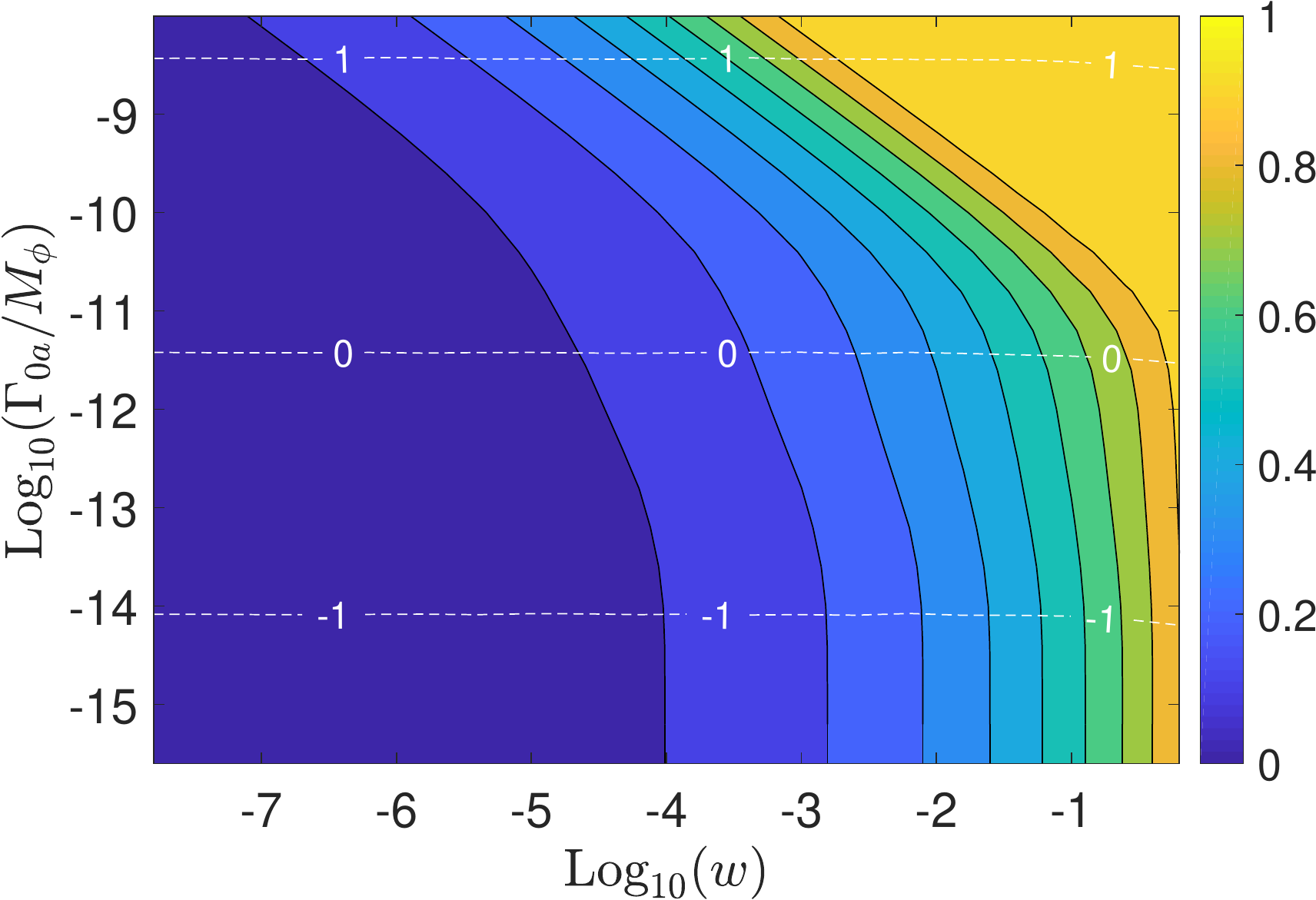}
\end{subfigure}
\begin{subfigure}{.5\textwidth}
\includegraphics[width=1.00\textwidth]{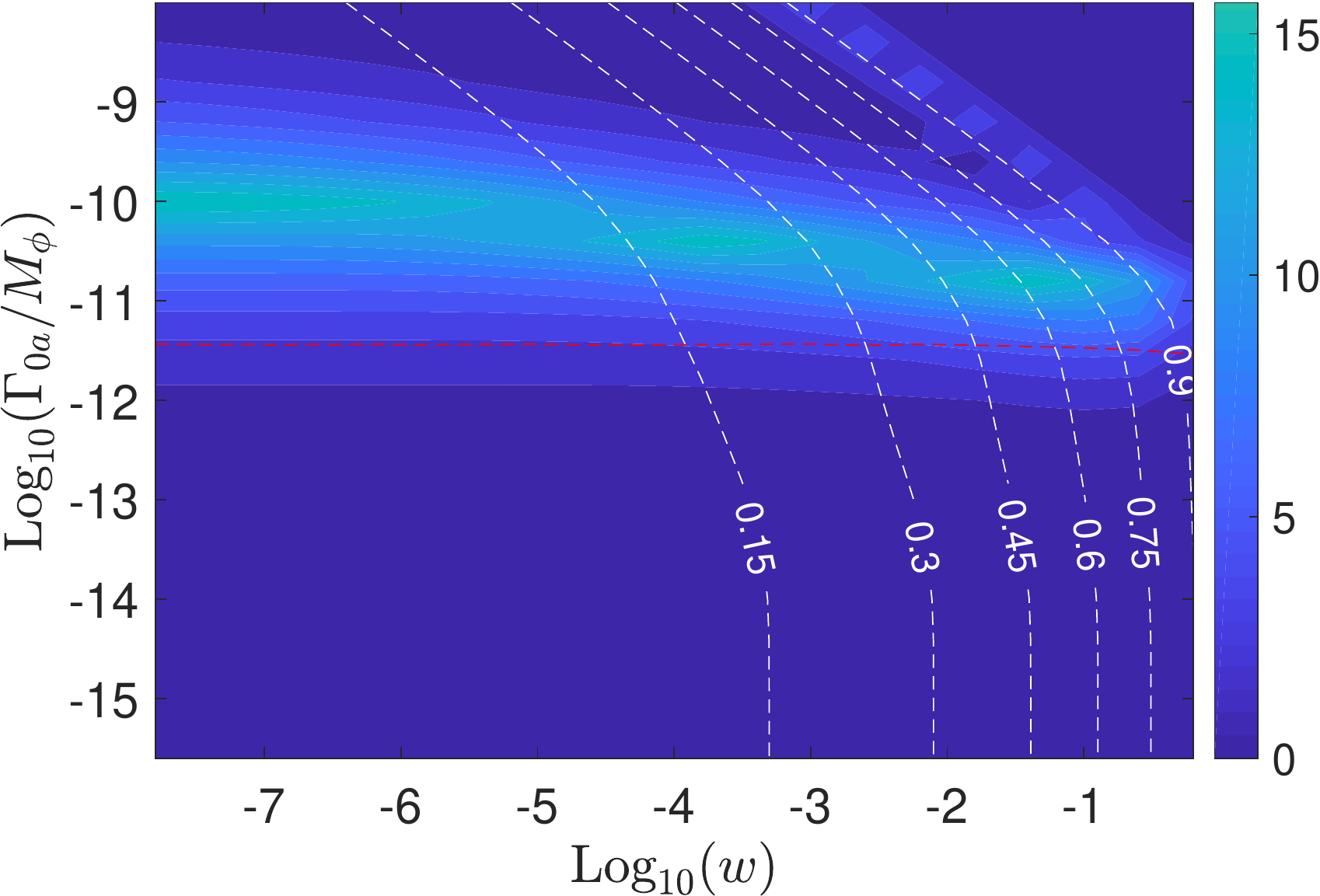}
\end{subfigure}
\caption{\emph{Left panel}: Contours of the final temperature ratio obtained numerically, $x_{f,n}=T_b/T_a$, shown in color, for an inflaton with Yukawa couplings to fermions in both  sectors. The white dashed lines show contours of $\log_{10}(T_{\textrm{rh}}/M\f)$. \emph{Right panel}: Contours of relative fractional discrepancy, $|x_{f,a}-x_{f,n}|/x_{f,n}$, where $x_{f,a}$ is the analytic estimate of eq.~\eqref{eq:x_predict} with  eq.~\eqref{eq:fermion_xfscat}, in percent. The white dashed contours depict $x_{f,n}$ and the red dashed line marks $T_{\textrm{rh}}=M\f$.}
\label{fig:xfscat_fermion}
\end{figure}

We can again obtain an analytic expression for the final temperature asymmetry due to inflaton-mediated scattering, as we did for scalars in section~\ref{sec:scattering_attractor}.
Using $\mathcal{C}_E$ from eq.~\eqref{eq:fermion_fit} and taking $x\rightarrow0$, we obtain
\begin{align}\label{eq:fermion_xfscat}
x_{sc}=1.19 \bigg(\frac{1}{M\f^2}\dfrac{\Gamma_{0,a}\Gamma_{0,b}}{\Gamma_{0,a}+\Gamma_{0,b}}\frac{{M}_{\textrm{Pl}}}{\alpha_b\sqrt{\alpha_a}}\bigg)^{1/4}.
\end{align}
The final temperature asymmetry can then be estimated using eq.~\eqref{eq:x_predict}, i.e., by comparing the
lower bounds from the scattering and reheating attractor solutions. In figure~\ref{fig:xfscat_fermion} we show  numerical final temperature ratios in the left panel and in the right panel compare our analytic estimate to the numerical results.   We again observe a transitional region around $T_{\rm rh}\sim \,\mathrm{few}\times M\f$ where both reheating and scattering are  important for determining the final value of $x_f$.  Note that the analytic estimate from the scattering attractor curve  has better agreement with the numerical results in the region near thermalization, $x_f\to 1$, than we saw for the scalar case; this is because the Fermi blocking of $C_E^f$ that occurs here is nowhere near as large an effect as the Bose enhancement we discussed in the previous subsection.

\subsubsection{Axionic couplings to gauge bosons}

We next consider a theory where a pseudo-scalar inflaton couples to gauge bosons in both sectors,
\begin{eqnarray}
\mathcal{L}_{\textrm{int}}=-\frac{1}{4\Lambda_a}\phi F_a^{\mu\nu}\tilde{F}_{a,\mu\nu}-\frac{1}{4\Lambda_b}\phi F_b^{\mu\nu}\tilde{F}_{b,\mu\nu}.
\end{eqnarray}
This interaction results in zero-temperature decay widths given by
\begin{eqnarray}\label{eq:gauge_boson_decays}
\Gamma_{0a,b}=\frac{1}{256\pi}\frac{M\f^3}{\Lambda_{a,b}^2}\sqrt{1-\frac{4m_{a,b}^2}{M\f^2}}\approx \frac{M\f^3}{256\pi{\Lambda}_{a,b}^2},
\end{eqnarray}
where $m_{a,b} \ll \Mp$ denotes the mass of the gauge fields, $A^{\nu}_{a,b}$. The $s$-channel  spin-summed amplitude for $A_a A_a \leftrightarrow A_b A_b$ scattering mediated by inflaton exchange is
\begin{align}\label{eqn:gaugeamp}
|\overline{\mathcal{M}}(s)|^2=\frac{4}{128\Lambda_a^2\Lambda_b^2} \bigg(1-\frac{4m_{a}^2}{M_{\phi}^2}\bigg)\bigg(1-\frac{4m_{b}^2}{M_{\phi}^2}\bigg)\frac{s^4}{(s-M^2_{\phi})^2+(\Gamma_{0a}+\Gamma_{0b})^2}.
\end{align}
In appendix~\ref{appendix:s-channel}, we derive the total energy transfer rate, $\mathcal{C}_E$ for this amplitude; see figure~\ref{fig:gauge_boson_analytical_compare}.   
When the temperature ratio between the two sectors is very small, $x \ll 1$, the temperature dependence of $\mathcal{C}_E$ is approximately given by
\begin{align}\label{eq:gauge_boson_collision}
\mathcal{C}_E=\frac{1}{64\pi^3}\times\begin{cases}
\dfrac{14.0}{2\pi^2}\dfrac{1}{\Lambda_a^2\Lambda_b^2} T_a^9 & \Lambda_a>T_a\gg M\f\\
\\
\dfrac{M\f^4}{\Lambda_a^2+\Lambda_b^2}T_a^3\Big[1.6\log\Big(\frac{T_a}{M\f}\Big)+1.3\Big]& T_a\gtrsim M_{\phi}
\\ 
\\
\dfrac{M\f^6}{\Lambda_a^2+\Lambda_b^2}\dfrac{T_a}{4} K_2\Big(\frac{M_{\phi}}{T_a}\Big) & T_a\lesssim M_{\phi}
\\ 
\\
\dfrac{7.1\times10^4}{2\pi^2}\dfrac{1}{\Lambda_a^2\Lambda_b^2} \dfrac{T_a^{13}}{M\f^4} & m_{a,b}\ll T_a\ll M_{\phi}.
\end{cases}
\end{align}
The steep rise in the collision term ($\mathcal{C}_E\propto T_a^9$ ) at high temperatures is a consequence of the high mass-dimension of the operators mediating the interaction. This behavior will be modified when $T_a\gtrsim \Lambda_a$ and the effective field theory breaks down.

\begin{figure}
\begin{subfigure}{.5\textwidth}
\includegraphics[width=1.00\textwidth]{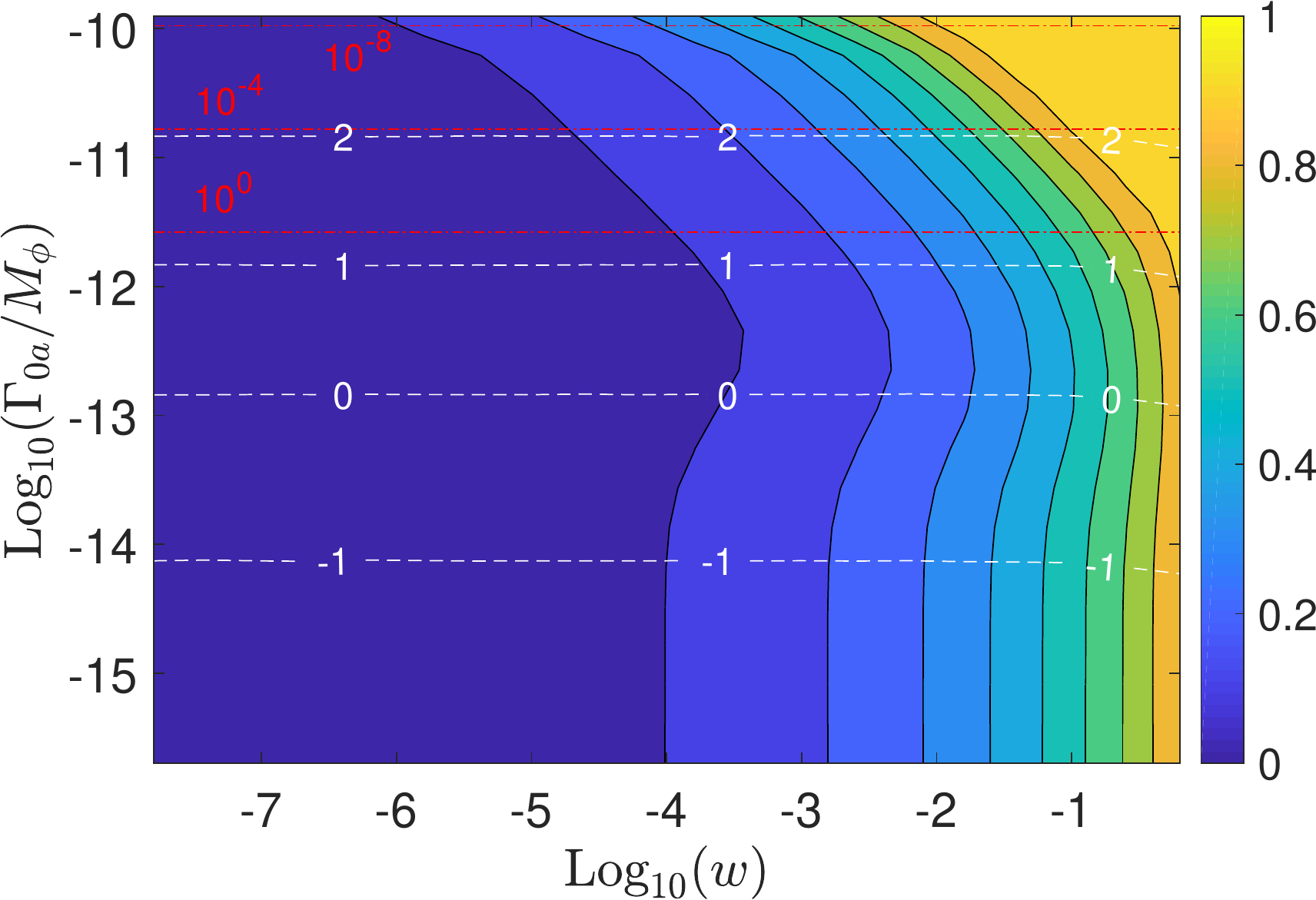}
\end{subfigure}
\begin{subfigure}{.5\textwidth}
\includegraphics[width=1.00\textwidth]{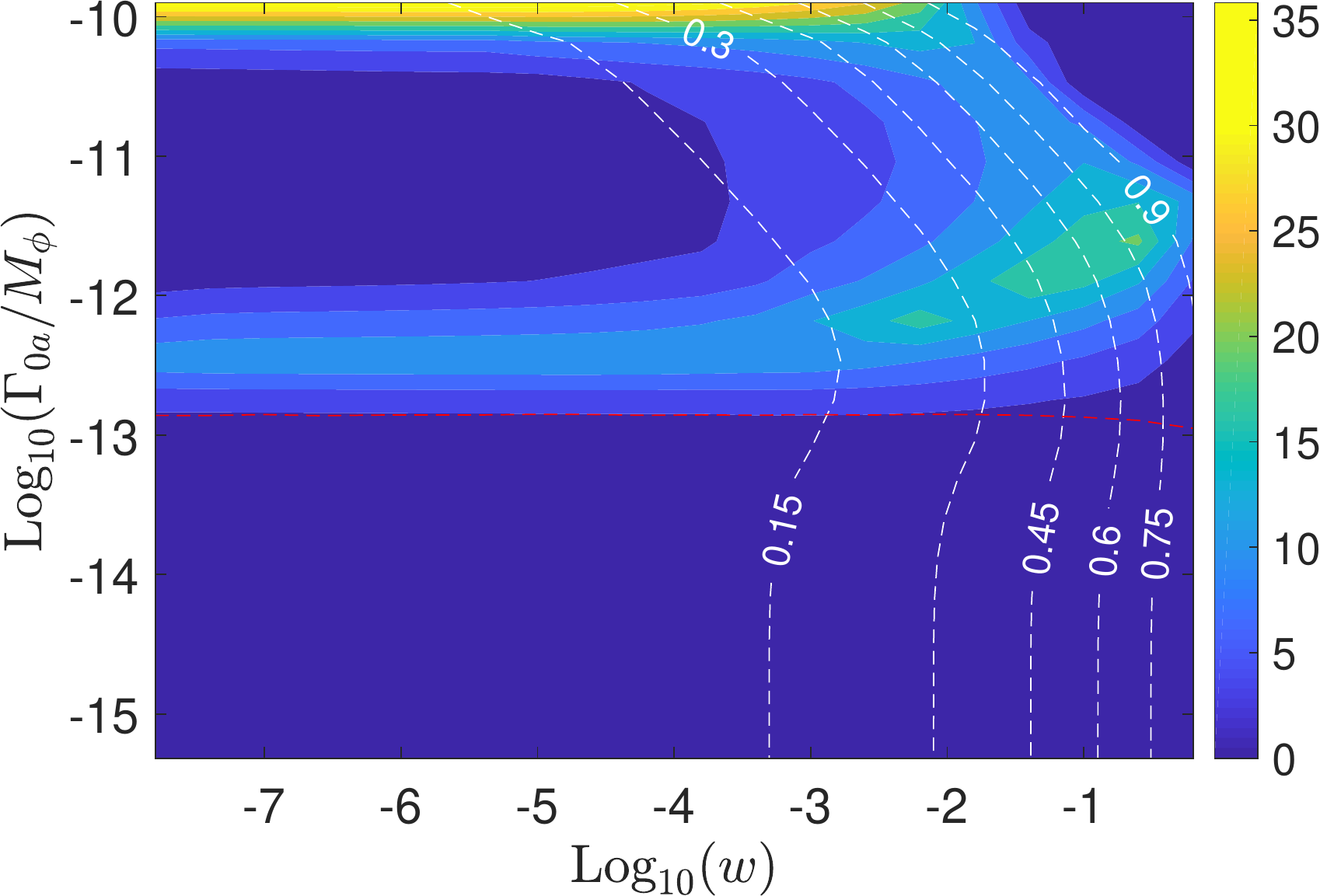}
\end{subfigure}
\caption{\emph{Left panel}: Contours of the final temperature ratio obtained numerically, $x_{f,n}=T_b/T_a$, shown in color, for an inflaton with axionic couplings to gauge bosons in both  sectors. The white dashed contours show  $\log_{10}(T_{\textrm{rh}}/M\f)$. The red dot-dashed lines mark the region where $T_{a,{\rm max}}=\Lambda_a$ for initial inflaton densities $\rho_{\phi,I}=10^{-8}M_{\rm Pl}^2M\f^2$, $10^{-4}M_{\rm Pl}^2M\f^2$ and $10^0M_{\rm Pl}^2M\f^2$. \emph{Right panel}: Contours of relative fractional discrepancy, $|x_{f,a}-x_{f,n}|/x_{f,n}$, where $x_{f,a}$ is the analytic estimate of eq.~\eqref{eq:x_predict} with eq.~\eqref{eq:gauge_boson_xfscat}, in percent. The white dashed contours depict $x_{f,n}$ and the red dashed line indicates where $T_{\textrm{rh}}=M\f$.}
\label{fig:xfscat_gauge_boson}
\end{figure}

Repeating the calculation from section~\ref{sec:scattering_attractor}, using $\mathcal{C}_E$ from eq.~\eqref{eq:gauge_boson_fit} with $x\rightarrow0$, we obtain the asymptotic
temperature asymmetry resulting from the scattering attractor curve,
\begin{align}\label{eq:gauge_boson_xfscat}
x_{sc}=1.49\bigg(\frac{1}{M\f^2}\dfrac{\Gamma_{0,a}\Gamma_{0,b}}{\Gamma_{0,a}+\Gamma_{0,b}}\frac{{M}_{\textrm{Pl}}}{\alpha_b\sqrt{\alpha_a}}\bigg)^{1/4}.
\end{align}
The final temperature asymmetry can then be estimated using eq.~\eqref{eq:x_predict}. In the left panel of figure~\ref{fig:xfscat_gauge_boson} we show the final temperature ratio determined by numerically solving the Boltzmann equations. In this section, we take $\rho_{\phi,I}=10^{-10}M\f^2M_{\rm Pl}^2$ in order to keep $T_{\rm max}<\Lambda_a$ in all of our parameter space,  thus ensuring that the effective field theory is valid throughout the entire evolution of the system.  Due to the attractor nature of the  Boltzmann equations describing reheating, larger values of $\rho_{\phi,I}$ do not change the final value of $x$ that one would compute for a given set of Lagrangian parameters. However,  changing $\rho_{\phi,I}$ does alter the maximum temperature attained (see eq.~\eqref{eq:boson_tmax}), and therefore if we require $T_{\rm max}<\Lambda_a, \Lambda_b$ then we are restricted to parameters that satisfy
\begin{align}
\frac{\Gamma_{0a}}{M_{\phi}}<\left(\frac{\alpha_a^2M\f^{6}}{(256\pi)^3M_{\rm Pl}^2\rho_{\phi,I}}\right)^{1/5}.
\end{align}
In the left panel of figure~\ref{fig:xfscat_gauge_boson} the red dot-dashed lines indicate where $T_{\rm max}=\Lambda_a$ for different values of $\rho_{\phi,I}$. Above those lines $T_{\rm max}> \Lambda_a$, and thus the early evolution of the system probes the theory above the cutoff. In the right panel of figure~\ref{fig:xfscat_gauge_boson} we compare our analytic estimate to the numerical result.

\begin{figure}
\begin{subfigure}{.33\textwidth}
\includegraphics[width=1.00\textwidth]{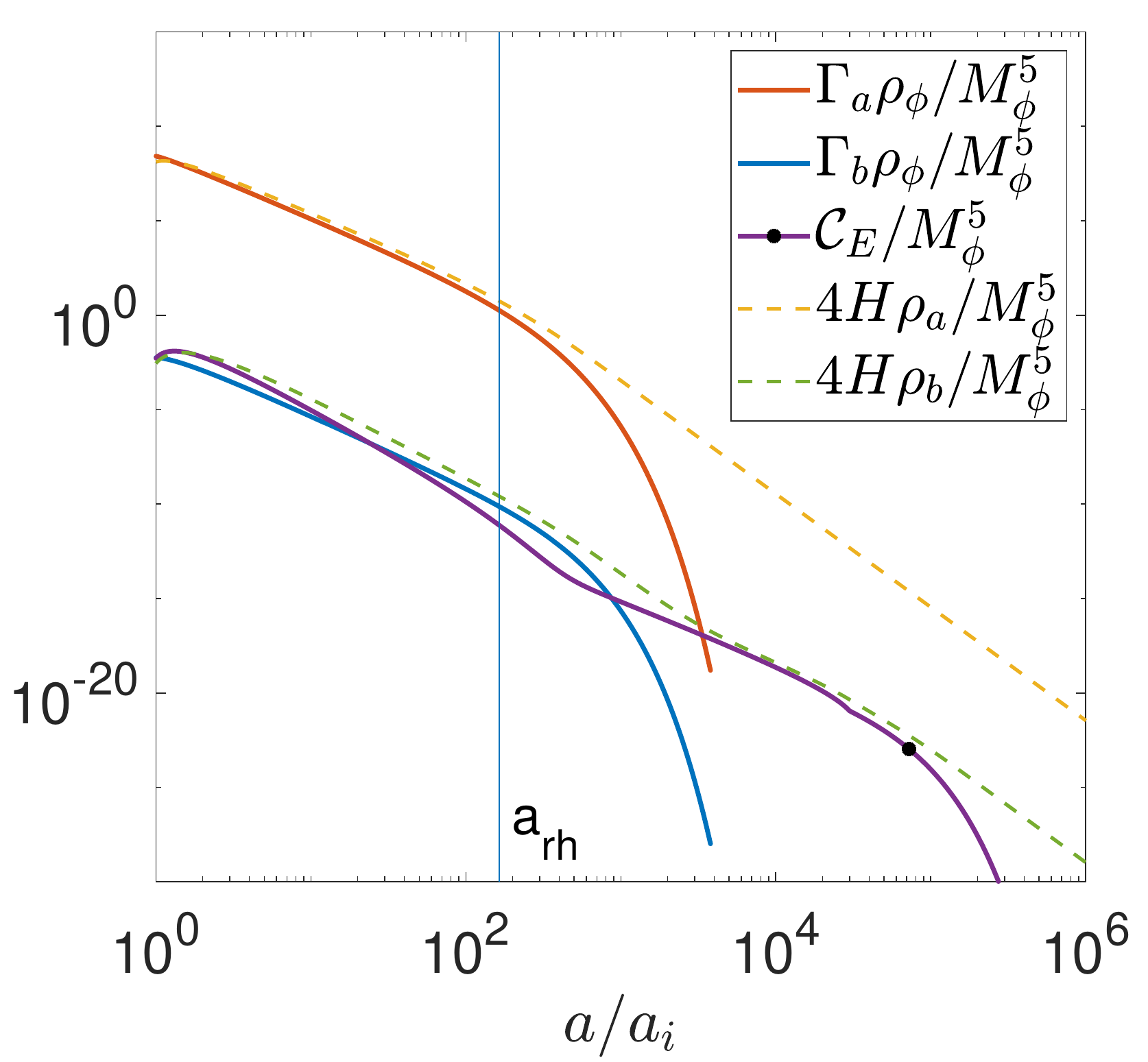}
\end{subfigure}\hfill
\begin{subfigure}{.33\textwidth}
\includegraphics[width=1.00\textwidth]{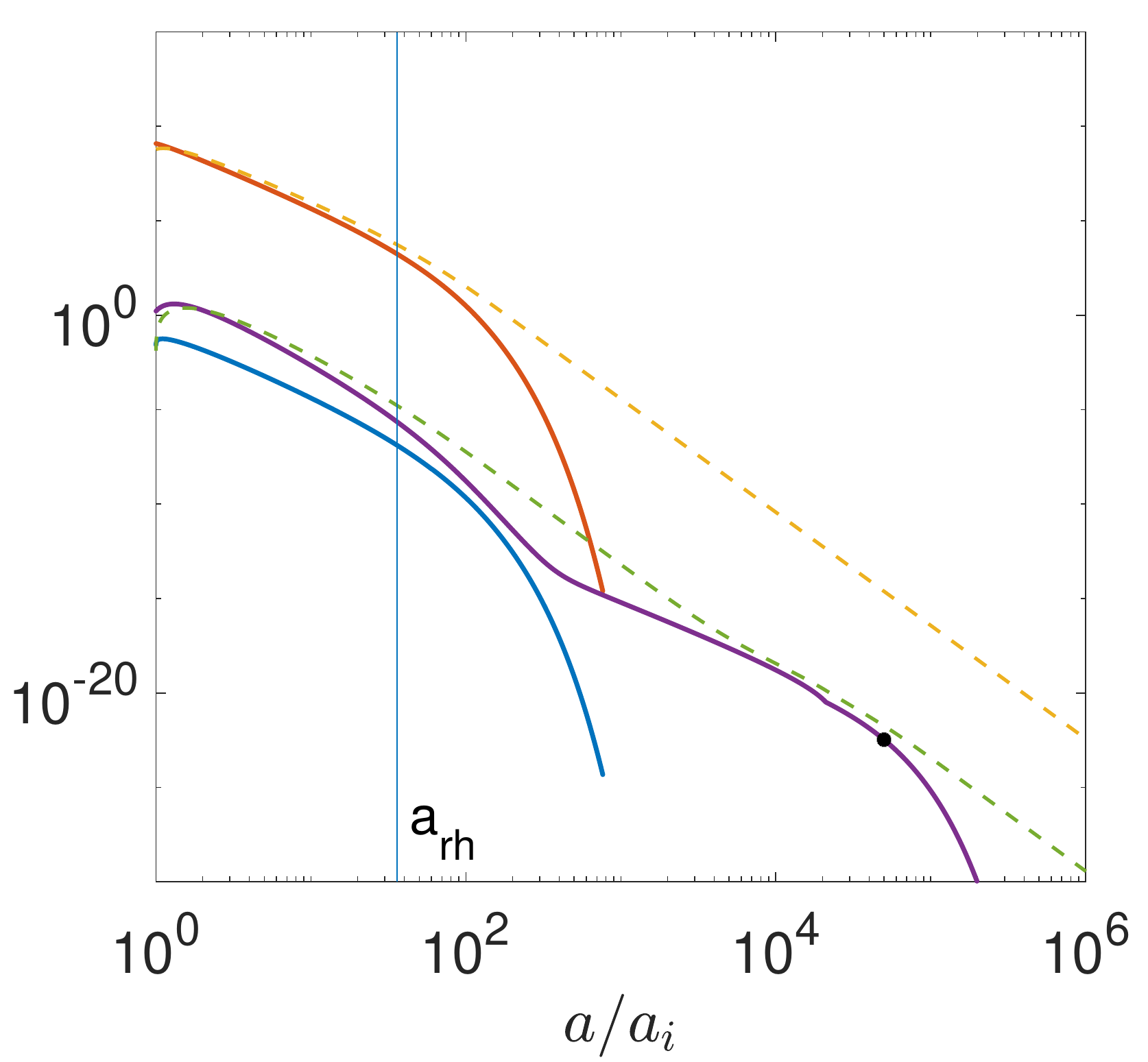}
\end{subfigure}\hfill
\begin{subfigure}{.33\textwidth}
\includegraphics[width=1.00\textwidth]{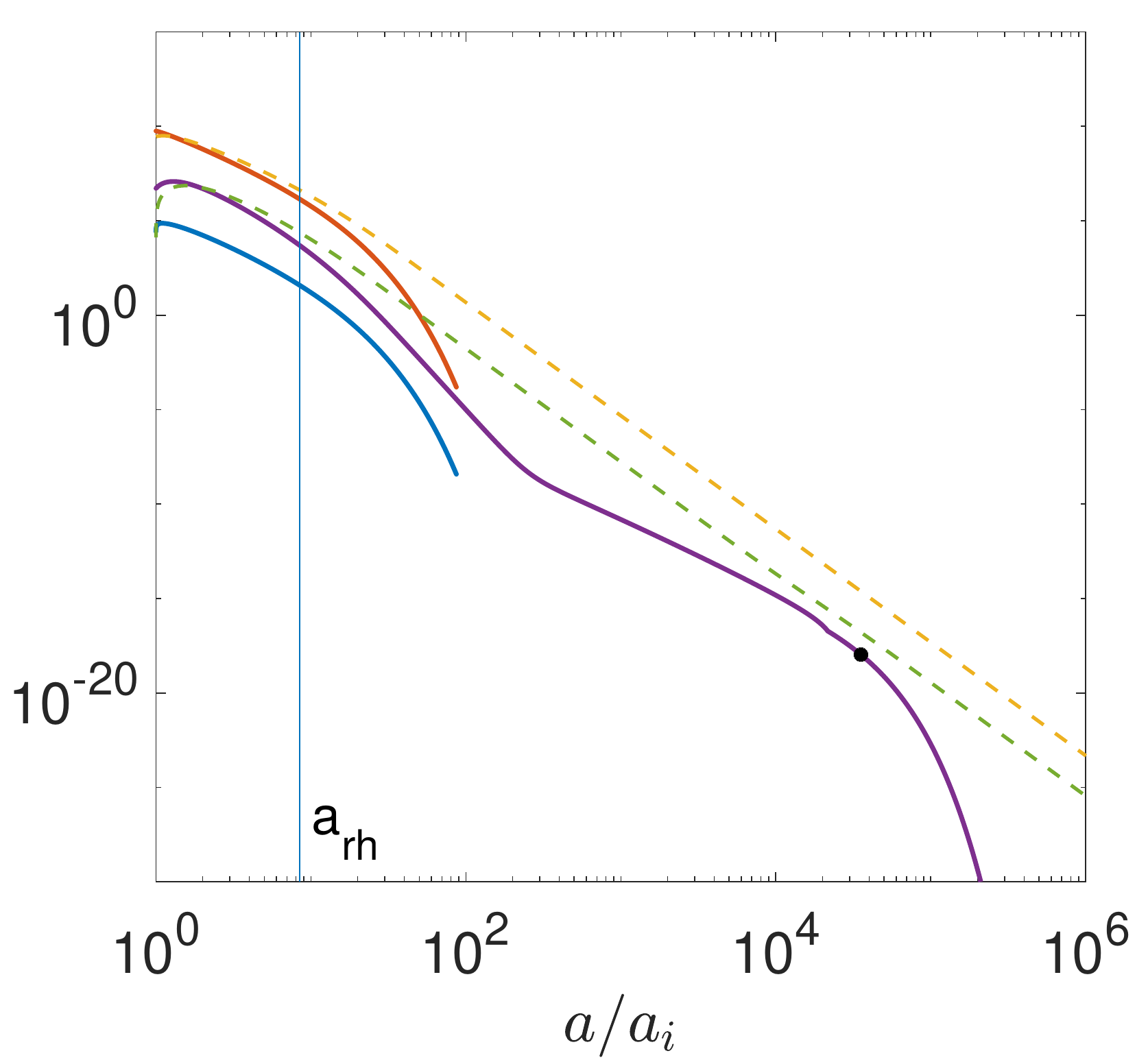}
\end{subfigure}
\caption{Comparison of the collision term with the inflaton decay terms into the two sectors. The plots are for parameters $\{\Gamma_{0a}=10^{-11}$, $w=10^{-8}\}$ (left), $\{\Gamma_{0a}=10^{-10.5}$, $w=10^{-8}\}$ (center) and $\{\Gamma_{0a}=10^{-10}$, $w=10^{-4}\}$ (right). The vertical blue line denotes the point where reheating occurs. In these plots $4H\rho_{a,b}$ serve as proxies for the temperatures of the two sectors. Since we have taken $\alpha_a=\alpha_b$, the temperature ratio is simply $x=(4H\rho_b/4H\rho_a)^{1/4}$.}
\label{fig:rates_UV}
\end{figure}

In the top left corner of the right panel of figure~\ref{fig:xfscat_gauge_boson}, large discrepancies between the analytic estimate and the numerical computation are  becoming  evident. In the same region in the left panel, the contours of fixed temperature asymmetry  are beginning to extend more deeply into the region of small $w$ than the previous examples. Both these features are the consequence of early (i.e.\ pre-reheating) thermalization, enabled by the UV-dominated energy transfer process ($\mathcal{C}_{E,UV}\propto T_a^9$) whose effects are not incorporated  into the analytic estimate in eq.~\eqref{eq:x_predict}. At sufficiently high temperatures, $T_a $, $\mathcal{C}_{E,UV}$  dominates over the energy dumped from the inflaton. This UV behavior can be seen in figure~\ref{fig:rates_UV}, which shows the various contributions to the evolution of the energy density in eq.~\eqref{eqn:boltzwints} as a function of scale factor. Because $\mathcal{C}_{E,UV}$ redshifts faster than $\Gamma_{0a,b}\rho\f$, the energy injection due to inflaton decays can exceed $\mathcal{C}_{E,UV}$ before reheating terminates, $\mathcal{C}_{E,UV}(T\rh)<\Gamma_{0,b}(T\rh)\rho\f$. When this occurs, (see, e.g.\ the left plot in figure~\ref{fig:rates_UV}), the temperature ratio at the end of reheating is the same as the one obtained due to the reheating attractor. Thus, asymmetric reheating overwhelms the collision term. However, when $\mathcal{C}_{E,UV}(T\rh)>\Gamma_{0,b}(T\rh)\rho\f$, the temperature ratio at the end of reheating, $x\rh$, is larger than the case without scattering, i.e.\ the result obtained from the reheating attractor. This deviation would not be reflected in the final temperature asymmetry if $x_{sc}$ is larger than this modified $x_{\rh}$ (center plot in figure~\ref{fig:rates_UV}). It is only when the modified $x_{\rm rh}$ due to $\mathcal{C}_{E,UV}(T\rh)$ is larger than $x_{sc}$ (right plot in figure~\ref{fig:rates_UV}), that the effects from $\mathcal{C}_{E,UV}$ impact the final temperature ratio as we see in the top left corner of figure~\ref{fig:xfscat_gauge_boson}. It is worth recalling that thermal effects beyond the scope of this paper, in particular Landau damping and thermal blocking, can be important for determining the duration and dynamics of reheating in the high-$T\rh$ regime where the effects from $\mathcal{C}_{E,UV}$ show up.  

\subsubsection{Mixed Yukawa and trilinear couplings}

Finally, we consider a theory in which inflaton has trilinear couplings to scalars in sector $a$ and Yukawa couplings to fermions in sector $b$,
\begin{eqnarray}
\mathcal{L}_{\textrm{int}}=\frac{1}{2}\mu_a\phi\chi_a\chi_a+y_b\phi\bar{\psi}_b\psi_b.
\end{eqnarray}
This interaction results in zero-temperature partial widths given by
\begin{eqnarray}
{\Gamma}_{0,a}\approx\frac{{\mu}_a^2}{32\pi M\f} , \qquad \text{and}\qquad {\Gamma}_{0,b}\approx\frac{y_b^2}{8\pi} M\f.
\end{eqnarray}
The spin-summed $s$-channel scattering amplitude between the two sectors is
\begin{align}
|\overline{\mathcal{M}}(s)|^2=2\mu_a^2y_b^2\bigg(1-\frac{4m_{b}^2}{s}\bigg)\frac{s}{(s-M^2_{\phi})^2+(\Gamma_{0a}+\Gamma_{0b})^2}.
\end{align}
Using this scattering amplitude we derive the total energy transfer rate, $\mathcal{C}_E$, given in eq.~\eqref{eq:fermion_boson_fit}; see figure \ref{fig:fermion_boson_analytical_compare}. The collision term is almost insensitive to $T_b$ except when $T_b\approx T_a$.  However, since the two sectors have different quantum statistics, the behavior of the collision term changes depending on which sector is hotter. When there is a large temperature asymmetry between the two sectors ($T_a\gg T_b$ or $T_b\gg T_a$), analogous to the cases considered above, the temperature dependence of $\mathcal{C}_E$ is approximately
\begin{align}
\mathcal{C}_E=\frac{1}{16\pi^3}\times\begin{cases}
\dfrac{4{\mu}_a^2y_b^2}{({\mu}_a/M\f)^2+4y_b^2}T_a^3\Big[1.6\log\Big(\frac{T_a}{M\f}\Big)+1.1\Big]& T_a> M_{\phi},\ T_b\ll T_a
\\ 
\\
-0.30\dfrac{4{\mu}_a^2y_b^2}{({\mu}_a/M\f)^2+4y_b^2}T_b^3& T_b> M_{\phi},\ T_b\gg T_a
\\ 
\\
\pm \dfrac{4{\mu}_a^2y_b^2}{({\mu}_a/M\f)^2+4y_b^2}M\f^2\dfrac{T_{a,b}}{4} K_2\Big(\frac{M_{\phi}}{T_{a,b}}\Big) & T_{a,b}\lesssim M_{\phi} ,\ T_{b,a}\gg T_{a,b}
\\ 
\\
\dfrac{31}{4\pi^2}{\mu}_a^2y_b^2 \dfrac{T_a^7}{M\f^4} & m_{a,b}\ll T_a\ll M_{\phi},\ T_b\ll T_a 
\\ 
\\
-\dfrac{21}{4\pi^2}{\mu}_a^2y_b^2 \dfrac{T_b^7}{M\f^4} & m_{a,b}\ll T_b\ll M_{\phi},\ T_a\ll T_b,
\end{cases}
\end{align}
where the minus signs appear when $T_b > T_a$, as consistent with our definition of the energy transfer term in eq.\ \eqref{eqn:boltzwints}.

\begin{figure}
\begin{subfigure}{.5\textwidth}
\includegraphics[width=1.00\textwidth]{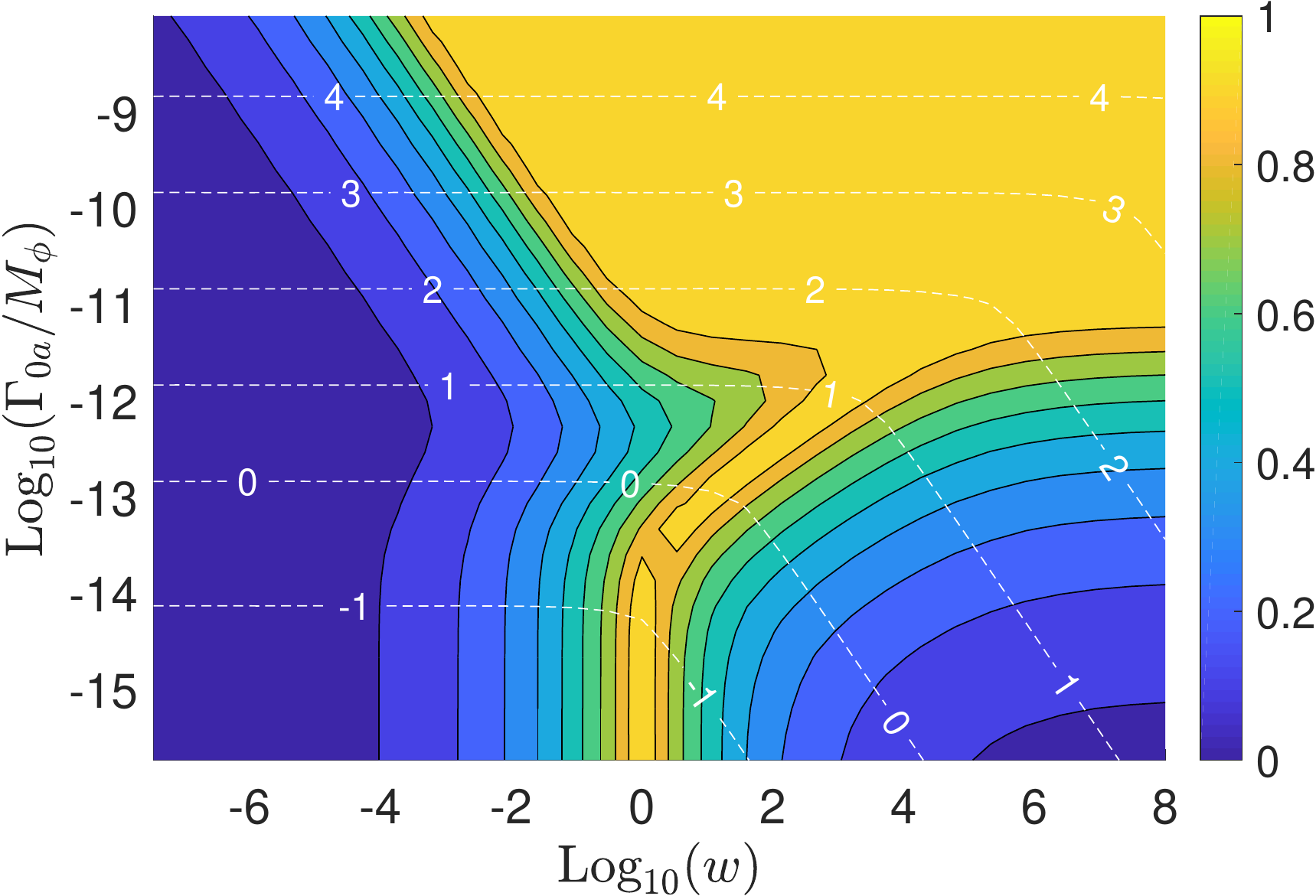}
\end{subfigure}
\begin{subfigure}{.5\textwidth}
\includegraphics[width=1.00\textwidth]{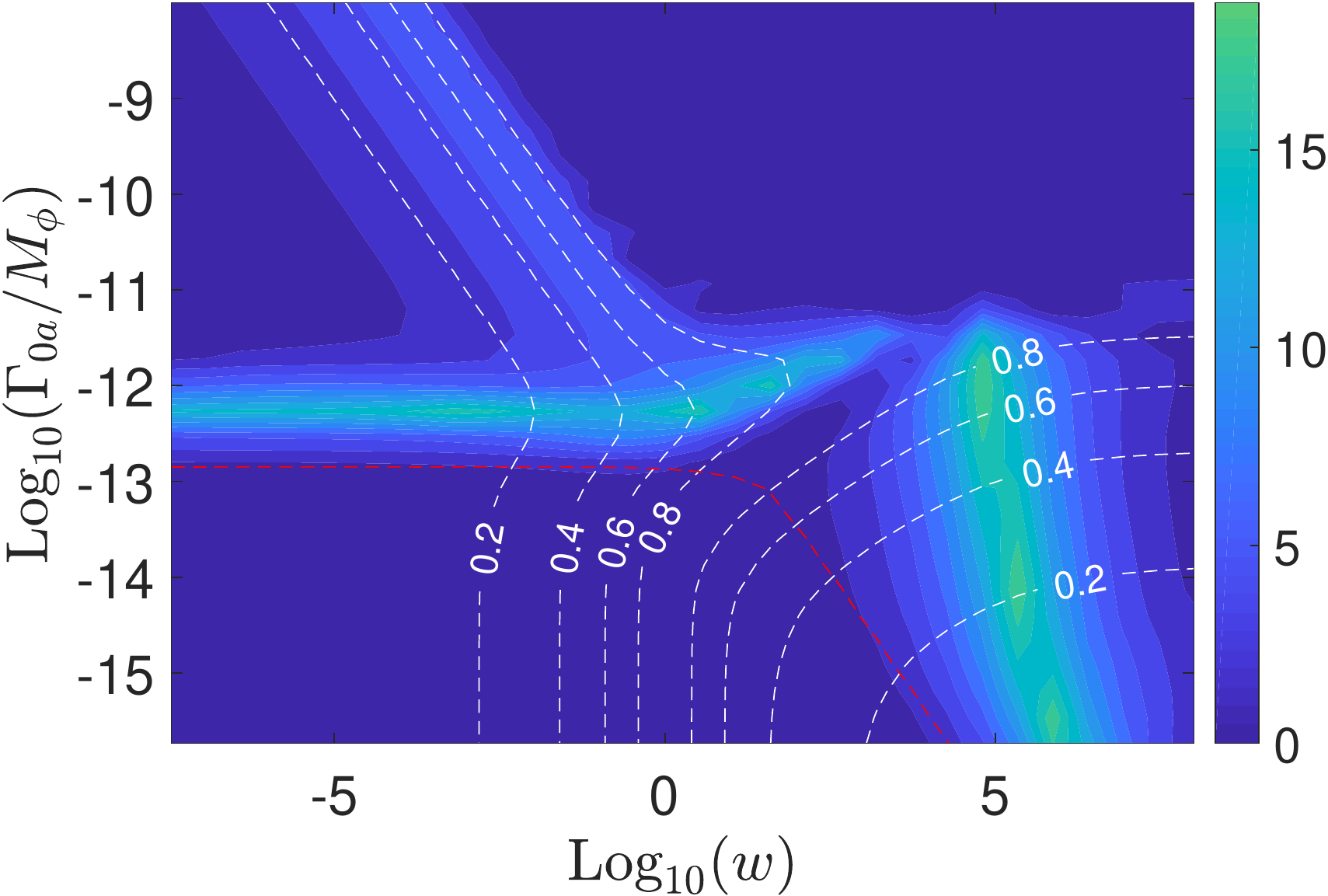}
\end{subfigure}
\caption{\emph{Left panel}: Contours of the final temperature ratio obtained numerically, $x_{f,n}=T_b/T_a$, shown in color, for an inflaton coupled to scalars in sector $a$ and fermions in sector $b$. The white dashed lines show contours of $\log_{10}(T_{\textrm{rh}}/M\f)$. \emph{Right panel}: Contours of relative fractional discrepancy, $|x_{f,a}-x_{f,n}|/x_{f,n}$, where $x_{f,a}$ is the analytic estimate of eq.~\eqref{eq:x_predict} with eq.~\eqref{eq:fermion_boson_xfscat}, in percent. The white dashed contours depict $x_{f,n}$ and the red dashed line marks $T_{\textrm{rh}}=M\f$.}
\label{fig:xfscat_fb}
\end{figure}

Determining the final temperature asymmetry due to inflaton-mediated scattering as in section~\ref{sec:scattering_attractor}, making use of $\mathcal{C}_E$ from eq.~\eqref{eq:fermion_boson_fit} with $x\rightarrow0$, we find
\begin{align}\label{eq:fermion_boson_xfscat}
x_{sc}= \bigg(\frac{1}{M\f^2}\dfrac{\Gamma_{0,a}\Gamma_{0,b}}{\Gamma_{0,a}+\Gamma_{0,b}}\frac{{M}_{\textrm{Pl}}}{\alpha_{\textrm{cold}}\sqrt{\alpha_{\textrm{hot}}}}\bigg)^{1/4} \times \begin{cases} 1.24 & T_b\ll T_a\\ 1.19 & T_b\gg T_a,\end{cases}
\end{align}
where $\alpha_{\textrm{hot}}$ ($\alpha_{\textrm{cold}}$) denotes the value of $\alpha = \pi^2 g_*/30$ corresponding to the hotter (colder) sector.  The final temperature asymmetry can then be estimated using eq.~\eqref{eq:x_predict}.
In the left panel of figure~\ref{fig:xfscat_fb} we show  numerical results for the final temperature ratio. In the right panel of figure~\ref{fig:xfscat_fb}  we show the  disagreement between our analytic estimate and the numerical result as a percentage of the numerical result.

\section{Summary and conclusion}\label{sec:conclusions}

Asymmetric reheating is a minimal way to populate dark sectors that are otherwise completely decoupled from the SM following inflation.  In this work, we have performed the first detailed analysis of perturbative asymmetric reheating. Specifically, by solving the Boltzmann equations describing the perturbative decay of the inflaton into two otherwise decoupled radiation sectors, we have studied in detail the resulting temperature asymmetries attained by the sectors. Scattering processes mediated by inflaton exchange couple the two sectors in the UV, and our self-consistent treatment takes into account the associated collision terms that transfer energy between the radiation sectors. Furthermore, we have carefully accounted for the effects of quantum statistics. At high temperatures (compared to the relevant mass scale in the problem, the inflaton mass) these quantum-statistical effects lead to important corrections in both the inflaton decay rate, as well as the inflaton-mediated scattering processes that transfer energy between the sectors.

The system of Boltzmann equations describing the evolution of the energy densities in the various sectors is a coupled set of three first-order non-linear differential equations, and a general analytic solution is not available. However, in this work we have demonstrated that the system can be accurately analyzed by making use of the attractor nature of its solutions. Broadly, we have identified two classes of quasi-static attractor solutions to which the energy density of the radiation bath evolves depending on the physical process that is dominating the evolution. In a broad range of parameter space and to a good approximation, at any given time the evolution is dominated by either 1) the energy injection from the decay of the inflaton, 2)  the transfer of energy between the sectors through inflaton-mediated scattering, or  3) the adiabatic expansion of the Universe. Case 1) leads to a \emph{reheating attractor curve},  case 2) yields a \emph{scattering attractor curve}, while in case 3) the radiation density simply redshifts as $\rho \propto a^{-4}$. As we have demonstrated, the utility of these attractor solutions is that they allow for a very accurate semi-analytic determination of the resulting temperature asymmetry between the sectors; the asymmetry is simply determined by the process which dominates the evolution at the latest time. 

Our results for the temperature asymmetries generated by asymmetric reheating are surprisingly universal across various coupling structures and particle types. The key property that determines the outcome of asymmetric reheating is the reheating temperature, $T\rh$ relative to the inflaton mass-scale, $M\f$, as follows:
\begin{itemize}
\item When $T\rh \ll   M\f/4$,  the temperature asymmetry is solely determined by perturbative reheating process. More specifically, when $T\rh< M\f/10$, the final temperature ratio is simply given by the ratio of the zero-temperature decay widths, $x=T_b/T_a=(\alpha_a\Gamma_{0b}/\alpha_b\Gamma_{0a})^{1/4}$. As the reheat temperature is increased (but still $< M\f$) quantum-statistical corrections to the inflaton decay width begin to significantly affect the final temperature asymmetry.   In this region asymmetric reheating can be achieved by quantum statistical effects alone, with otherwise identical couplings.

\item When $T\rh \gg M\f/4$, the final temperature asymmetry is determined solely by inflaton-mediated scattering. Inflaton-mediated energy transfer between the sectors falls off exponentially when the temperature of the hotter sector falls below $T_a<M_{\phi}/4$ due to the $s$-channel scattering process going off-resonance. If the radiation sectors have not thermalized by this time, the colder sector is populated by a  freeze-in like process where its final density (or equivalently the temperature ratio $x=T_b/T_a$) is primarily determined by the collision term and the Hubble rate at $T_a=M_{\phi}/4$,
\begin{align*}
x_{sc}\sim \Big(\frac{\alpha_a\mathcal{C}_E}{H\alpha_b\rho_a}\Big)^{1/4}_{T_a=M_{\phi}/4}.
\end{align*}
Because the collision term  $\mathcal{C}_E(T_a=M_{\phi}/4)$ is largely insensitive to the inflaton coupling to the hotter sector as well as (at $T_a=M\f/4$) the quantum statistics of the interacting particles, the final temperature ratio is determined solely by the coupling strength of the colder sector irrespective of its particle identity. 
\end{itemize}

In the region $T\rh \sim M\f/4$, both reheating and scattering are important in determining the final temperature asymmetry.    We find that the final temperature asymmetry, as a function of $T\rh$ and the ratio of zero temperature partial widths $w$ depends on the inflaton mass only through $T\rh/M\f$.  However, lower inflaton masses allow for the consistent realization of higher values of $T/M_\phi$ prior to reheating, which can be particularly important for models where the inflaton couples to the radiation baths through non-renormalizeable interactions (as in the axionic coupling to gauge bosons considered here).

The primary goal of this paper was to analyze, in detail, the temperature evolution of two otherwise-decoupled radiation sectors during and after asymmetric reheating,
 but along the way we obtained a number of other novel results. We found novel power laws describing the evolution of radiation baths during reheating at  temperatures larger than the inflaton mass scale, when quantum statistics are important.  We developed methods to derive closed form (approximate) analytic expressions for energy transfer rates between two relativistic particles at different temperatures via $s$-channel interactions mediated by a massive scalar field.  Finally, we derived reduced integral-expressions for energy-transfer rate between two relativistic sectors at different temperatures via $t$-channel interactions. 

The analytic estimates of the final temperature ratio developed here for two-sector reheating can be straightforwardly extended to $N$-sector reheating scenarios \cite{Arkani-Hamed:2016rle}.  In such cosmologies, for each of the sub-dominant sectors, the dominant energy injection from scattering  is the collision term determined by the hottest sector. Provided the expansion rate is dominated by a single component (either the inflaton, or a single dominant  radiation bath), to a very good approximation, the subdominant sectors are insensitive to each others presence.

In this work, we limited our analysis to perturbative reheating, ignoring the effects from 1) incomplete internal thermalization in the sectors during early reheating, 2) thermal modifications to the inflaton decays from collective effects, such as thermal blocking or Landau damping, 3) back-scatterings into inflaton quanta and 4) preheating. As long as these effects do not significantly alter the final reheating temperature obtained from perturbative reheating, our results for the final temperature asymmetry  remain robust. Even in scenarios where such effects do significantly alter the reheat temperatures,  the scattering attractor curve provides a strict upper bound to the temperature asymmetry between the sectors, $x\geq x_{sc}$ (see eq.\ \eqref{eq:x_predict}), as long as reheating occurs before inflaton-mediated scattering drops off resonance, $T_{\textrm{rh}}>M_{\phi}/4$. We leave the further study of  temperature asymmetries under these potentially disruptive effects to future work.  Another possible extension of this work is to study scenarios that include large asymmetries in the number of degrees of freedom in the two sectors. In such a scenario, the sector with the higher temperature could have a sub-dominant energy density, a possibility we explicitly ignored in this work.

\acknowledgments

We thank Yanou Cui and Scott Watson for useful discussions. The work of PA and PR was supported in part by NASA Astrophysics Theory Grant NNX17AG48G. The work of JS and PR was supported in part by DOE Early Career grant DE-SC0017840.

\appendix

\setcounter{tocdepth}{2}

\section{Cosmological attractor solutions}\label{appendix:attractor}

In this section we provide a general discussion of an important class of solutions to the cosmic Boltzmann equations that exhibit extended periods of quasi-static evolution.  Such quasi-static equilibria can occur in any scenario where a quantity is fed by an external source at a faster rate than it is diluted by cosmic expansion.  These quasi-static equilibria are {\it attractor solutions} in a sense that we make precise in this appendix.  Our primary interest here is in the energy density contained in a thermal radiation bath, where notable examples of such attractor solutions include the $T\propto a^{-3/8}$ evolution of a radiation bath during (classical, perturbative) reheating \cite{Chung:1998rq} and the $T\propto a^{-3/4}$ behavior of a radiation bath fed by out-of-equilibrium renormalizeable scattering processes \cite{Chu:2011be,egs}.  Another important class of examples is realized by various models of freeze-in dark matter \cite{McDonald:2001vt,Hall:2009bx}, where the relevant quantity is the number density of  DM. 

We begin with a representative Boltzmann equation for the energy density of a radiation sector, given by
\begin{align}
\dd{\rho}{t}+4H\rho &= \mathcal{C}_E ,
\end{align}
where $\mathcal{C}_E$ is the net energy input into the sector from an external source. We rewrite this equation in the form
\begin{align}
\label{eq:rhodot_attractor}
a\dd{\rho}{a}+4\rho &=\frac{\mathcal{C}_E}{H}\equiv F(\rho,a).
\end{align}
 By defining new variables as
 \begin{align}\label{eq:m_n}
 \lambda(a)=\dfrac{\rho(a)}{F(\rho(a),a)},&& p(a)=\dfrac{\partial \ln F}{\partial \ln\rho}, && q(a)=\dfrac{\partial \ln F}{\partial \ln a},
 \end{align}
we can further modify eq.~\eqref{eq:rhodot_attractor} to yield
\begin{align}
\dd{\ln\lambda}{\ln a} =(1-p)\frac{1}{\lambda}-4(1-p)-q.
\end{align}
This equation dictates how the ratio of $H\rho/\mathcal{C}_E$ evolves depending on the functional behavior of $F(\rho,a)=\mathcal{C}_E/H$ encoded in $p,q$. Now note that for $p<1$ and  $q>-4(1-p)$,
\begin{align}\label{eq:lambda}
\lambda=\frac{1}{4+\frac{q}{1-p}}
\end{align}
is a stable attractor solution for this equation, provided that $p$ and $q$ slowly vary with $a$ ($\frac{\partial \ln (p,q)}{\partial \ln a}\ll1$). 
This solution is an attractor: radiation baths initially below this steady-state solution rise up very rapidly to meet it, while radiation baths initially above it redshift as $\rho\propto a^{-4}$ until the attractor solution is attained.
Thus, the quasi-static behavior of $\rho$ can be found by simply solving the equation
\begin{align}\label{eq:rho_attractorsol}
\rho (a) =\frac{1}{4+\frac{q}{1-p}}\frac{\mathcal{C}_E}{H}.
\end{align}
In cases of cosmological interest $F$ very frequently has power law dependence on $\rho$ and $a$, thus making $\lambda$ a fixed and readily computable constant (usually of $O(1)$). In such cases, the relevant power law describing the temperature evolution can then be quickly obtained by solving $\rho\ \propto \mathcal{C}/H$.

When $q<-4(1-p)$, there is no attractor solution (as $\lambda$ is always positive) and $\lambda$ increases uncontrollably. This corresponds to the cases when $\mathcal{C}_E$ is falling faster than the redshifting of the energy density, and the evolution of the radiation bath is approximately adiabatic. On the other hand, when $p>1$, the attractor solution (when it exists) is not stable. If $\mathcal{C}_E$ ever came to dominate in this scenario then it would lead to an indefinite explosive rise in $\rho$ due to the positive feedback from $\mathcal{C}_E$. The subsequent solution can be obtained by simply solving $\dot{\rho}=\mathcal{C}_E$.

We can perform an analogous exercise for number density.  The Boltzmann equation we start with here is
\begin{align}
\dd{n}{t}+3Hn  =\mathcal{C},
\end{align}
and, defining
\begin{align}\label{eq:q_p}
 \kappa(a)=\dfrac{n(a)}{F(n,a)},&& p(a)=\dfrac{\partial \ln F}{\partial \ln n}, && q(a)=\dfrac{\partial \ln F}{\partial \ln a},
 \end{align}
we may rewrite this equation as
\begin{align}
\dd{\ln\kappa}{\ln a} =(1-p)\frac{1}{\kappa}-3(1-p)-q.
\end{align}
Then the attractor solution is given by
\begin{align}\label{eq:kappa}
\kappa=\frac{1}{3+\frac{q}{1-p}},
\end{align}
or
\begin{align}\label{eq:n_attractorsol}
n(a)=\frac{1}{3+\frac{q}{1-p}}\frac{\mathcal{C}}{H}.
\end{align}

\subsection*{Example: perturbative reheating of bosons at high temperatures}
As an example, we apply the above formalism to reheating a thermal bath of bosons at temperatures $T\gg M\f$. In this regime,
\begin{align}
\Gamma(T)=4{\Gamma}_0\frac{T}{M\f} && {H}=\frac{1}{\sqrt{3}{M}_{\textrm{Pl}}}\sqrt{{\rho}_{\phi,I}}(a/a_I)^{-3/2}
\end{align}
The corresponding quantity $F(\rho,a)$ is thus given by 
\begin{align}
F(\rho,a) = \frac{\Gamma \rho_{\phi,I} a^{-3}}{{H}}=4\sqrt{3}{M}_{\textrm{Pl}}\frac{1}{\alpha^{1/4}}{\Gamma}_0\frac{{\rho}^{1/4}}{M\f}\sqrt{{\rho}_{\phi,I}}(a/a_I)^{-3/2},
\end{align}
where again $\alpha = \pi^2 g_*/30$, 
giving $p=1/4$ and $q=-3/2$ (eq.~\eqref{eq:m_n}).  Eq.~\eqref{eq:rho_attractorsol} then gives
\begin{align}
{\rho} &=2 \sqrt{3} \frac{M_{\textrm{Pl}}\Gamma_0}{M\f} \left(\frac{\rho}{\alpha}\right)^{1/4}\sqrt{{\rho}_{\phi,I}}\left(\frac{a}{a_I}\right)^{-3/2}.
\end{align}
Solving for the temperature evolution then yields  $ {T}(a)=\Big(2\sqrt{3}{M}_{\textrm{Pl}}{\Gamma}_0\sqrt{\rho_{\phi,I}}/(\alpha M\f)\Big)^{1/3}(a/a_I)^{-1/2}$, exactly the expected asymptotic
behavior.

\section{Preheating and the Bose power law}\label{appendix:preheating}

In this appendix we demonstrate that it is possible to realize a radiation bath following the Bose power law $T\propto a^{-1/2}$ of eq.~\eqref{eq:boson_power_law} during perturbative reheating.
 We focus on a theory with an inflaton, $\phi$, coupled to a scalar field, $\chi$, via the trilinear coupling 
\begin{eqnarray}
\mathcal{L}_{\textrm{int}}=\frac{1}{2}\mu\phi\chi^2.
\end{eqnarray}
This model can experience broad resonance preheating for sufficiently large values of $\mu$ and sufficiently large inflaton oscillations $\Phi=\sqrt{\langle \phi^2 \rangle}$, in which case energy is effectively drained from the inflaton condensate before perturbative reheating can occur.  The condition that no broad resonances are present in the theory is
\begin{align}\label{eq:mu_phi_preheat}
\mu\Phi<M^2\f.
\end{align} 
When this condition is satisfied, preheating is inefficient and perturbative reheating dominates the properties of the radiation bath \cite{Traschen:1990sw, Kofman:1994rk, Dufaux:2006ee}.  

During perturbative reheating, for a given value of $\mu$, the inflaton amplitude $\Phi$ uniquely specifies the temperature of the matter sector. Using the reheating attractor solution eq.~\eqref{eq:attractor_propto} along with the quantum statistics correction to the inflaton decay width eq.~\eqref{eq:decay_widths}, we obtain for the radiation bath
\begin{align}\label{eq:T_equation}
\alpha T^4=\frac{1}{4-\frac{3/2}{1-p}}\times\frac{{\mu}^2}{32\pi}\left(\frac{e^{\frac{\Mp}{2T}}+ 1}{e^{\frac{\Mp}{2T}}- 1}\right)\sqrt{\frac{3}{2}}\Phi M_{\rm Pl},
\end{align}
where
\begin{align}
p=\frac{1}{2}\frac{e^{\frac{\Mp}{2T}}\left(\frac{\Mp}{2T}\right)}{e^{\frac{\Mp}{T}}- 1}.
\end{align}
Eq.\ \eqref{eq:T_equation} can be solved to yield $T$ as a function of $\Phi$. At high and low temperatures, the above relation simplifies to
\begin{align}
T=\begin{cases}
\Big(\frac{\sqrt{3}}{16\sqrt{2}\pi\alpha}\frac{\mu^2}{M\f}M_{\textrm{Pl}}\Phi\Big)^{1/3} & T\gg M\f \\
\Big(\frac{\sqrt{3}}{80\pi\alpha}\mu^2M_{\textrm{Pl}}\Phi\Big)^{1/4}  & T\ll M\f. 
\end{cases}
\end{align}
Eq.~\eqref{eq:T_equation} is valid as long as $\Gamma(T)< H$, or
\begin{align}\label{eq:mu_phi_reheat}
\frac{{\mu}^2}{32M\f\pi}\left(\frac{e^{\frac{\Mp}{2T}}+ 1}{e^{\frac{\Mp}{2T}}- 1}\right)&<\frac{M\f\Phi}{\sqrt{6}M_{\rm Pl}}.
\end{align}

\begin{figure}
\begin{subfigure}{.33\textwidth}
  \includegraphics[width=1\linewidth,left]{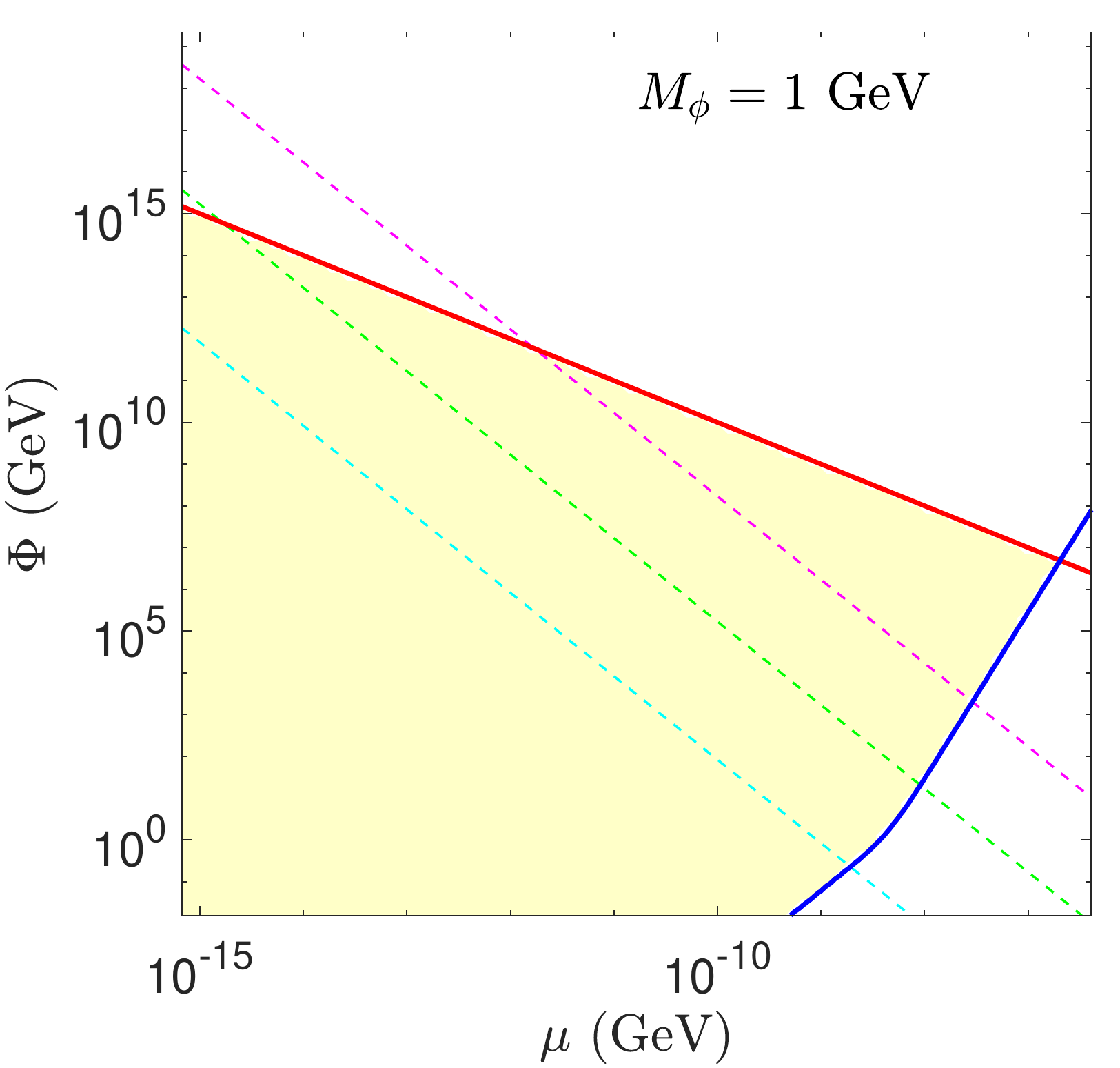}
\end{subfigure}\hfill
\begin{subfigure}{.33\textwidth}
  \includegraphics[width=1\linewidth,right]{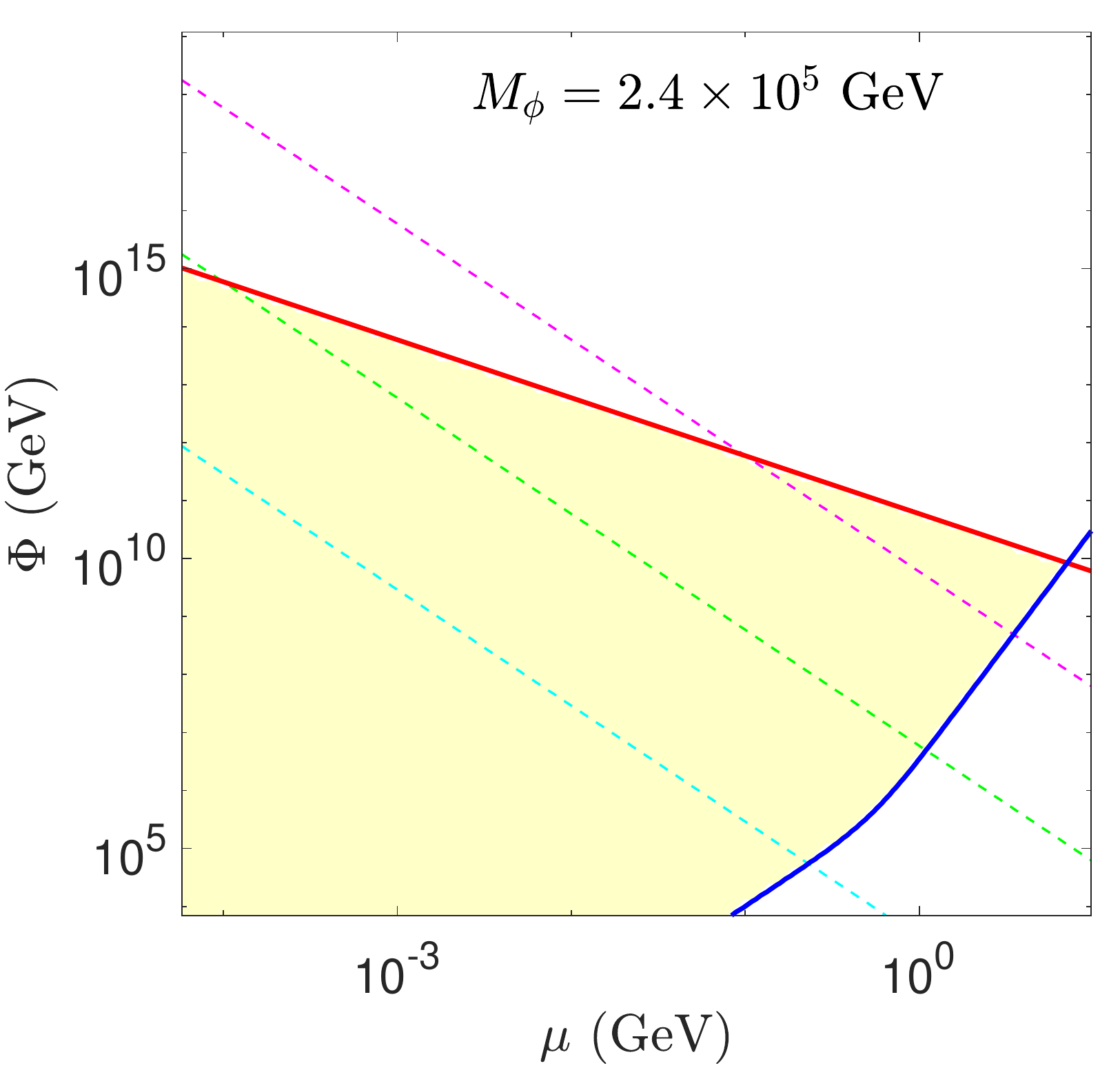}
\end{subfigure}\hfill
\begin{subfigure}{.33\textwidth}
  \includegraphics[width=1\linewidth,center]{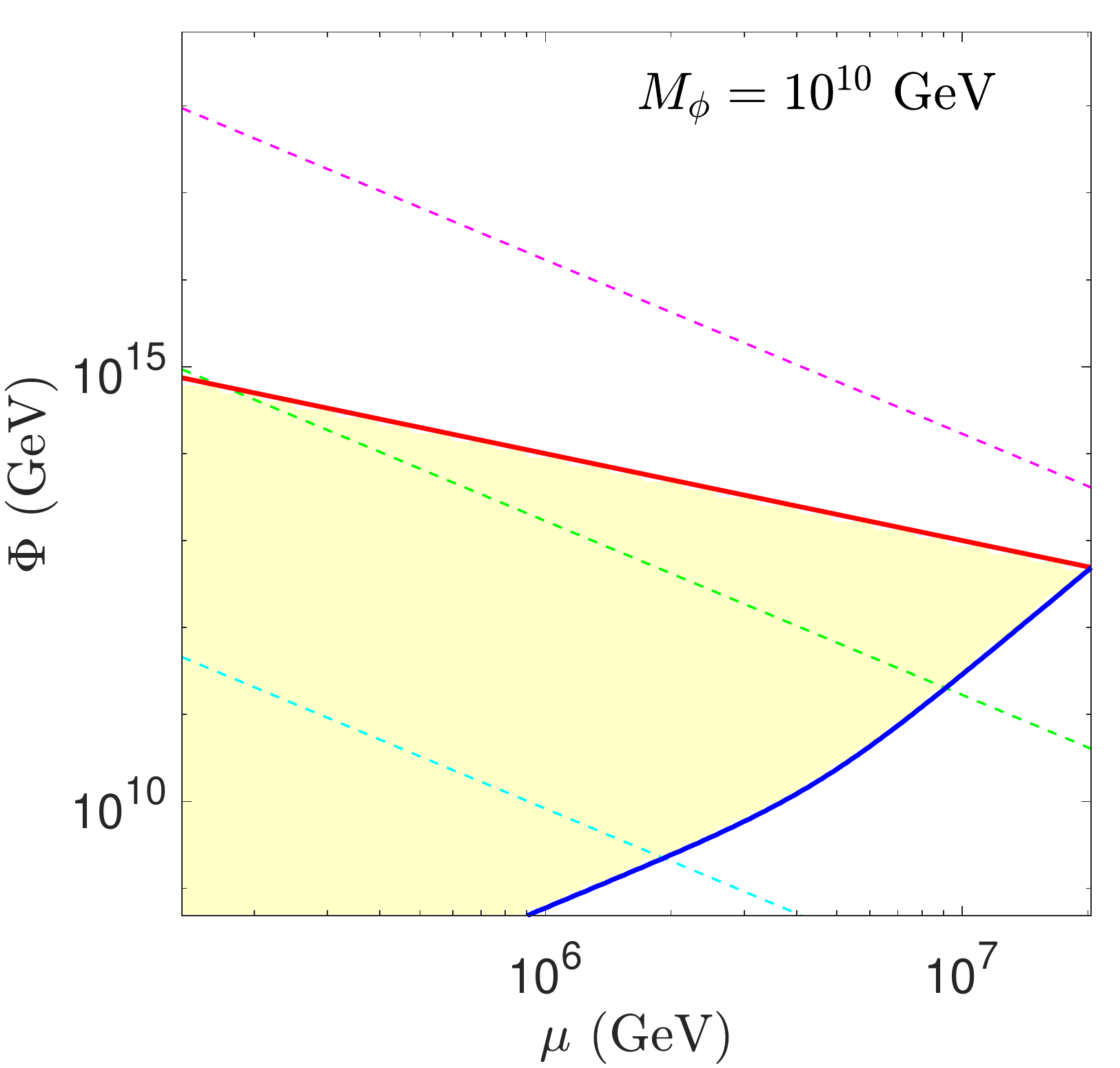}
\end{subfigure}
\caption{Reheating parameter space for $M\f=1$ GeV (left), $M\f=2.4\times 10^5$ GeV (center) and $M\f=10^{10}$ GeV (right). If the inflaton amplitude at the end on inflation is in the region above the red line, then the universe undergoes by broad resonance preheating, eq. \eqref{eq:mu_phi_preheat}. If the inflaton amplitude at the end of inflation is in the yellow shaded region, then the effects of preheating are sub-dominant and the Universe undergoes perturbative reheating. Perturbative reheating ends on the blue line, eq. \eqref{eq:mu_phi_reheat}. The dotted lines represent constant temperature contours during perturbative reheating calculated using eq.~\eqref{eq:T_equation}, $T=10M\f$ (magenta), $T=M\f$ (green) and $T=0.1M\f$ (cyan).}
\label{fig:preheat_parameter_space}
\end{figure}

In figure~\ref{fig:preheat_parameter_space} we show the resulting parameter space for perturbative reheating for three different values of $M\f$.  We show the equalities corresponding to broad resonance preheating (eq.\ \eqref{eq:mu_phi_preheat}) and the end of perturbative reheating (eq. \eqref{eq:mu_phi_reheat}) in red and blue respectively; the yellow shaded region represents the region where perturbative reheating dominates the evolution of the radiation bath.  We further show contours of $T$, from eq.~\eqref{eq:T_equation}. Above $T=M\f$, the matter sector realizes the $T\propto a^{-1/2}$ power law.  We thus observe that for all three mass points, there is some region of parameter space where reheating is dominated by perturbative processes and the radiation bath realizes the bosonic power law.  Lower inflaton masses enable the radiation bath to reach higher temperatures during perturbative reheating.

For a given value of $\mu$, the theory may avoid preheating if the inflaton amplitude at the end of inflation is below the red line in figure \ref{fig:preheat_parameter_space}. As perturbative reheating occurs, the inflaton amplitude decreases due to redshifting in an expanding universe. This redshifting corresponds to traversing downward in the $\mu-\Phi$ parameter space.  The temperature of the radiation bath decreases correspondingly along this trajectory.   This downward trajectory continues until we reach the blue line and reheating occurs.

\section{Collision terms for $s$-channel processes}\label{appendix:s-channel}
In this section we derive the total energy transfer rate between two relativistic species at different temperatures, mediated by a massive scalar field.  We extend the analysis of references \cite{Birrell:2014uka} and  \cite{Adshead:2016xxj} to determine the non-equilibrium energy transfer between two sectors at different temperatures. Further, we  develop a new procedure to  evaluate the phase space integral for a $t$-channel process in appendix~\ref{appendix:t-channel}, and confirm explicitly that  the contribution from $t$-channel processes is always  negligible compared to the $s$-channel process. Finally, for some specific theories where the interactions between the two relativistic species is mediated by a massive scalar field, we find closed-form analytic  approximations for the total energy transfer rate.

We start by considering the forward energy transfer for the process
$$ 1\ +\ 2\ \rightarrow\ 3\ +\ 4,$$
where 1, 2, 3, and 4 represent the particles participating in the scattering process. The forward collision term for this process is given by
\begin{align}\label{eq:full_collision_forward}
\mathcal{C}_E^f=\int \frac{d^3p_1}{2E_1(2\pi)^3}\frac{d^3p_2}{2E_2(2\pi)^3}\frac{d^3p_3}{2E_3(2\pi)^3}\frac{d^3p_4}{2E_4(2\pi)^3} 
(2\pi)^4\delta^4(p_1+p_2-p_3-p_4)|\overline{\mathcal{M}}|^2 S \nonumber\\ \times(E_1+E_2)f_1(U\cdot p_1)f_2(U\cdot p_2)(1\pm f_3(U\cdot p_3))(1\pm f_4(U\cdot p_4)). 
\end{align}
Here $f_i$ is the distribution function for the $i^{th}$ particle, $U$ is the four-velocity of the frame in which we are calculating the collision term, $|\overline{\mathcal{M}}|^2$ is the spin-summed scattering amplitude of the process, and $S$ includes any identical particle factors.

This phase space integral simplifies in the center-of-mass (CM) frame, where $U$ is non-trivial. For simplicity we shift to the  variables
\begin{align}
p= & p_1+p_2, \quad  p'=p_3+p_4,\nonumber\\
q= & p_1-p_2 , \quad  q'=p_3-p_4.
\end{align}
In terms of these variables, the Mandelstam variables are $s=p^2$, $t=(q-q')^2/4$ and $u=(q+q')^2/4$. In the CM frame, $p=(\sqrt{s},0,0,0)$ and consequently $U=\frac{1}{\sqrt{s}}(\sqrt{\vp^2+s},0,0,\vp)$, where $\vp$ is the spatial component of $p$ in the frame $U=(1,0,0,0)$.  Reference~\cite{Birrell:2014uka} shows that the 12-dimensional phase-space integral of eq.~\eqref{eq:full_collision_forward} can be reduced to 
\begin{align}\label{eq:gamma_temp0}
\mathcal{C}_E^f=&\frac{1}{256(2\pi)^8}\int 4\pi \vp^2d\vp\int ds\frac{4rr'}{s} S\nonumber\\ &\times\int\bigg[\int|\overline{\mathcal{M}}|^2(1\pm f_3(U\cdot p_3))(1\pm f_4(U\cdot p_4))d\theta'dy'\bigg]f_1(U\cdot p_1)f_2(U\cdot p_2)d\theta dy,
\end{align}
where $r$ is the magnitude of $\vec q$ and $y=\cos\phi, \theta$ give the direction of $\vec q$ with respect to $\vec p$, while $r'$, $y'$, and $\theta'$  denote the corresponding quantities for $q'$.
The spatial and temporal magnitudes of $q$ and $q'$ are given by
\begin{align}
r= & \frac{1}{\sqrt{s}}\sqrt{(s-(m_1+m_2)^2)(s-(m_1-m_2)^2)}, \quad q^0=\frac{m_1^2-m_2^2}{\sqrt{s}},\\
r'= & \frac{1}{\sqrt{s}}\sqrt{(s-(m_3+m_4)^2)(s-(m_3-m_4)^2)}, \quad q'^0=\frac{m_3^2-m_4^2}{\sqrt{s}}. 
\end{align}

For scenarios where the scattering amplitude is  a function only of $s$,  eq.~\eqref{eq:gamma_temp0} further simplifies to
\begin{align}\label{eq:gamma_temp}
\mathcal{C}_E^f=&\frac{ S}{64(2\pi)^5}\int_0^{\infty} d\vp\vp^2\int_{s_0}^{\infty} ds |\overline{\mathcal{M}}(s)|^2\frac{rr'}{s} \bigg[\int_{-1}^1(1\pm f_3(U\cdot p_3))(1\pm f_4(U\cdot p_4))dy'\bigg]\nonumber\\ &\times\bigg[\int_{-1}^1f_1(U\cdot p_1)f_2(U\cdot p_2)dy\bigg]
\end{align}
where $s_0=\textrm{max}((m_1+m_2)^2,(m_3+m_4)^2)$.  To evaluate these final integrals, we need to specify the distribution functions.  We take particles 1 and 2 to be of species $a$ at a temperature $T_a$ and particles 3 and 4 of species $b$ at temperature $T_b$.   Inserting the corresponding equilibrium distribution functions, we obtain 
\begin{align}
\mathcal{C}_E^f=\frac{T_aT_b S}{4(2\pi)^5}\int_0^{\infty}\int_{s_0}^{\infty} d\vp\ ds\ |\overline{\mathcal{M}}(s)|^2\frac{\exp\bigg(\dfrac{\sqrt{\vp^2+s}}{T_b}\bigg)}{\bigg[\exp\bigg(\dfrac{\sqrt{\vp^2+s}}{T_a}\bigg)-1\bigg]\bigg[\exp\bigg(\dfrac{\sqrt{\vp^2+s}}{T_b}\bigg)-1\bigg]}\nonumber \\
\times\log\Bigg(\frac{\exp\Big(\dfrac{\sqrt{\vp^2+s}+\beta_a\vp}{2T_a}\Big)+\zeta_a}{\exp\Big(\dfrac{\sqrt{\vp^2+s}}{2T_a}\Big)+\zeta_a\exp\Big(\dfrac{\beta_a\vp}{2T_a}\Big)}\Bigg)\log\Bigg(\frac{\exp\Big(\dfrac{\sqrt{\vp^2+s}+\beta_b\vp}{2T_b}\Big)+\zeta_b}{\exp\Big(\dfrac{\sqrt{\vp^2+s}}{2T_b}\Big)+\zeta_b\exp\Big(\dfrac{\beta_b\vp}{2T_b}\Big)}\Bigg).
\end{align}
Here $\zeta_{a,b}=\pm 1$ depending on whether the respective particles are fermions/bosons and
\begin{align}
\beta_{a,b}=\sqrt{1-\frac{4m_{a,b}^2}{s}}.
\end{align}

Next, we scale out the temperature of the hotter sector, $T_a$, by defining 
\begin{align}
\tilde{s}=\frac{s}{T_a^2},\qquad  \tilde{p}=\frac{\vp}{T_a}, \qquad \tilde{m}_{a,b}=\frac{m_{a,b}}{T_a}, \quad \text{and}\quad x=\frac{T_b}{T_a} \leq1. 
\end{align}
This isolates the temperature dependence in the integral, which becomes
\begin{align}
\mathcal{C}_E^f=\frac{ S}{4(2\pi)^5}xT_a^5\int_0^{\infty}\int_{s_0/T_a^2}^{\infty} d\tilde{p}\ d\tilde{s}\ |\overline{\mathcal{M}}(\s)|^2\frac{\exp(\frac{1}{x}\sqrt{\tilde{p}^2+\tilde{s}})}{\bigg[\exp(\frac{1}{x}\sqrt{\tilde{p}^2+\tilde{s}})-1\bigg]\bigg[\exp(\sqrt{\tilde{p}^2+\tilde{s}})-1\bigg]}\nonumber \\
\times\log\Bigg(\frac{\exp\Big(\frac{1}{2}(\sqrt{\tilde{p}^2+\tilde{s}}+\beta_a\tilde{p})\Big)+\zeta_a}{\exp\Big(\frac{1}{2}\sqrt{\tilde{p}^2+\tilde{s}}\Big)+\zeta_a\exp\Big(\frac{1}{2}\beta_a\tilde{p}\Big)}\Bigg)\log\Bigg(\frac{\exp\Big(\frac{1}{2x}(\sqrt{\tilde{p}^2+\tilde{s}}+\beta_b\tilde{p})\Big)+\zeta_b}{\exp\Big(\frac{1}{2x}\sqrt{\tilde{p}^2+\tilde{s}}\Big)+\zeta_b\exp\Big(\frac{1}{2x}\beta_b\tilde{p}\Big)}\Bigg).
\end{align}
The temperature $T_a$ enters the integrand only through ${\mathcal{M}}(T_a^2 \s)$ and $\tilde{m}_{a,b}$. 

The total collision term describing net energy transfer  is
\begin{align}
\mathcal{C}_E&= S'xT_a^5\int_0^{\infty}\int_{s_0/T_a^2}^{\infty} d\tilde{p}\ d\tilde{s}\ |\overline{\mathcal{M}}(\s)|^2\frac{\exp(\frac{1}{x}\sqrt{\tilde{p}^2+\tilde{s}})-\exp(\sqrt{\tilde{p}^2+\tilde{s}})}{\bigg[\exp(\frac{1}{x}\sqrt{\tilde{p}^2+\tilde{s}})-1\bigg]\bigg[\exp(\sqrt{\tilde{p}^2+\tilde{s}})-1\bigg]}\\
&\times\log\Bigg(\frac{\exp\Big(\frac{1}{2}(\sqrt{\tilde{p}^2+\tilde{s}}+\beta_a\tilde{p})\Big)+\zeta_a}{\exp\Big(\frac{1}{2}\sqrt{\tilde{p}^2+\tilde{s}}\Big)+\zeta_a\exp\Big(\frac{1}{2}\beta_a\tilde{p}\Big)}\Bigg)\log\Bigg(\frac{\exp\Big(\frac{1}{2x}(\sqrt{\tilde{p}^2+\tilde{s}}+\beta_b\tilde{p})\Big)+\zeta_b}{\exp\Big(\frac{1}{2x}\sqrt{\tilde{p}^2+\tilde{s}}\Big)+\zeta_b\exp\Big(\frac{1}{2x}\beta_b\tilde{p}\Big)}\Bigg)\nonumber \\ \label{eq:collision}
&\equiv S'xT_a^5\int_0^{\infty}\int_{s_0/T_a^2}^{\infty} d\tilde{p}\ d\tilde{s}\ |\overline{\mathcal{M}}(\s)|^2 D(\s,\p,x,\tilde{m}_{a,b}),
\end{align}
where $ S'=S/(4(2\pi)^5)$.  Given ${\mathcal{M}}, \zeta_a$ and $\zeta_b$, we can now numerically evaluate this integral to obtain the total energy transfer.

Even though we are interested in the regime where all external particles are relativistic, $\tilde{m}_{a,b}\ll1$, we have explicitly retained $\tilde{m}_{a,b} \neq 0$  in the integral.  Retaining finite masses can be important for regulating bosonic scattering amplitudes: for bosons ($\zeta=-1$), the integrand diverges as $\s, \p \rightarrow0$ when $\tilde{m} = 0$, reflecting the zero-momentum singularity in the Bose-Einstein distribution.  However, if $|\overline{\mathcal{M}}(\s)|^2$ is finite at $\s\rightarrow0$ then the divergence vanishes after integration over $\s,\p$ and the collision term remains finite as $\tilde m_{a,b}\to 0$.\footnote{In the \textit{scattering} collision term, the cancellation of the divergence depends on summing both forward and backward processes; the forward \textit{scattering} collision term alone retains a logarithmic dependence on  $\tilde{m}$.  In general one expects thermal self-energies to regulate this behavior when $T_{a,b} \gg m_{a,b}$; see also \cite{Evans:2017kti}.}  In our calculations below we can thus freely work in the limit $\tilde m_{a,b}\to 0$; however, in numerical work we retain small finite external particle masses for ease of computation.

In the subsequent sections we present analytic estimates for the collision term given in eq.~\eqref{eq:collision} in various theories.  We consider the full collision term, i.e., both forward and backward contributions.

\subsection{Trilinear scalar couplings}\label{appendix:s_scalar}

We consider two scalar species, $\chi_a$ and $\chi_b$, interacting via
\begin{eqnarray}
\mathcal{L}_{\textrm{int}}=\frac{1}{2}\mu_a\phi\chi_a^2+\frac{1}{2}\mu_b\phi\chi_b^2.
\end{eqnarray}
Here $\phi$ is a massive scalar (inflaton) mediator with mass $M\f$.
As both the coupled fields are scalars (and hence bosons), we take $\zeta_{a,b}=-1$ and $ S'=1/(16(2\pi)^5)$.

The scattering amplitude for the s-channel process in this theory, for $m_{a,b}\ll M\f$, is given by
\begin{equation}\label{eq:matrix_boson}
|\overline{\mathcal{M}}(\s)|^2=\frac{\mu_a^2\mu_b^2}{(s-M^2_{\phi})^2+\Big(\Gamma_{0a}+\Gamma_{0b}\Big)^2}, \quad \Gamma_{0a,b} = \frac{\mu_{a,b}^2}{32\pi M\f}.
\end{equation}
For $\mu_{a,b}\ll M\f$ we can approximate the scattering amplitude as \cite{Adshead:2016xxj}
\begin{eqnarray}\label{eq:matrix_delta}
|\overline{\mathcal{M}}(\s)|^2\approx 32\pi^2\frac{w\mu_a^2}{w+1}\frac{1}{T_a^2}\delta(\s-\Mf^2)+\Theta(\Mf^2-\s)\frac{{\mu}^4_aw}{M\f^4}.
\end{eqnarray}
where $w=\Gamma_{0b}/\Gamma_{0a}=\br{\mu}_b^2/\br{\mu}_a^2$.
To analytically estimate the behavior of $\mathcal{C}_E$, we combine the simplified form of the scattering amplitude given in eq.~\eqref{eq:matrix_delta} along with approximations $\Mf\ll1$ and $\Mf\gg1$ at high and low temperatures respectively.

\paragraph{High temperature limit, $T_a\gg M\f$.} In the high-temperature limit $\Mf\rightarrow0$,  the contribution to the integral in eq.~\eqref{eq:collision} from the $\Theta$ function term in eq.~\ref{eq:matrix_delta} is dwarfed by the contribution from the Dirac delta term.  Subsequently, in the high temperature limit we can to good approximation retain only the Dirac delta portion, giving
\begin{align}\label{eq:highT_delta}
\mathcal{C}_{\textrm{high-T}}= S'xT_a^5\int_0^{\infty}d\tilde{p}\ 32\pi^2\frac{w\mu_a^2}{w+1}\frac{1}{T_a^2}D(\Mf^2,\p,x,0).
\end{align} 
To evaluate the above integral we separate it into two domains: $\p<0.1$ and $\p>0.1$. In the latter region we approximate $\p\gg\Mf$ to give
\begin{align}
\int_{0.1}^{\infty}d\tilde{p}\ D(\Mf^2,\p,x,0)\bigg|_{\p\gg\Mf}&\approx\int_{0.1}^{\infty}d\tilde{p}\ \frac{\exp(\p/x)-\exp(\p)}{[\exp(\p)-1][\exp(\p/x)-1]}\Big[\log^2\big(1/\Mf^2\big)\nonumber\\ &+\log\Big(64\p^2\sinh(\p/2)\sinh(\p/(2x))x\Big)\log\big(1/\Mf^2\big)\nonumber\\ &+\log\Big(8\p\sinh(\p/2)\Big)\log\Big(8\p\sinh(\p/(2x))x\Big)\Big]\nonumber\\
&=\frac{1}{x}\Big(Y_1(x)\log^2\Big(\frac{T_a}{M\f}\Big)+Y_2(x)\log\Big(\frac{T_a}{M\f}\Big)+Y_3(x)\Big)
\end{align}
where
\begin{eqnarray}
Y_1(x)&=&4x\int_{0.1}^{\infty}d\tilde{p}\frac{\exp(\p/x)-\exp(\p)}{[\exp(\p)-1][\exp(\p/x)-1]}=9.01x+4x^2\log(e^{0.1/x}-1)\nonumber\\
&\xrightarrow{x<0.1}&\approx 9.4x\label{eq:Y1}\\
Y_2(x)&=& 2x\int_{0.1}^{\infty}d\tilde{p}\frac{\exp(\p/x)-\exp(\p)}{[\exp(\p)-1][\exp(\p/x)-1]}\log\Big(64\p^2\sinh(\p/2)\sinh(\p/(2x))x\Big)\nonumber\\
&\xrightarrow{x<0.1}&\approx 0.71x+1.6+4.7x\log(x)\label{eq:Y2}\\
Y_3(x)&=&x\int_{0.1}^{\infty}d\tilde{p}\frac{\exp(\p/x)-\exp(\p)}{[\exp(\p)-1][\exp(\p/x)-1]}\log\Big(8\p\sinh(\p/2)\Big)\log\Big(8\p\sinh(\p/(2x))x\Big)\nonumber\\
&\xrightarrow{x<0.1}&\approx 3.2x-0.82x\log(x)+1.3.\label{eq:Y3}
\end{eqnarray}

To evaluate the integral in the region $\p<0.1$ we first consider the case where $T_a, T_b \gg M\f$, allowing us to approximate $\Mf\ll x\ll 1$. Next, note that the integrand $D(\Mf^2,\p,x,0,0)$ is peaked near $\p\sim \Mf$. Near this peak we can use the approximation $\p\ll x$. Assuming the contribution from the peak dominates the integral, we extend the approximation $\p\ll x$ to the entire integration range $\p\in (0,0.1)$, yielding
\begin{align}\label{eq:domain_p_less_1}
\int_0^{0.1}d\tilde{p}\ D(\Mf^2,\p,x,0)\bigg|_{\p,\Mf\ll x}&\approx \int_0^{0.1}d\tilde{p}\frac{ (1-x)}{\sqrt{\p^2+\Mf^2}}
\log^2\Bigg(\frac{\sqrt{\tilde{p}^2+\Mf^2}+\tilde{p}}{\sqrt{\tilde{p}^2+\Mf^2}-\tilde{p}}\Bigg)\nonumber\\
&\approx \frac{4}{3}(1-x)\log^3\Big(0.2\frac{T_a}{M\f}\Big).
\end{align}
In the case $T_b\ll M\f\ll T_a$ the assumptions we used above no longer hold.  One can instead use the approximations $\p,\Mf\ll1$ along with $e^{\Mf/x}\gg1$ to simplify the integral and show that its contribution is always dwarfed by the contribution from $\p>0.1$. For brevity we do not show the calculations here. 
Thus we can neglect contributions from $\p < 0.1$ in eq.~\eqref{eq:highT_delta}  when $T_b<M\f$\footnote{In fact, even when $\Mf\ll x$, the contributions from the $p<0.1$ integral remains sub-dominant until extremely large temperatures, $T_a\geq 10^5M\f$.}.   We find empirically that using  eq.~\eqref{eq:domain_p_less_1} for all $T_b$  helps improve the agreement between the analytic estimate and the full numerical calculation for $T_a$ as low as  $T_a\sim M\f$. Thus we approximate the full collision term at high temperatures as
\begin{align}\label{eq:scalar_highT}
{\mathcal{C}}_{\textrm{high-T}}\approx&  S'32\pi^2\frac{{\mu}^2_aw}{w+1}{T}_a^3\bigg[\frac{4}{3}(1-x)x\log^3\Big(0.2\frac{T_a}{M\f}\Big)+Y_1(x)\log^2\Big(\frac{T_a}{M\f}\Big)\nonumber\\&+Y_2(x)\log\Big(\frac{T_a}{M\f}\Big)+Y_3(x)\bigg].
\end{align}
From the asymptotic behavior at small $x$ we see that ${\mathcal{C}}_{\textrm{high-T}}$ is largely insensitive to $T_b$. At extremely small $x$, the logarithmic term is dominant. However, at large temperatures $T\gtrsim 10^2M\f$, as $x$ increases to $x\sim0.5$, the higher powers of the logarithm take over and enhance the collision term by roughly two orders of magnitude. This enhancement is due to the Bose enhancement of the forward energy transfer. As $x$ further increases towards unity, the backward collision term starts catching up to forward collision term, eventually completely cancelling it at $x=1$. 

\begin{figure}
\begin{subfigure}{.5\textwidth}
\includegraphics[width=1.00\textwidth]{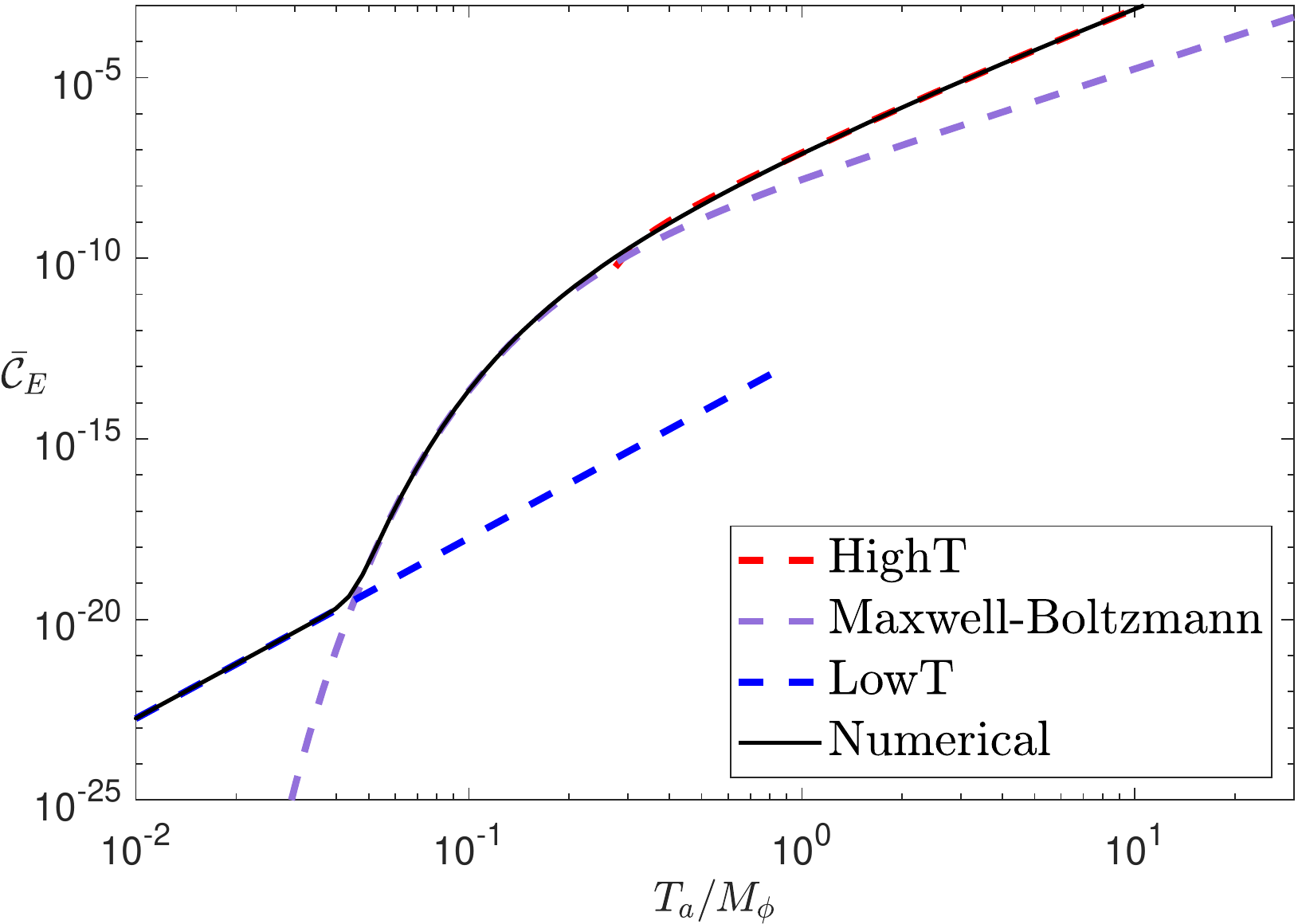}
\end{subfigure}
\begin{subfigure}{.5\textwidth}
\includegraphics[width=1.00\textwidth]{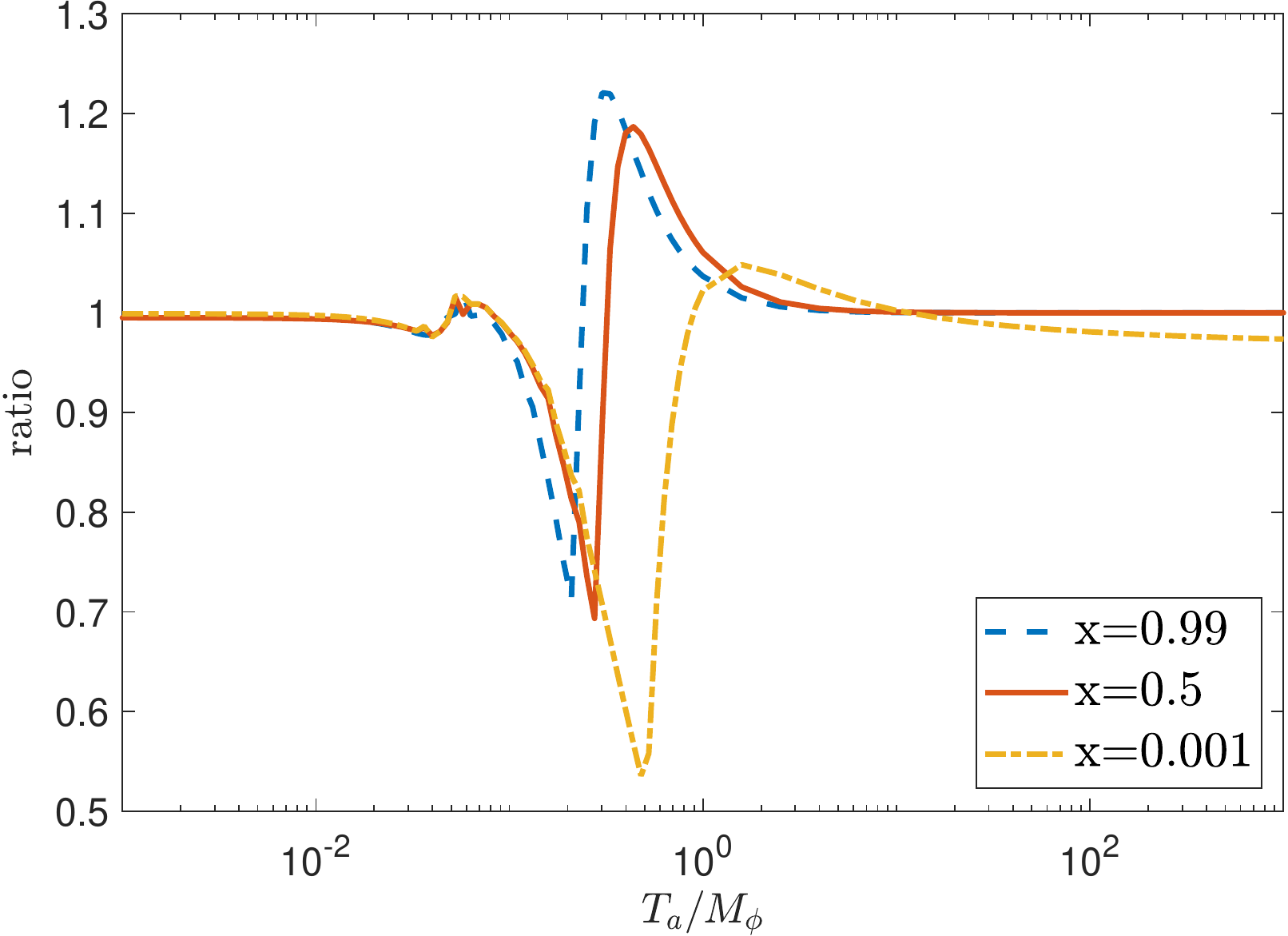}
\end{subfigure}
\caption{\textit{Left Panel:} Total energy transfer collision term $\bar{\mathcal{C}}_E=\mathcal{C}_E/M\f^5$ as a function of $T_a/M\f$ at fixed temperature ratio $x=T_b/T_a=0.5$. The black solid line corresponds to $\mathcal{C}_E$ numerically evaluated using eq.~\eqref{eq:collision} with matrix element given in eq.~\eqref{eq:matrix_boson}, red dashed line corresponds to $\mathcal{C}_{\textrm{high-T}}$ (eq.~\eqref{eq:scalar_highT}),   purple dashed line corresponds to $\mathcal{C}_{\textrm{MB}}$ (eq.~\eqref{eq:scalar_MB}) and the blue dashed line corresponds to $\mathcal{C}_{\textrm{low-T}}$ (eq.~\eqref{eq:scalar_lowT}). \textit{Right Panel:} The ratio of the analytic approximation to $\mathcal{C}_E$ (eq.~\eqref{eq:scalar_fit}) with the full numerical value
as a function of $T_a/M\f$ for $x=0.99,0.5,0.001$.  Results are shown for $\mu_a=0.01M\f$,  $\mu_b/\mu_a=0.5$, and $m_{a,b}=10^{-8}M\f$.}
\label{fig:scalar_boson_analytical_compare}
\end{figure}

In the left panel of figure~\ref{fig:scalar_boson_analytical_compare}, we compare our high temperature estimate with the numerically evaluated collision term.  The Bose-Einstein enhancement over the classical Maxwell-Boltzmann result is clearly visible at high temperatures.

\paragraph{Intermediate temperatures, $\bm{T_a\leq M\f}$.} As $\Mf$ begins to exceed unity, the Dirac delta contribution to the matrix element ensures that the integral of  eq.~\eqref{eq:collision} has support dominantly at $\s=\Mf>1$.   However, here the phase space distribution functions, contributing through the factor $D$, are exponentially suppressed. This Boltzmann suppression causes the collision term to fall sharply.   In other words, in the intermediate temperature regime  the integral receives its  dominant contribution from an energy scale much larger than either temperature, which means that to excellent approximation the scattering here can be described using classical statistics.

Using classical statistics, the overall integral over $\p$ (eq.~\eqref{eq:highT_delta}) can be performed exactly,
\begin{align}\label{eq:scalar_MB}
{\mathcal{C}}_{\textrm{MB}}&=  S' 32\pi^2M\f^2\dfrac{{\mu}_a^2w}{(w+1)}\dfrac{{T}_a}{4}\Big(K_2\Big(\frac{M\f}{{T}_a}\Big)-xK_2\Big(\frac{M\f}{x{T}_a}\Big)\Big),
\end{align}
where $K_2$ is the modified Bessel function of second kind. Again, we can see that at small $x$ the collision term becomes insensitive to variations in the colder sector. As the temperatures fall further below the inflaton mass, the collision term becomes Boltzmann-suppressed.

\paragraph{{Low temperature limit, $\bm{m_{a,b}\ll T\ll M\f}$}.} In the low-temperature regime, the integral is dominated by off-shell inflaton scattering, described by the Heaviside term in  eq.~\eqref{eq:matrix_delta}. Thus at low temperatures we need to evaluate
\begin{eqnarray}
{\mathcal{C}}_{\textrm{low-T}}&\approx&  S'\frac{xT_a^5}{M\f^4}{\mu}^4_aw\int_0^{\infty}\int_{0}^{\Mf^2}d\tilde{p}\ d\tilde{s} D(\s,\p,x,0).
\end{eqnarray}
As $D$ is exponentially suppressed at large values of $\s$, we can take the upper limit of the $\s$ integral to  infinity with negligible errors. Both the integrand and the limits of integration thus become independent of temperature, giving 
\begin{align}\label{eq:scalar_lowT}
{\mathcal{C}}_{\textrm{low-T}}&\approx S'\frac{T_a^5}{M\f^4}{\mu}^4_aw\bigg[x\int_0^{\infty}\int_{0}^{\infty}d\tilde{p}\ d\tilde{s} D(\s,\p,x,0)\bigg],\nonumber\\
&\equiv  S'\frac{T_a^5}{M\f^4}{\mu}^4_awf(x)\xrightarrow{x<0.1} 7.9 S'\frac{T_a^5}{M\f^4}{\mu}^4_aw.
\end{align}
Again, as required, we find that the energy transfer function becomes insensitive to the colder sector as $x\rightarrow0$.

\paragraph{Total collision term.}

To get a complete analytic estimate of $\mathcal{C}_E$ over all temperature ranges we combine the analytic estimates as
\begin{align}\label{eq:scalar_fit}\nonumber
\br{\mathcal{C}}_E(T_a) & =  \mathcal{C}_{\textrm{low-T}}\Theta(M\f-T_a)  +\br{\mathcal{C}}_{\textrm{high-T}}\Theta({T}_a-M\f)+\br{\mathcal{C}}_{\textrm{MB}}\Theta(0.2M\f-{T}_a) \\ & \quad +\textrm{max}(\br{\mathcal{C}}_{\textrm{MB}},\br{\mathcal{C}}_{\textrm{high-T}})\Theta(M\f-{T}_a)\Theta({T}_a-0.2M\f), 
\end{align}
where $\br{\mathcal{C}}_{\textrm{high-T}},\br{\mathcal{C}}_{\textrm{MB}}$ and $\br{\mathcal{C}}_{\textrm{low-T}}$ are given in eq.~\eqref{eq:scalar_highT}, \eqref{eq:scalar_MB} and \eqref{eq:scalar_lowT}. The Heaviside functions ensure that each function contributes only in its region of validity. This function describes the collision term for all temperature ranges as long as the scattering particles remain relativistic ($T_{a,b}\gg m_{a,b}$).

In the left panel of figure~\ref{fig:scalar_boson_analytical_compare}, we compare the analytic approximations to the energy transfer collision term derived in high-, low-, and intermediate temperature regions with the exact numerical value. For illustrative purposes, each approximation is shown over a range larger than that taken in eq.\ \eqref{eq:scalar_fit}.  Together these approximations accuratelly describe the behavior of $\mathcal{C}$ in their respective regions.  In the right panel of figure~\ref{fig:scalar_boson_analytical_compare}, we compare the ratio of our analytic approximation, eq.~\eqref{eq:scalar_fit}, to the numerically evaluated collision term using the full matrix element of eq.~\eqref{eq:matrix_boson}. The largest deviation occurs during the transition from $\mathcal{C}_{\textrm{high-T}}$ to $\mathcal{C}_{\textrm{MB}}$ between $M_{\phi}/4<T_a<M\f$ and is of the order $\sim 50\%$.

\subsection{Yukawa couplings}\label{appendix:s_fermions}
We next consider two Dirac fermions, $\psi_a$ and $\psi_b$, interacting with a scalar inflaton $\phi$ via
\begin{eqnarray}
\mathcal{L}_{\textrm{int}}=y_a\phi\bar{\psi}_a\psi_a+y_b\phi\bar{\psi}_b\psi_b.
\end{eqnarray}
In this case we have $\zeta_{a,b}=1$ and $ S'=1/(4(2\pi)^5)$.

The $s$-channel scattering amplitude in this theory, for $m_{a,b}\ll M\f$, is given by
\begin{equation}\label{eq:matrix_fermion}
|\overline{\mathcal{M}}(s)|^2=4y_a^2y_b^2\bigg(1-\frac{4m_a^2}{s}\bigg)\bigg(1-\frac{4m_b^2}{s}\bigg)\frac{s^2}{(s-M^2_{\phi})^2+\Big(\Gamma_{0a}+\Gamma_{0b}\Big)^2},
\end{equation}
where
\begin{align}
 \Gamma_{0a,b} = \frac{y_{a,b}^2M\f}{8\pi}.
\end{align}
For small $y_{a,b}$ the scattering amplitude can be approximated as \cite{Adshead:2016xxj}
\begin{align}\label{eq:matrix_delta_fermion}\nonumber
|\overline{\mathcal{M}}(\s)|^2\approx  8\pi^2\frac{4y_a^2w}{w+1}\Mf^2\delta(\s-\Mf^2)&+\Theta(\Mf^2-\s)\bigg(1-\frac{4\tilde{m}_a^2}{\s}\bigg)\bigg(1-\frac{4\tilde{m}_b^2}{\s}\bigg) 4y_a^4w\frac{\s^2}{\Mf^4}\\ +\Theta(\s-\Mf^2) 4y_a^4w.
\end{align}
where $w=\Gamma_{0,b}/\Gamma_{0,a}=y_b^2/y_a^2$.

To  estimate $\mathcal{C}_E$ analytically, we combine the simplified form of the scattering amplitude given in eq.~\eqref{eq:matrix_delta_fermion} along with the approximations $\Mf\ll1$ and $\Mf\gg1$ at high and low temperatures respectively. Moreover, since for fermions the contribution of the distribution functions to the integrand, $D$, is regular at $\s,\p=0$, the limit ${m}_{a,b}=0$ does not need any special attention.

\paragraph{{High temperature limit, $\bm{T_a\gg M\f}$}.} In the high temperature limit only the Dirac delta term and the second Heaviside theta term in eq.~\eqref{eq:matrix_delta_fermion} contribute to the integral. The collision term then becomes
\begin{align}\label{eq:fermion_highT_delta}
{\mathcal{C}}_{\textrm{high-T}}&\approx S'xT_a^5\bigg[8\pi^2\frac{4y_a^2w}{w+1}\Mf^2\int_0^{\infty}d\tilde{p}\ D(\Mf^2,\p,x,0)+4y_a^4w\int_0^{\infty}d\tilde{p}\int_{\Mf^2}^{\infty}d\s D(\s,\p,x,0)\bigg].
\end{align}
Note that as $\s\rightarrow0$ the integrand $D(\s,\p,x,0)$ asymptotes to a finite value over all $\p$. Thus, we can safely approximate $\Mf=0$ in the integrand, making the integrals independent of $T_a$,
\begin{align}\label{eq:fermion_highT}
{\mathcal{C}}_{\textrm{high-T}}=S'\bigg[8\pi^2M\f^2\frac{4y_a^2w}{w+1}V_1(x)T_a^3+4y_a^4wV_2(x)T_a^5\bigg].
\end{align}
where,
\begin{align}
V_1(x)&=x\int_0^{\infty}d\tilde{p}\ D(0,\p,x,0)
\xrightarrow{x<0.1} 0.29\\
V_2(x)&=x\int_0^{\infty}d\tilde{p}\int_{0}^{\infty}d\s D(\s,\p,x,0)\xrightarrow{x<0.1} 3.0
\end{align}
We can check that  at small $x$ we see that $\br{\mathcal{C}}_{\textrm{high-T}}$ is in this limit  insensitive to $T_b$, and at $x=1$ all these functions go to zero as backward energy transfer exactly balances the forward energy transfer.  The collision term at very high temperatures in this case is not sensitive to the inflaton mass.

\paragraph{Intermediate regime, $\bm{T_a\lesssim M\f}$.} 
For $T_a\sim M\f$ the Dirac delta part of the scattering amplitude will dominate the collision term. As discussed in section~\ref{appendix:s_scalar} above for scalars, the collision term can be well approximated using Maxwell-Boltzmann statistics as the temperature drops below the inflaton mass scale, $T_a<M\f$. Thus, the collision term can be simply written as
\begin{align}\label{eq:fermion_MB}
{\mathcal{C}}_{\textrm{MB}}&= S'8\pi^2M\f^4\frac{4y_a^2w}{w+1}\dfrac{{T}_a}{4}\Big(K_2\Big(\frac{M\f}{{T}_a}\Big)-xK_2\Big(\frac{M\f}{x{T}_a}\Big)\Big),
\end{align}
where $K_2$ is the modified Bessel function of the second kind. 

\paragraph{Low temperature regime, $\bm{m_{a,b}\ll T\ll M\f}$.} 
In the low temperature regime, the integral is dominated by the $\Theta(\Mf^2-\s)$ term. Just as for scalars, we can  to a good approximation replace $\Mf\rightarrow\infty$ in the limit of integration. This yields
\begin{align}\label{eq:fermion_lowT}
{\mathcal{C}}_{\textrm{low-T}}&\approx S'\frac{4T_a^9}{M\f^4}y_a^4w\bigg[x\int_0^{\infty}\int_{0}^{\infty}d\tilde{p}\ d\tilde{s}\ \s^2D(\s,\p,x,0)\bigg],\nonumber\\
&\xrightarrow{x<0.1} 1.4\times 10^3S'y^4_aw4\frac{T_a^9}{M\f^4}.
\end{align}

\paragraph{Total collision term.}
\begin{figure}
\begin{subfigure}{.5\textwidth}
\includegraphics[width=1.00\textwidth]{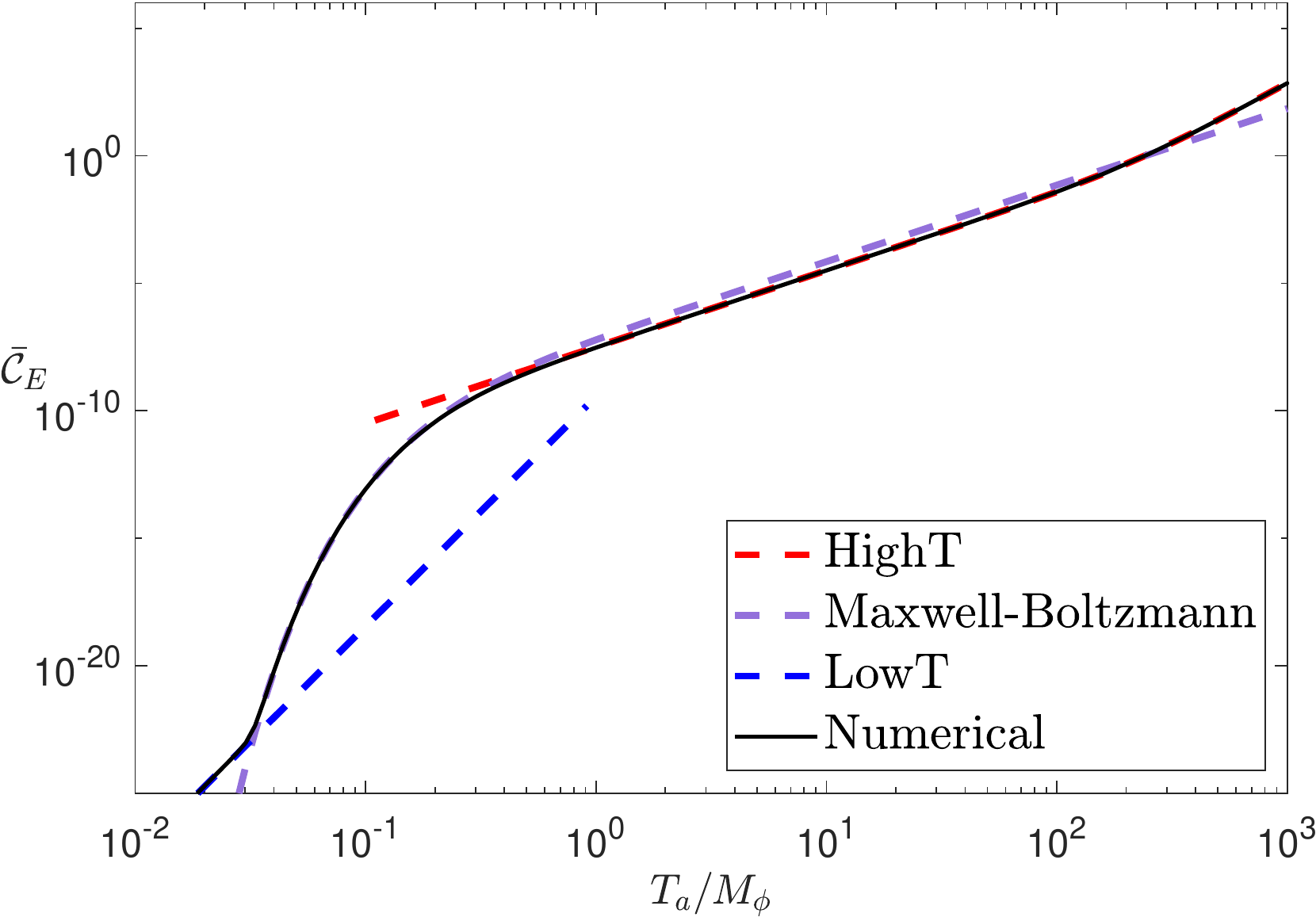}
\end{subfigure}
\begin{subfigure}{.5\textwidth}
\includegraphics[width=1.00\textwidth]{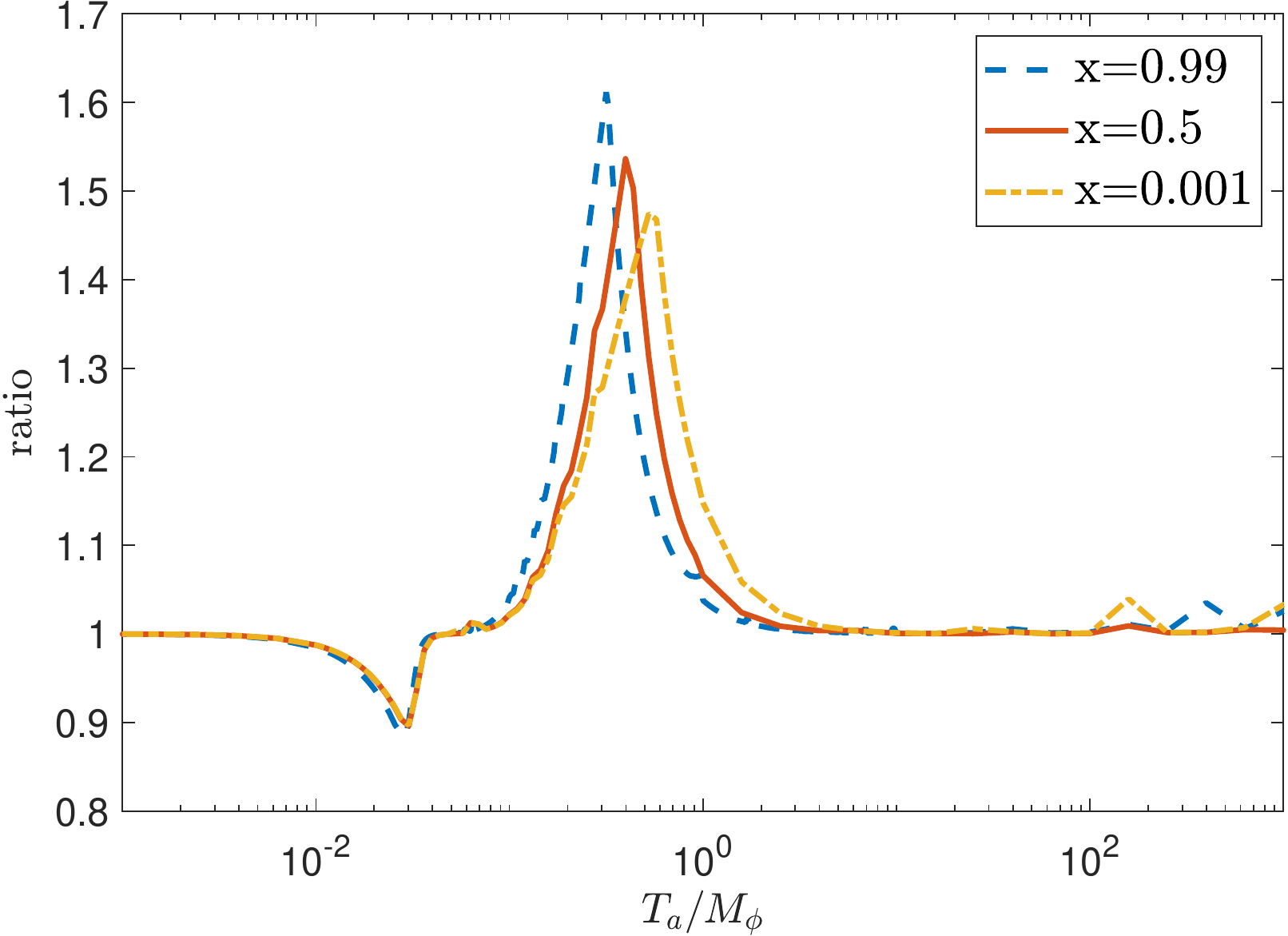}
\end{subfigure}
\caption{\textit{Left Panel:} Energy transfer collision term $\mathcal{C}_E/M\f^5$ as a function of $T_a/M\f$ for fixed $x=T_b/T_a=0.5$. The black solid line corresponds to $\mathcal{C}_E$ numerically evaluated using eq.~\eqref{eq:collision} and scattering amplitude as given in eq.~\eqref{eq:matrix_fermion}, the red dashed line corresponds to $\mathcal{C}_{\textrm{high-T}}$ (eq.~\eqref{eq:fermion_highT}), the purple dashed line corresponds to $\mathcal{C}_{\textrm{MB}}$ (eq.~\eqref{eq:fermion_MB}) and the blue dashed line corresponds to $\mathcal{C}_{\textrm{low-T}}$ (eq.~\eqref{eq:fermion_lowT}). \textit{Right Panel:} Ratio of the analytic approximation to the total collision term as given in eq.~\eqref{eq:fermion_fit} to the numerical value as a function of $T_a/M\f$ for fixed temperature ratios $T_b/T_a=x=0.99,0.5,0.001$. Results are shown for $y_a=0.01$, $y_b/y_a=0.5$, and  $m_{a,b}=10^{-8}M\f$.}
\label{fig:fermion_analytical_compare}
\end{figure}
We combine the analytic estimates derived above to approximate the collision term for all temperatures as
\begin{align}\label{eq:fermion_fit}
\br{\mathcal{C}}_E=&\br{\mathcal{C}}_{\textrm{low-T}}\Theta(M\f-{T}_a)+\br{\mathcal{C}}_{\textrm{MB}}\Theta(0.2M\f-{T}_a)+\textrm{min}(\br{\mathcal{C}}_{\textrm{MB}},\br{\mathcal{C}}_{\textrm{high-T}})\Theta(M\f-{T}_a)\Theta({T}_a-0.2M\f)\nonumber\\&+\br{\mathcal{C}}_{\textrm{high-T}}\Theta({T}_a-M\f).
\end{align}
where $\br{\mathcal{C}}_{\textrm{high-T}},\br{\mathcal{C}}_{\textrm{MB}}$, and $\br{\mathcal{C}}_{\textrm{low-T}}$ are as described in eq.~\eqref{eq:fermion_highT},\eqref{eq:fermion_MB} and \eqref{eq:fermion_lowT}.
Eq.\ \eqref{eq:fermion_fit} predicts the collision term for all temperature ranges as long as the particles remain relativistic ($T_{a,b}\gg m_{a,b}$). 

In left panel of figure~\ref{fig:fermion_analytical_compare}, we compare our analytic estimate of the energy transfer collision term with the exact numerical value. Together these approximations accurately model the behavior of $\mathcal{C}_E$ in their respective regions. In the right panel of figure~\ref{fig:fermion_analytical_compare}, we compare the ratio of our analytic approximation, eq.~\eqref{eq:fermion_fit}, to the collision term numerically evaluated (eq.~\eqref{eq:collision}) with the scattering amplitude in eq.~\eqref{eq:matrix_fermion}. The largest deviation occurs during the transition from $\mathcal{C}_{\textrm{high-T}}$ to $\mathcal{C}_{\textrm{MB}}$ between $M_{\phi}/4<T_a<M\f$ and is of the order $\sim 50\%$.

\subsection{Axionic couplings to  gauge bosons}
Next, we consider two (Abelian) gauge bosons interacting with a (pseudoscalar) inflaton $\phi$ via
\begin{eqnarray}
\mathcal{L}_{\textrm{int}}=-\frac{1}{4\Lambda_a}\phi F_a^{\mu\nu}\tilde{F}_{a,\mu\nu}-\frac{1}{4\Lambda_b}\phi F_b^{\mu\nu}\tilde{F}_{b,\mu\nu}.
\end{eqnarray}
In this case we have $\zeta_{a,b}=-1$ and $S'=1/(16(2\pi)^5)$.

The $s$-channel scattering amplitude in this theory, for $m_{a,b}\ll M\f$, is given by
\begin{align}\label{eq:matrix_gauge_boson}
|\overline{\mathcal{M}}(s)|^2=& \frac{4}{128\Lambda_a^2\Lambda_b^2}\frac{s^4}{(s-M^2_{\phi})^2+\Big(\Gamma_{0a}+\Gamma_{0b}\Big)^2},  
\end{align}
where
\begin{align}
\Gamma_{0a,b}= & \frac{M\f^3}{256\pi \Lambda_{a,b}^2}.
\end{align}
For $\br{\Lambda}_{a,b}\gg1$, we approximate the scattering amplitude as \cite{Adshead:2016xxj}
\begin{align}\label{eq:matrix_delta_gauge_boson}
|\overline{\mathcal{M}}(\s)|^2\approx&  \frac{2\pi^2}{{\Lambda}_a^2}\frac{4w}{w+1}\frac{M\f^4}{T_a^2}\delta(\s-\tilde{M}^2_{\phi})+ \frac{4wT_a^8}{128{\Lambda}_a^4M\f^4}\Theta(\tilde{M}^2_{\phi}-\s)\s^4+\frac{4wT_a^4}{128{\Lambda}_a^4}\Theta(\s-\Mf^2) \s^2,
\end{align}
where $w=\Gamma_{0b}/\Gamma_{0a}=\Lambda_a^2/\Lambda_b^2$.

To  estimate $\mathcal{C}_E$ analytically, we combine the simplified scattering amplitude given in eq.~\eqref{eq:matrix_delta_gauge_boson} along with high and low-temperature approximations in the limits $\Mf \ll1$ and $\Mf\gg1$  respectively.

\paragraph{{High temperature limit, $\bm{T_a\gg M\f}$}.} At high temperatures both the Dirac delta contribution and the Heaviside theta term $\propto {\tilde s}^2$ contribute importantly to the integral. Subsequently we approximate the scattering amplitude as
\begin{align}\label{eq:gauge_boson_highT_delta}
{\mathcal{C}}_{\textrm{high-T}}=S'{xT_a^5}\bigg[&\frac{2\pi^2}{{\Lambda}_a^2}\frac{4w}{w+1}\frac{M\f^4}{T_a^2}\int_0^{\infty}d\tilde{p}\ D(\Mf^2,\p,x,0)+\frac{4wT_a^4}{128{\Lambda}_a^4}\int_0^{\infty}d\tilde{p}\int_{\Mf^2}^{\infty}d\s\ \s^2 D(\s,\p,x,0)\bigg].
\end{align}
In the above equation we have already assumed $\tilde{m}_{a,b}=0$. The first integral on the RHS is exactly the same as that evaluated for scalars at high temperatures. In the second integral, the integrand vanishes as $s\rightarrow 0$, allowing us to freely take $\Mf\approx 0$. This yields
\begin{align}\nonumber\label{eq:gauge_boson_highT}
{\mathcal{C}}_{\textrm{high-T}}=S'\bigg[&\frac{2\pi^2M\f^4}{{\Lambda}_a^2}\frac{4w}{w+1}T_a^3\bigg(\frac{4}{3}(1-x)x\log^3\Big(\frac{T_a}{M\f}\Big)+Y_1(x)\log^2\Big(\frac{T_a}{M\f}\Big)+Y_2(x)\log\Big(\frac{T_a}{M\f}\Big)\\ &+Y_3(x)\bigg)+ \frac{4w}{{\Lambda}_a^4}Z(x)T_a^9\bigg],
\end{align}
where
\begin{align}
Z(x)&=\frac{x}{128}\int_0^{\infty}d\tilde{p}\int_{0}^{\infty}d\s\ \s^2D(\s,\p,x,0)
\xrightarrow{x<0.1}\approx 14.
\end{align}
Here the $Y_i$ are defined in eq.~\eqref{eq:Y1}, \eqref{eq:Y2} and \eqref{eq:Y3}.

\begin{figure}
\begin{subfigure}{.5\textwidth}
\includegraphics[width=1.00\textwidth]{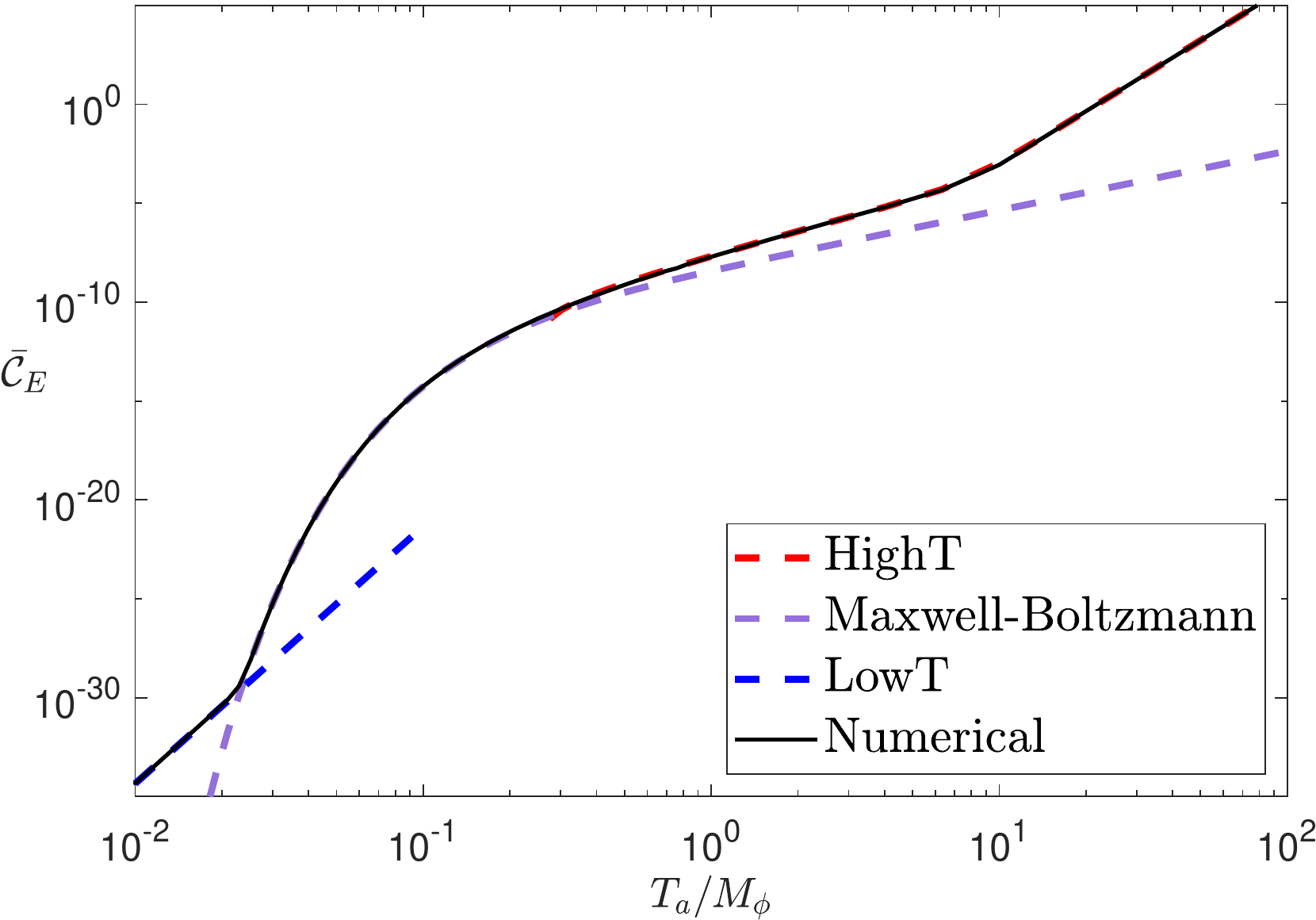}
\end{subfigure}
\begin{subfigure}{.5\textwidth}
\includegraphics[width=1.00\textwidth]{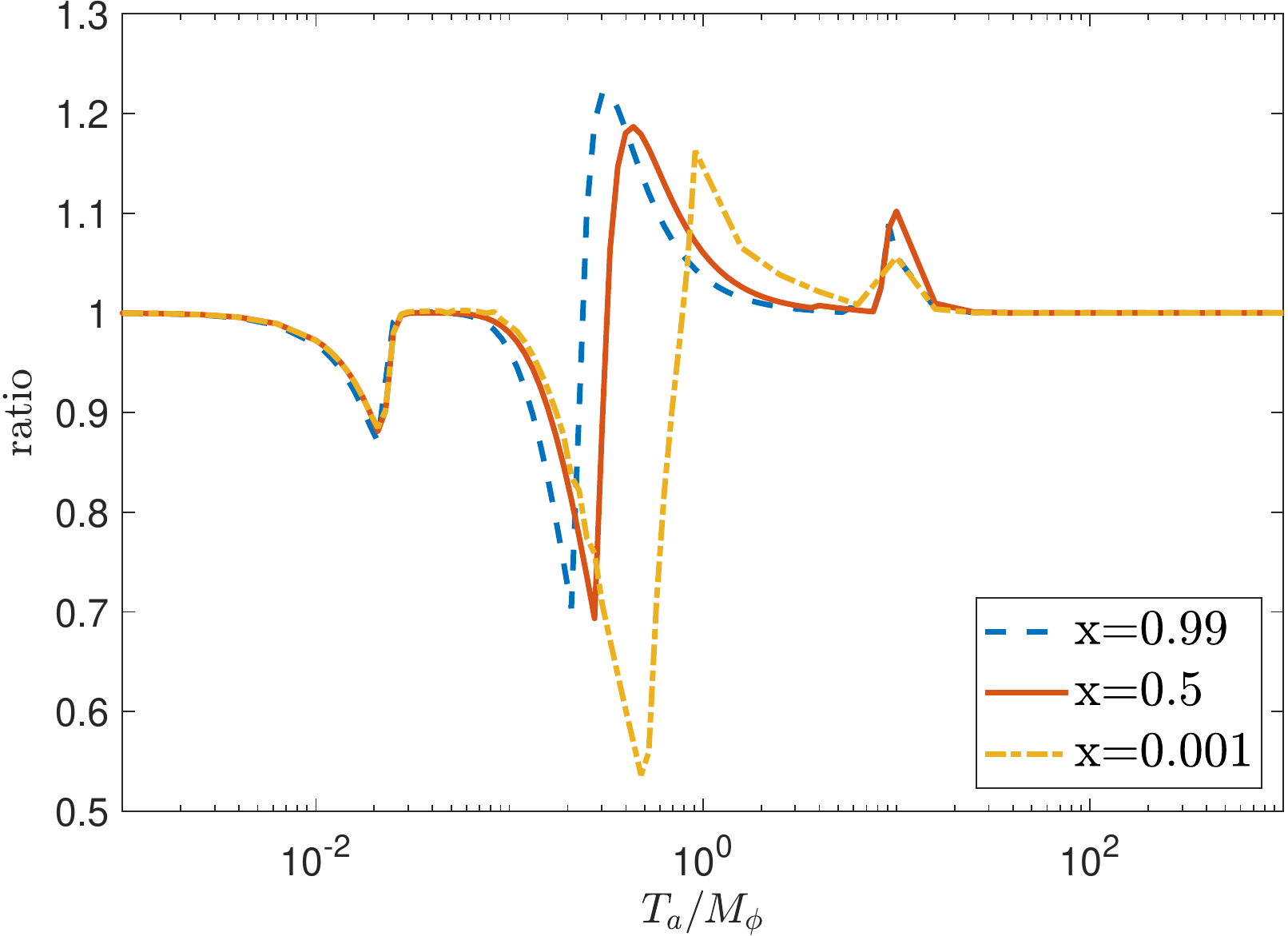}
\end{subfigure}
\caption{\textit{Left Panel:} Total  energy transfer collision term $\mathcal{C}_E/M\f^5$  as a function of $T_a/M\f$ at fixed $x=T_b/T_a=0.5$.  The black solid line corresponds to $\mathcal{C}_E$ numerically evaluated using eq.~\eqref{eq:collision} with scattering amplitude given by eq.~\eqref{eq:matrix_gauge_boson}, the red dashed line corresponds to $\mathcal{C}_{\textrm{high-T}}$ (eq.~\eqref{eq:gauge_boson_highT}),  the purple dashed line corresponds to $\mathcal{C}_{\textrm{MB}}$ (eq.~\eqref{eq:gauge_boson_MB}) and the blue dashed line corresponds to $\mathcal{C}_{\textrm{low-T}}$ (eq.~\eqref{eq:gauge_boson_lowT}). \textit{Right Panel:} Ratio of our analytic  estimate as given in eq.~\eqref{eq:gauge_boson_fit} to the numerically evaluated collision term as a function of $T_a/M\f$ for fixed temperature ratios $T_b/T_a=x=0.99,0.5,0.001$. Here we fix $\Lambda_a=100M\f$, $\frac{1/\Lambda_b}{1/\Lambda_a}=0.5$ and $m_{a,b}=10^{-8}M\f$.}
\label{fig:gauge_boson_analytical_compare}
\end{figure}

In the left panel of figure~\ref{fig:gauge_boson_analytical_compare} we compare this high temperature estimate with the numerically evaluated collision term. 

\paragraph{Intermediate temperatures, $\bm{T_a\leq M\f}$.}
In the intermediate regime near $T_a\sim M\f$, integral can again be well approximated with Maxwell-Boltzmann distribution. Thus the collision term can be simply written as
\begin{align}\label{eq:gauge_boson_MB}
{\mathcal{C}}_{\textrm{MB}}&= S'\frac{2\pi^2M\f^6}{{\Lambda}_a^2}\frac{4w}{w+1}\dfrac{{T}_a}{4}\Big(K_2\Big(\frac{M\f}{{T}_a}\Big)-xK_2\Big(\frac{M\f}{x{T}_a}\Big)\Big),
\end{align}
where $K_2$ is the modified Bessel function of the second kind. 

\paragraph{Low temperature limit $\bm{m_{a,b}\ll T\ll M\f}$}
In this regime, the  $\Theta(\Mf^2-\s)$ term dominates in eq.~\eqref{eq:collision}.  We can again take  $\Mf\rightarrow\infty$ in the limit of integration. This yields
\begin{align}\label{eq:gauge_boson_lowT}
{\mathcal{C}}_{\textrm{low-T}}&\approx S'T_a^{13}\frac{4w}{M\f^4{\Lambda}_a^4}\bigg[\frac{x}{128}\int_0^{\infty}\int_{0}^{\infty}d\tilde{p}\ d\tilde{s}\ \s^4D(\s,\p,x,0)\bigg],\nonumber\\
&\xrightarrow{x<0.1} 7.1\times 10^4S'\frac{4w}{M\f^4{\Lambda}_a^4}T_a^{13}.
\end{align}

\paragraph{Total collision term.}
We combine the analytic estimates in the following manner, 
\begin{align}\label{eq:gauge_boson_fit}
\br{\mathcal{C}}_E=\br{\mathcal{C}}_{\textrm{low-T}}\Theta(M\f-{T}_a)+\textrm{max}(\br{\mathcal{C}}_{\textrm{MB}},\br{\mathcal{C}}_{\textrm{high-T}})\Theta(M\f-{T}_a)+\br{\mathcal{C}}_{\textrm{high-T}}\Theta({T}_a-M\f).
\end{align}
where $\br{\mathcal{C}}_{\textrm{high-T}},\br{\mathcal{C}}_{\textrm{MB}}$, and $\br{\mathcal{C}}_{\textrm{low-T}}$ are as described in eq.~\eqref{eq:gauge_boson_highT},\eqref{eq:gauge_boson_MB} and \eqref{eq:gauge_boson_lowT}. This function  describes the collision term for all temperature ranges as long as the scattering particles remain relativistic ($T_{a,b}\gg m_{a,b}$).

In the left panel of figure~\ref{fig:gauge_boson_analytical_compare}, we compare our analytic estimate derived in the three regions with the numerical value.   In the right panel of figure~\ref{fig:gauge_boson_analytical_compare}, we compare the ratio of our analytic  estimate, eq.~\eqref{eq:gauge_boson_fit}, to the numerically evaluated collision term (eq.~\eqref{eq:collision}) with scattering amplitude given in eq.~\eqref{eq:matrix_gauge_boson}. The largest deviation occurs during transition from $\mathcal{C}_{\textrm{high-T}}$ to $\mathcal{C}_{\textrm{MB}}$ between $M_{\phi}/4<T_a<M\f$ and is of the order $\sim 50\%$.

\subsection{Mixed Yukawa and scalar trilinear couplings}
In this case we consider a Dirac fermion $\psi$ and a scalar field $\chi$ that interact with the inflaton $\phi$ via 
\begin{eqnarray}
\mathcal{L}_{\textrm{int}}=\frac{1}{2}\mu_a\phi\chi_a\chi_a+y_b\phi\bar{\psi}_b\psi_b.
\end{eqnarray}
Note that sector $a$ is not necessarily hotter in this scenario. For this theory  $\zeta_{a}=-1$, $\zeta_{b}=+1$, and $S'=1/(8(2\pi)^5)$.

The spin-summed  $s$-channel scattering amplitude in this theory, for $m_{a,b}\ll M\f$, is given by
\begin{align}\label{eq:matrix_fermion_boson}
|\overline{\mathcal{M}}(s)|^2=2\mu_a^2y_a^2\bigg(1-\frac{4m_{b}^2}{s}\bigg)\frac{s}{(s-M^2_{\phi})^2+(\Gamma_{0a}+\Gamma_{0b})^2},
\end{align}
where
\begin{align}
\Gamma_{0a}=\frac{\mu_a^2}{32\pi M\f}, \qquad \Gamma_{0b}=\frac{y_b^2M\f}{8\pi}.
\end{align}
For $y,\br{\mu}\ll1$ we approximate the scattering amplitude as \cite{Adshead:2016xxj}
\begin{align}\label{eq:matrix_delta_fermion_boson}
|\overline{\mathcal{M}}(\s)|^2\approx 16\pi^2{\mu}_a^2\frac{w}{w+1}\frac{1}{T_a^2}\delta(\s-\tilde{M}^2_{\phi})+\frac{1}{2}\bigg(1-\frac{4m_{b}^2}{s}\bigg)\Theta(\tilde{M}^2_{\phi}-\s) \frac{\mu_a^4w}{M\f^6}T_a^2\s,
\end{align}
where $w=\Gamma_{0b}/\Gamma_{0a}=4y_b^2M\f^2/{\mu}_a^2$.

To  estimate the behavior of $\mathcal{C}_E$ analytically, we combine the simplified form of the scattering amplitude given in eq.~\eqref{eq:matrix_delta_fermion_boson} along with the high- and low-temperature approximations $\Mf\ll1$ and $\Mf\gg1$.  

\paragraph{{High temperature limit, $\bm{T\gg M\f}$}.}
When the temperature of the hotter sector, $T=\textrm{max}(T_{a},T_{b})$, is larger than the inflaton mass, the Dirac-delta term of the scattering amplitude dominates in the integral in eq.~\eqref{eq:collision}, giving
\begin{align}\label{eq:fermion_boson_highT_delta}
{\mathcal{C}}_{\textrm{high-T}}= S'xT_a^5\bigg[16\pi^2{\mu}_a^2\frac{w}{w+1}\frac{1}{T_a^2}\int_0^{\infty}d\tilde{p}\ D(\Mf^2,\p,x,0)\bigg].
\end{align}
As the scalars $\chi_a$ follow BE distribution, $D(\Mf^2,\p,x,0)$  has a pole as $\Mf^2\rightarrow 0$. We cannot simply approximate $\Mf=0$, and we proceed analogously to the case of section \ref{appendix:s_scalar} and split the integral as,
\begin{align}
\int_{0}^{\infty}d\tilde{p}\ D(\Mf^2,\p,x,0)&=\int_{0}^{\Mf}d\tilde{p}\ D(\Mf^2,\p,x,0)\bigg|_{\p,\Mf\ll 1}+\int_{\Mf}^{\infty}d\tilde{p}\ D(\Mf^2,\p,x,0)\bigg|_{\p\gg\Mf}.
\end{align}
The integrand in the first term on RHS vanishes for small $\p$. Hence, for $\Mf\ll1$ the first integral can be ignored. Expanding in $\p\gg \Mf$ limit, the second integral yields
\begin{align}
\int_{0}^{\infty}d\tilde{p}\ D(\Mf^2,\p,x,0)\approx& \int_{\Mf}^{\infty}d\tilde{p}\ \frac{\exp(\p/x)-\exp(\p)}{[\exp(\p)-1][\exp(\p/x)-1]}\Big[\log(8\p \sinh(\p/2))\log(\cosh(\p/2/x))\nonumber\\&+2\log\Big(\frac{1}{\Mf}\Big)\log(\cosh(\p/2/x))\Big].
\end{align}
As the integrand above diverges only logarithmically as $\p\rightarrow0$, the integral is insensitive to its lower limit, which can be replaced with $0$ with negligible error. The collision term at high temperatures can then be simply written as
\begin{align}\label{eq:fermion_boson_highT}
{\mathcal{C}}_{\textrm{high-T}}&=S'16\pi^2\frac{{\mu}_a^2w}{w+1}T_a^3\Big(W_1(x)\log\Big(\frac{T_a}{M\f}\Big)+W_2(x)\Big),
\end{align}
where
\begin{align}\nonumber
W_1(x)& = 2x\int_{0}^{\infty}d\tilde{p}\frac{\exp(\p/x)-\exp(\p)}{[\exp(\p)-1][\exp(\p/x)-1]}\log(\cosh(\p/2/x))
\longrightarrow  \begin{cases} 1.6 & x<0.1 \\ -0.48x^2 & x>10 \end{cases}\\
W_2(x)& =x\int_{0}^{\infty}d\tilde{p}\frac{\exp(\p/x)-\exp(\p)}{[\exp(\p)-1][\exp(\p/x)-1]}\log(8\p \sinh(\p/2))\log(\cosh(\p/2/x))\nonumber\\
&\qquad  \longrightarrow  \begin{cases} 1.1 & x<0.1 \\ -0.30x^3 & x>10 \end{cases}
\end{align}
Note that the high temperature collision term is approximately insensitive to the colder sector in this case. The collision term's logarithmic sensitivity on the inflaton mass depends on whether or not the scalars are hotter than the fermions. However, in both the cases the collision term is IR-sensitive due to the dependence on inflaton mass.

\begin{figure}

\begin{subfigure}{.5\textwidth}
\includegraphics[width=1.00\textwidth]{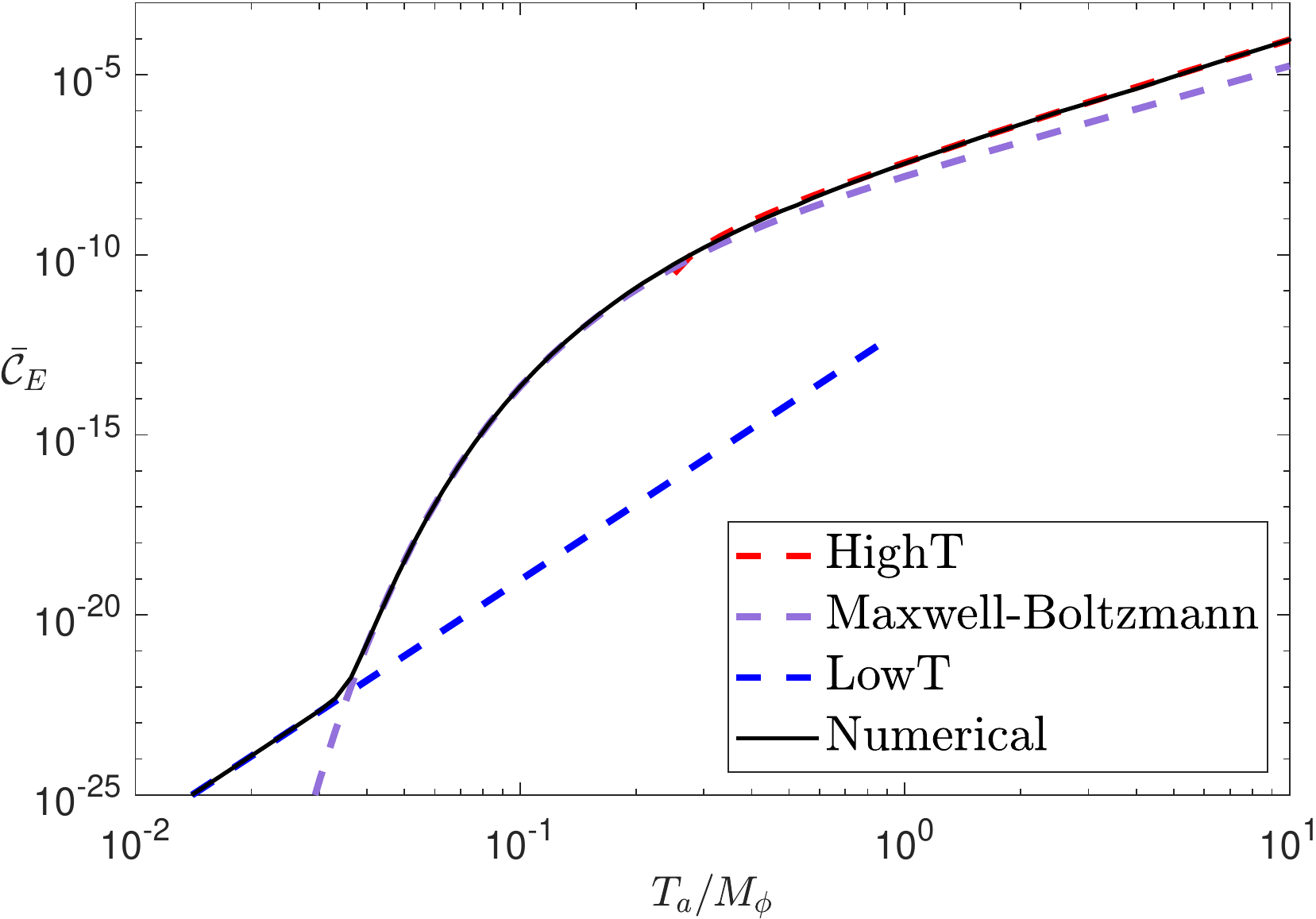}
\end{subfigure}
\begin{subfigure}{.5\textwidth}
\includegraphics[width=1.00\textwidth]{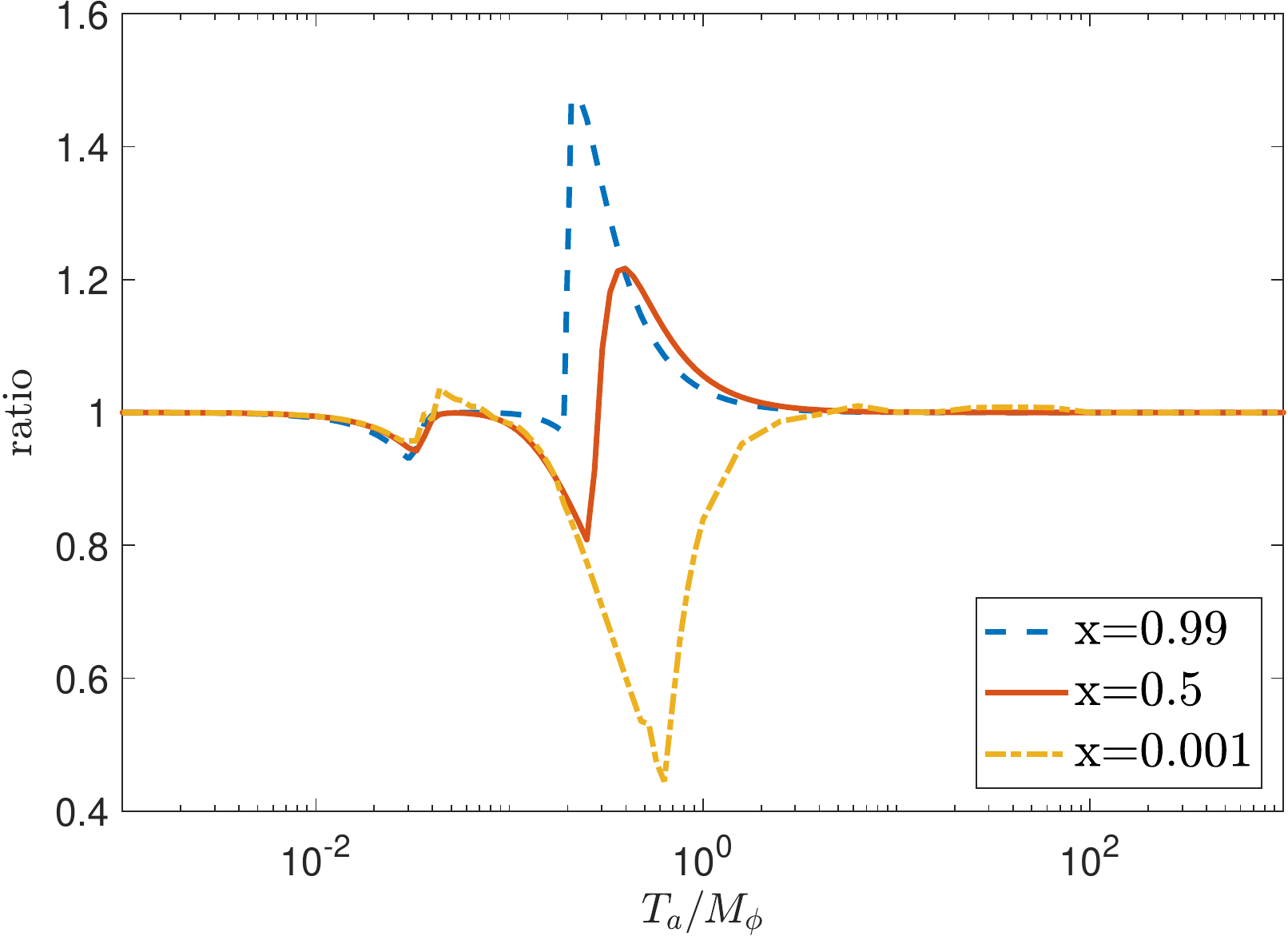}
\end{subfigure}

\begin{subfigure}{.5\textwidth}
\includegraphics[width=1.00\textwidth]{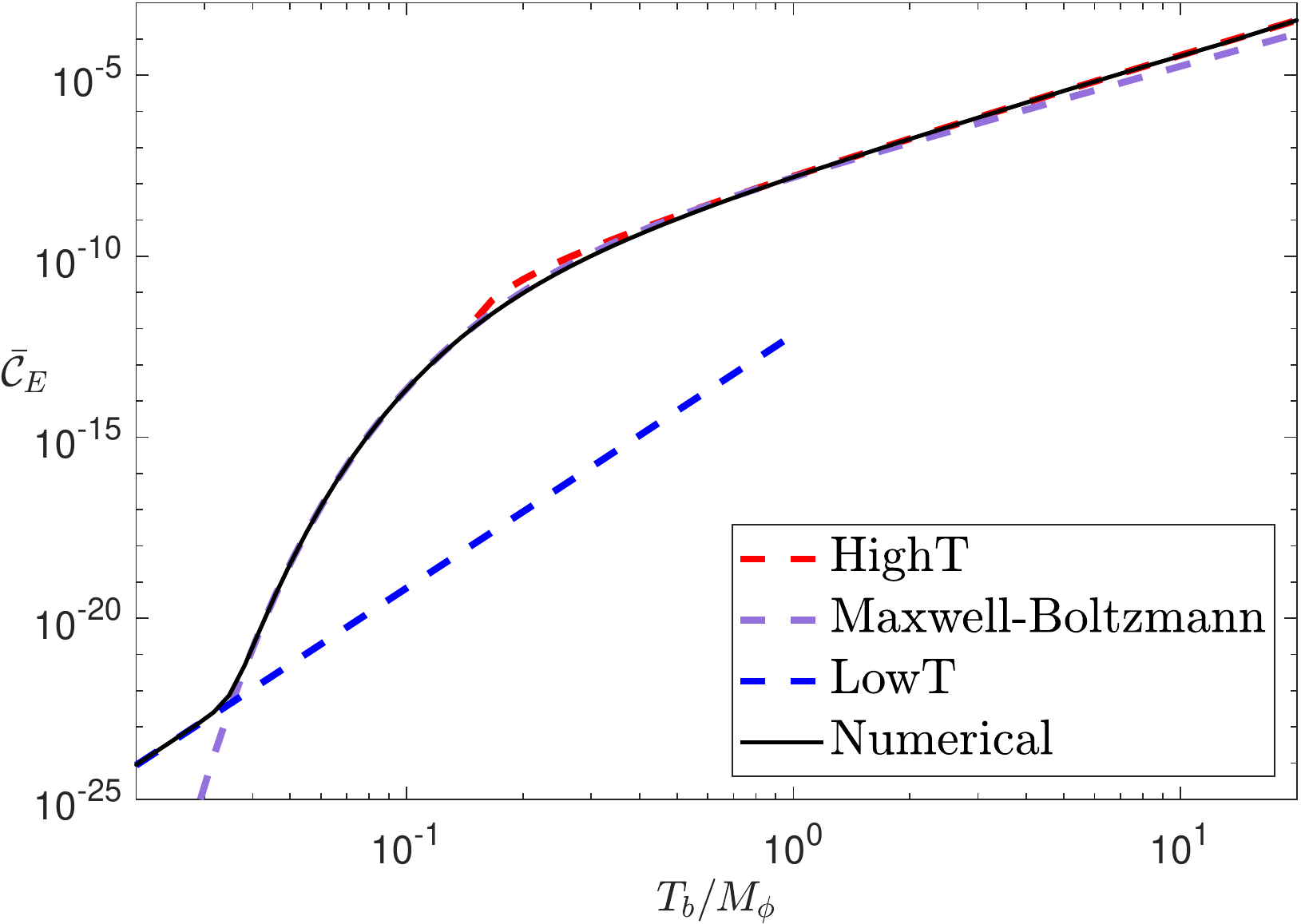}
\end{subfigure}
\begin{subfigure}{.5\textwidth}
\includegraphics[width=1.00\textwidth]{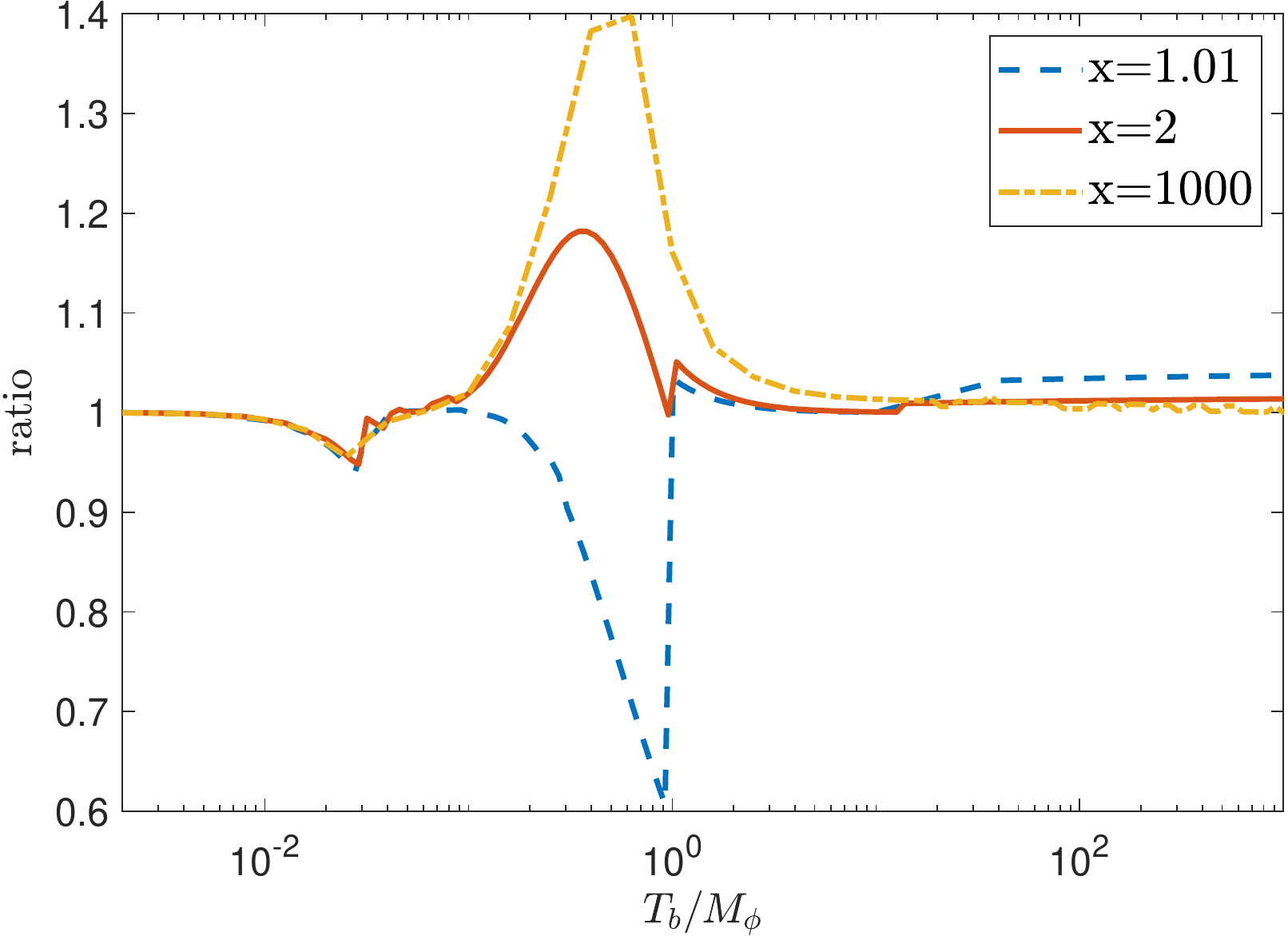}
\end{subfigure}
\caption{\textit{Left Panel:} Magnitude of total energy transfer collision term $\bar{\mathcal{C}}_E=\textrm{abs}(\mathcal{C}_E)/M\f^5$ as a function of $T=\textrm{max}(T_{a},T_{b})$ at fixed temperature ratio $x=T_b/T_a=0.5$ (top left) and $x=2$ (bottom left). The collision term is always positive for the top left panel and always negative for the bottom left.
 The black solid line corresponds to $\mathcal{C}$ numerically evaluated using eq.~\eqref{eq:collision}, the red dashed line corresponds to $\mathcal{C}_{\textrm{high-T}}$ (eq.~\eqref{eq:fermion_boson_highT}), the purple dashed line corresponds to $\mathcal{C}_{\textrm{MB}}$ (eq.~\eqref{eq:fermion_boson_MB}) and the blue dashed line corresponds to $\mathcal{C}_{\textrm{low-T}}$ (eq.~\eqref{eq:fermion_boson_lowT}). \textit{Right Panel:} 
Ratio of the analytic approximation to $\mathcal{C}_E$ (eq.~\eqref{eq:fermion_boson_fit}) to the full numerical value
as a function of $T=\textrm{max}(T_{a},T_{b})$ for fixed temperature ratios $T_b/T_a=x=0.99,0.5,0.001$ (top right) and $x=1.01,1,1000$ (bottom right). Results are shown for $\mu_a=0.01M\f$,  $2y_bM\f/\mu_a=0.5$, and $m_{a,b}=10^{-8}M\f$.}
\label{fig:fermion_boson_analytical_compare}
\end{figure}

In left panels of figure~\ref{fig:fermion_boson_analytical_compare} we compare our high temperature estimate with the numerically evaluated collision term. 

\paragraph{Intermediate temperatures, $\bm{T\leq M\f}$.} For temperatures near $T\sim M\f$ the Dirac delta part of the scattering amplitude dominates the behavior of collision term. As discussed above, in this region, the distribution functions are well approximated by Maxwell-Boltzmann distributions as temperature drops below the inflaton mass scale, $T<M\f$. The collision term can therefore be simply computed as,
\begin{align}\label{eq:fermion_boson_MB}
{\mathcal{C}}_{\textrm{MB}}&= S' 16\pi^2\dfrac{{\mu}_a^2w}{(w+1)}\dfrac{M\f^2}{4}\Big(T_aK_2\Big(\frac{M\f}{{T}_a}\Big)-T_bK_2\Big(\frac{M\f}{{T}_b}\Big)\Big),
\end{align}
where $K_2$ is the modified Bessel function of second kind. Again, we can see that at small/large $x$ the collision term becomes insensitive to variations in the colder sector.

\paragraph{Low temperature limit, $\bm{m_{\chi,\psi}\ll T\ll M\f}$.}
In the low-temperature regime, the contribution from the Dirac delta part of the matrix element falls below the one from $\Theta(\tilde{M}^2_{\phi}-\s)$ term in the integral in eq.~\eqref{eq:collision}. Just like in scalar case, to a good approximation we can replace $\Mf\rightarrow\infty$ in the integral limit. This yields
\begin{align}\label{eq:fermion_boson_lowT}
{\mathcal{C}}_{\textrm{low-T}}&\approx S'\frac{T_a^7}{M\f^6}{\mu}_a^4w\Bigg[\frac{x}{2}\int_0^{\infty}\int_{0}^{\infty}d\tilde{p}\ d\tilde{s}\ \s D(\s,\p,x,0)\Bigg]
 \longrightarrow  \begin{cases} 31\, S'\frac{\mu^4_aw}{M\f^6}T_a^7 & x<0.1 \\ -21\, S'\frac{\mu^4_aw}{M\f^6}T_b^7 & x>10 .\end{cases}
\end{align}
Similar to our high temperature estimate, we find that the energy transfer function becomes insensitive to the colder sector as $x\rightarrow0,\infty$.

\paragraph{Total collision term.}
To get an analytic estimate of $\mathcal{C}_E$ over all temperature ranges we combine the analytic estimates as
\begin{align}\label{eq:fermion_boson_fit}
\br{\mathcal{C}}_E=&\br{\mathcal{C}}_{\textrm{low-T}}\Theta(M\f-{T})+\br{\mathcal{C}}_{\textrm{MB}}\Theta(0.2M\f-{T})+\textrm{max}(\br{\mathcal{C}}_{\textrm{MB}},\br{\mathcal{C}}_{\textrm{high-T}})\Theta(M\f-{T})\Theta({T}-0.2M\f)\nonumber\\&+\br{\mathcal{C}}_{\textrm{high-T}}\Theta({T}-M\f),
\end{align}
where $T=\textrm{max}(T_{a},T_{b})$ and $\br{\mathcal{C}}_{\textrm{high-T}},\br{\mathcal{C}}_{\textrm{MB}}$ and $\br{\mathcal{C}}_{\textrm{low-T}}$ are as described in eq.~\eqref{eq:fermion_boson_highT}, \eqref{eq:fermion_boson_MB} and \eqref{eq:fermion_boson_lowT}. 

In the left panels of figure~\ref{fig:fermion_boson_analytical_compare}, we compare our analytic estimate with the exact numerical value.  In the right panels of figure~\ref{fig:fermion_boson_analytical_compare}, we compare the ratio of our analytic fit, eq.~\eqref{eq:fermion_boson_fit}, with  the numerical evaluation of the full expression (eq.~\eqref{eq:collision}) with scattering amplitude given in eq.~\eqref{eq:matrix_delta_fermion_boson}. The largest deviation occurs during the transition from $\mathcal{C}_{\textrm{high-T}}$ to $\mathcal{C}_{\textrm{MB}}$ between $M_{\phi}/4<T<M\f$ and is of the order $\sim 50\%$.

\section{Collision terms for $t$-channel processes}\label{appendix:t-channel}
\addtocontents{toc}{\setcounter{tocdepth}{1}} 
In this section we calculate the energy transfer rate between two relativistic thermal bath via $t$-channel scatterings and explicitly demonstrate that its contribution will always be dwarfed by the $s$-channel.

For the $2\to2$ scattering process $ 1\ +\ 2\ \rightarrow\ 3\ +\ 4$, where 1, 2, 3, and 4 represent the particles participating in the scattering process, the forward collision term for a $t$-channel process is given by
\begin{align}\label{eq:t_forward_scatter}
\mathcal{C}_E^f=\int\prod_i\bigg(\frac{d^4p_i}{(2\pi)^3}\delta(p_i^2-m_i^2)\Theta(p_i^0)\bigg)
(2\pi)^4\delta^4(p_1+p_2-p_3-p_4)|\overline{\mathcal{M}}|^2S\ (p_1^0-p_3^0) \nonumber\\ f_1(U\cdot p_1)f_2(U\cdot p_2)(1\pm f_3(U\cdot p_3))(1\pm f_4(U\cdot p_4)). 
\end{align}
Here $f_i$ is the momentum space distribution function for the $i^{th}$ particle, $U$ is the 4-velocity of the frame in which we are calculating the collision term, $|\overline{\mathcal{M}}|^2$ is the spin summed scattering amplitude of the process, $S$ includes the identical particle factors of the process. Unlike the $s$-channel calculation, we do not transform to the center of mass frame, keeping $U=(1,0,0,0)$. To simplify the integral, we perform the following change of variables:  
\begin{align}
p=& p_1-p_3,\quad p'=p_2-p_4,\\
q=&p_1+p_3,\quad q'=p_2+p_4.
\end{align}
Correspondingly the Mandelstam variables will be given by $s=(q+q')^2/4$, $t=p^2$ and $u=(q-q')^2/4$. These substitutions also simplify the phase space element in eq.~\eqref{eq:t_forward_scatter} to, 
\begin{multline}
dI=\prod_i\bigg(\frac{d^4p_i}{(2\pi)^3}\delta(p_i^2-m_i^2)\Theta(p_i^0)\bigg)
(2\pi)^4\delta^4(p_1+p_2-p_3-p_4)\rightarrow \nonumber\\
\frac{(2\pi)^4}{2^8}\ \frac{d^4p}{(2\pi)^4}\frac{d^4q}{(2\pi)^4}\delta((p+q)^2/4-m_1^2)\Theta(p^0+q^0)\delta((p-q)^2/4-m_3^2)\Theta(p^0-q^0)\nonumber \\
\times\frac{d^4p'}{(2\pi)^4}\frac{d^4q'}{(2\pi)^4}\delta((p'+q')^2/4-m_2^2)\Theta(p'^0+q'^0)\delta((p'-q')^2/4-m_4^2)\Theta(p'^0-q'^0) (2\pi)^4\delta^4(p+p').
\end{multline}
Integrating out $p'$ and combining the conditions in the step functions we find
\begin{align}\label{eq:t_dI}
dI=&\frac{(2\pi)^4}{2^8}\frac{d^4p}{(2\pi)^4}\bigg(\frac{d^4q}{(2\pi)^4}\delta((p+q)^2/4-m_1^2)\delta((q-p)^2/4-m_3^2)\Theta(q^0-|p^0|)\bigg)\nonumber\\
& \times\bigg(\frac{d^4q'}{(2\pi)^4}\delta((p+q')^2/4-m_2^2)\delta((q'-p)^2/4-m_4^2)\Theta(q'^0-|p^0|)\bigg)\nonumber\\
\equiv &\frac{d^4p}{2^8}\ dI_{13}\ dI_{24}.
\end{align}
Here $dI_{13}$ and $dI_{24}$ are  phase space elements identical up to the different masses of the scattering particles. We first focus on simplifying the $dI_{13}$ phase space element  and then swap the masses to get the result for $dI_{24}$.

The delta functions in $dI_{13}$ impose the conditions
\begin{eqnarray}\label{eq:delta_conditions}
q\cdot p&&=m_1^2-m_3^2\equiv \Delta_{13}\\
p^2+q^2&&=4\frac{(m_1^2+m_3^2)}{2}\equiv 4m_{13}^2
\end{eqnarray}

Taking the $z$-axis to be along the spatial component of $p$, the above equations can be re-expressed as
\begin{align}
q^0&=\frac{q_z\vp}{p^0}+\frac{\Delta_{13}}{p^0},\\
|\vec{q}_{xy}|^2&=p^2\bigg(1-\frac{q_z^2}{(p^0)^2}\bigg)-4m_{13}^2+\frac{\Delta_{13}^2}{(p^0)^2}+\frac{2q_z\vp\Delta_{13}}{(p^0)^2},
\end{align}
where $\vec{q}_{xy}$ is the component of $\vec{q}$ along $x-y$ plane.
Hence, the delta function in $dI_{13}$ can be transformed to impose conditions on $q^0$ and $|\vec{q}_{xy}|$. The corresponding Jacobian for transforming the Dirac delta conditions is
\begin{eqnarray}
J&=&- \frac{|\vec{q}_{xy}|}{2}p^0.
\end{eqnarray}

Subsequently, using the Dirac delta conditions, we integrate over the $|\vec{q}_{xy}|$ and $q^0$ to yield,
\begin{eqnarray}
dI_{13}&=&-\frac{2}{p^0}\Theta\bigg(\frac{q_z\vp}{p^0}+\frac{\Delta_{13}}{p^0}-|p^0|\bigg)\Theta\bigg(p^2\bigg(1-\frac{q_z^2}{(p^0)^2}\bigg)-4m_{13}^2+\frac{\Delta_{13}^2}{(p^0)^2}+\frac{2q_z\vp\Delta_{13}}{(p^0)^2}\bigg)d\theta_{xy} \frac{dq_z}{(2\pi)^4}\nonumber\\
&\equiv&-\frac{2}{p^0}Q(q_z,m_1,m_3,p)d\theta_{xy} \frac{dq_z}{(2\pi)^4}
\end{eqnarray}
where $\theta_{xy}$ denotes the angle between $\vec{q}_{xy}$ and $x$ axis.

The measure $dI_{24}$ has the same form as $dI_{13}$, except with the substitutions $m_{1,3}\to m_{2,4}$ and $q\to q'$. Inserting these simplified phase-space elements back in eq.~\eqref{eq:t_forward_scatter}, using eq.~\eqref{eq:t_dI} and making use of the fact that the scattering amplitude only depends on $t=p^2$, we get
\begin{align}\label{eqn:tresult}
\mathcal{C}_E^f&=\frac{16\pi S}{2^8(2\pi)^{6}}\int dp^0\Big(\frac{\vp}{p^0}\Big)^2d\vp|\overline{\mathcal{M}}(t)|^2p^0\\ \times\nonumber&\bigg[\int Q(q_z,m_1,m_3,p) f_1(p_1^0)(1\pm f_3(p_3^0)) dq_z\bigg]\bigg[\int Q(q_z',m_2,m_4,p)f_2(p_2^0)(1\pm f_4(p_4^0))dq'_z\bigg]\\
\equiv&\frac{16\pi S}{2^8(2\pi)^{6}}\int|\overline{\mathcal{M}}(t)|^2 \Big(\frac{\vp}{p^0}\Big)^2\ p^0\ dp^0\ d\vp\ F_{13}(\vp,p^0)\ F_{24}(\vp,p^0).
\end{align}
In the first step above we have integrated over the phase space angles $\theta_{xy}$ and $\theta_{xy}'$. The result  in eq.\ \eqref{eqn:tresult} holds for any process whose scattering amplitude is only a function of $t$.

Eq.\ \eqref{eqn:tresult} can be further simplified if particles 1 and 3 are identical (henceforth, particle $a$), and  and similarly particle 2 and 4 (particle $b$). With these assumptions, $F_{a,b}=F_{13,24}$ simplifies to
\begin{align}
F_{a,b}=\Theta(-p^2)\int_{-\infty}^{\infty}\Theta\Big(p^0(q_z-\beta_{a,b}p^0)\Big)f_{a,b}(\frac{q^0+p^0}{2})\Big(1\pm f_{a,b}(\frac{q^0-p^0}{2})\Big)\bigg|_{q^0=q_z\vp/p^0} dq_z
\end{align}
where $\beta_{a,b}=\sqrt{1-\frac{4m_{a,b}^2}{p^2}}$.
Furthermore, if  particles $a$ and $b$ are thermally distributed with temperatures $T_a$ and $T_b$, respectively,  the integral in $F_{a,b}$ can be evaluated exactly to yield
\begin{align}\label{eq:F_t-channel}
F_{a,b}=&-\Theta(-p^2)\zeta_{a,b}2 T_{a,b}\frac{|p^0|}{\vp}\frac{\Theta(p^0)+e^{|p^0|/T_{a,b}}\Theta(-p^0)}{e^{|p^0|/T_{a,b}}-1}\log\bigg(\frac{\exp(\frac{\beta_{a,b}\vp+|p_0|}{2T_{a,b}})+\zeta_{a,b}}{\exp(\frac{\beta_{a,b}\vp+|p_0|}{2T_{a,b}})+\zeta_{a,b}e^{|p_0|/T_{a,b}}}\bigg),
\end{align}
where $\zeta_{a,b}=\pm1$, depending on whether particle $a,b$ is a fermion or boson respectively.
Substituting this result into the forward collision term and converting the integral over $p^0$ to an integral over $t$ using
\begin{align}
\int_0^{\infty} dp^0&=\int_{0}^{\infty} dp^0\int_{-\infty}^{\infty} \delta(t-p^2) dt=\int_{-\infty}^{\infty}dt\ \frac{\Theta(\vp^2+t)}{2\sqrt{\vp^2+t}}\bigg|_{p^0=\sqrt{\vp^2+t}},
\end{align}
we find
\begin{multline}\label{eq:t_sc}
\mathcal{C}_E^f=\frac{ST_aT_b\zeta_a\zeta_b}{2^4(2\pi)^{5}}\int_0^{\infty}d\vp \int_{-\vp^2}^{0}dt\ |\overline{\mathcal{M}}(t)|^2 \frac{e^{\sqrt{\vp^2+t}/T_b}-e^{\sqrt{\vp^2+t}/T_a}}{(e^{\sqrt{\vp^2+t}/T_b}-1)(e^{\sqrt{\vp^2+t}/T_a}-1)}\\
\times\log\Bigg(\frac{\exp\bigg(\dfrac{\sqrt{\vp^2+t}+\beta_a\vp}{2T_a}\bigg)+\zeta_a}{\exp\bigg(\dfrac{\sqrt{\vp^2+t}+\beta_a\vp}{2T_a}\bigg)+\zeta_a\exp\bigg(\dfrac{\sqrt{\vp^2+t}}{T_a}\bigg)}\Bigg)\\ \times\log\Bigg(\frac{\exp\bigg(\dfrac{\sqrt{\vp^2+t}+\beta_b\vp}{2T_b}\bigg)+\zeta_b}{\exp\bigg(\dfrac{\sqrt{\vp^2+t}+\beta_b\vp}{2T_b}\bigg)+\zeta_b\exp\bigg(\dfrac{\sqrt{\vp^2+t}}{T_b}\bigg)}\Bigg).
\end{multline}
In the case where sectors $a$ and $b$ have thermalized, i.e. $T_a=T_b$, the forward collision term is identically zero, as expected from a $t$-channel process. The backward collision term has the same form, except with an overall minus sign. The total collision term, $\mathcal{C}_E=\mathcal{C}_E^f-\mathcal{C}_E^b$, is therefore twice the forward rate. Finally, we make the integrand dimensionless by pulling overall factors of $T_a^5$ to give 
\begin{multline}\label{eq:t_collision}
\mathcal{C}_E=\frac{S'T_a^5x\zeta_a\zeta_b}{2}\int_0^{\infty}d\p \int_{-\p^2}^{0}d\tl\ |\overline{\mathcal{M}}(\tl)|^2 \frac{e^{\sqrt{\p^2+\tl}/x}-e^{\sqrt{\p^2+\tl}}}{(e^{\sqrt{\p^2+\tl}/x}-1)(e^{\sqrt{\p^2+\tl}}-1)}\\
\times\log\Bigg(\frac{e^{\frac{1}{2}(\sqrt{\p^2+\tl}+\beta_a\p)}+\zeta_a}{e^{\frac{1}{2}(\sqrt{\p^2+\tl}+\beta_a\p)}+\zeta_ae^{\sqrt{\p^2+\tl}}}\Bigg)\log\Bigg(\frac{e^{\frac{1}{2x}(\sqrt{\p^2+\tl}+\beta_b\p)}+\zeta_b}{e^{\frac{1}{2x}(\sqrt{\p^2+\tl}+\beta_b\p)}+\zeta_be^{\frac{1}{x}\sqrt{\p^2+\tl}}}\Bigg),
\end{multline}
where $S'=S/4/(2\pi)^5$, $x=T_b/T_a$ and a tilde over a variable indicates that it has  been made dimensionless by dividing by the appropriate power of  $T_a$.

\subsection{Trilinear scalar couplings}
In this section we evaluate the  $t$-channel  contribution to the collision term and compare it with the $s$-channel rate for the theory in which the two sectors interact via
\begin{align}
\mathcal{L}_{\textrm{int}}=\frac{1}{2}\mu_a\phi\chi_a^2+\frac{1}{2}\mu_b\phi\chi_b^2.
\end{align}
The $t$-channel scattering amplitude for this theory is given by
\begin{align}\label{eq:matrix_scalar_t}
|\overline{\mathcal{M}}(t)|^2=\frac{\mu_a^2\mu_b^2}{(t-M\f^2)^2}=\frac{{\mu}_a^4w}{M\f^4(\tl/\Mf^2-1)^2}.
\end{align}
\begin{figure}
\centering
\includegraphics[width=0.75\textwidth]{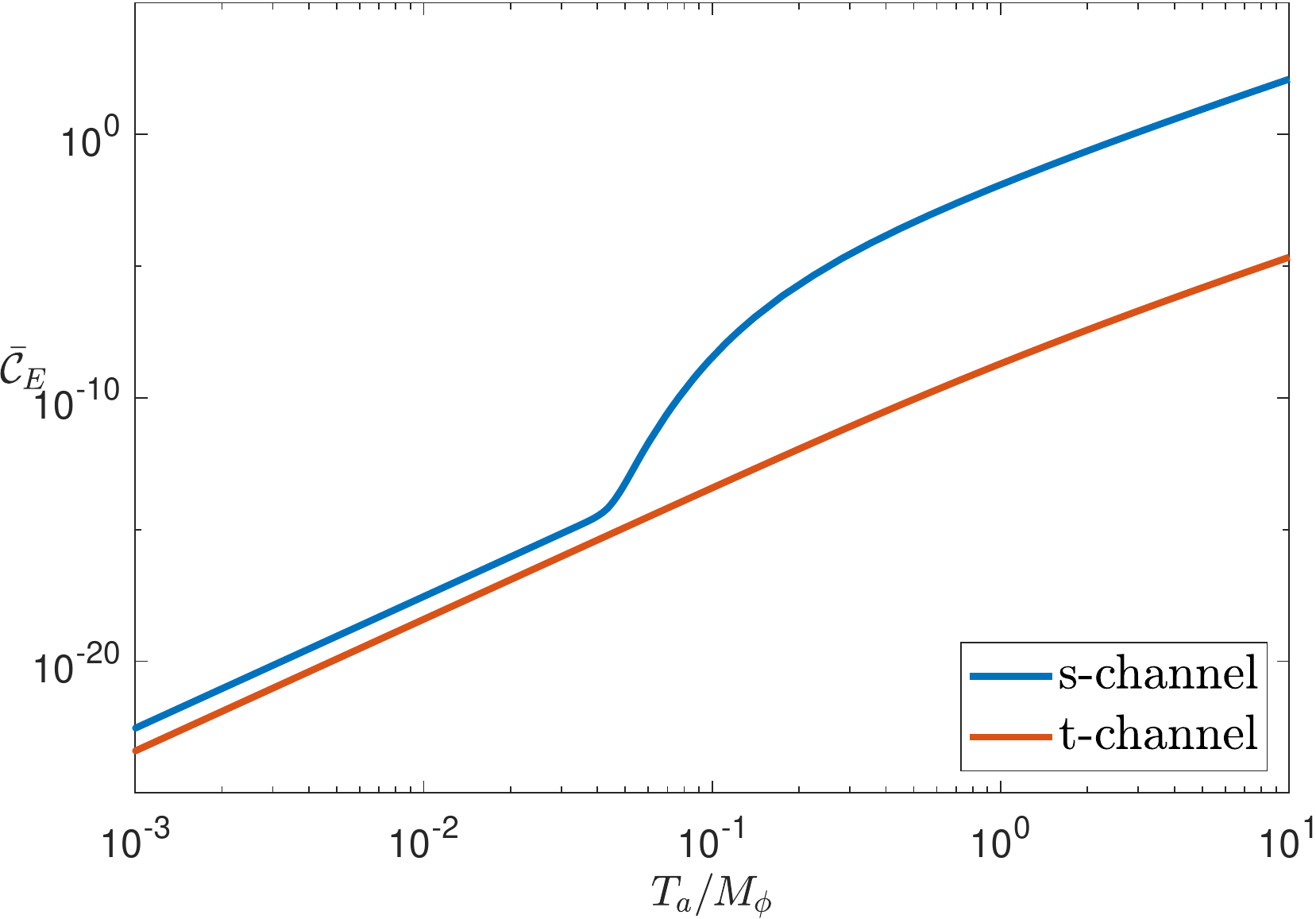}
\caption{Total collision term $\br{\mathcal{C}}_E=\mathcal{C}_E/M\f^5$ as a function of $T_a/M\f$ at fixed $x=T_b/T_a=0.5$, for $\mu_a=0.01M\f$, $k=\frac{\mu_b}{\mu_a}=0.5$, and $m_{a,b}=10^{-8}M\f$ and $S'=1$.  The blue line corresponds to the numerical $s$-channel contribution to $\mathcal{C}_E$ of eq.~\eqref{eq:collision} with scattering amplitude given by eq.~\eqref{eq:matrix_boson} while the red line shows the numerical $t$-channel contribution to $\mathcal{C}_E$ (eq.~\eqref{eq:t_collision}) with scattering amplitude as given in eq.~\eqref{eq:matrix_scalar_t}.}
\label{fig:scalar_boson_st_compare_T}
\end{figure}
In figure~\ref{fig:scalar_boson_st_compare_T} we compare the resulting $t$-channel contribution to the collision term with the $s$-channel  contribution for this theory at fixed $x=T_b/T_a$. The $s$-channel contribution always dominates over the $t$-channel contribution, both below the resonance as well as above. 

At low temperatures, $T_a\ll M\f$, we can approximate the scattering amplitude as a constant just like we did for $s$-channel,
\begin{align}
|\overline{\mathcal{M}}(t)|^2\approx \frac{\mu_a^4w}{M\f^4}.
\end{align}
The temperature dependence in this region can then be isolated,
\begin{align}\label{eq:t_scalar_lowT}
\mathcal{C}_{\textrm{low-T}}=&S'T_a^5\frac{\mu_a^4w}{M\f^4}\frac{x}{2}\int_0^{\infty}d\p \int_{-\p^2}^{0}d\tl\ \frac{e^{\sqrt{\p^2+\tl}/x}-e^{\sqrt{\p^2+\tl}}}{(e^{\sqrt{\p^2+\tl}/x}-1)(e^{\sqrt{\p^2+\tl}}-1)}\nonumber\\
&\times\log\Bigg(\frac{e^{\frac{1}{2}(\sqrt{\p^2+\tl}+\beta_a\p)}-1}{e^{\frac{1}{2}(\sqrt{\p^2+\tl}+\beta_a\p)}-e^{\sqrt{\p^2+\tl}}}\Bigg)\log\Bigg(\frac{e^{\frac{1}{2x}(\sqrt{\p^2+\tl}+\beta_b\p)}-1}{e^{\frac{1}{2x}(\sqrt{\p^2+\tl}+\beta_b\p)}-e^{\frac{1}{x}\sqrt{\p^2+\tl}}}\Bigg),\nonumber\\
\equiv&S'T_a^5\frac{\mu_a^4w}{M\f^4}f_t(x).
\end{align}
The low temperature behavior derived above is similar to the one we had in $s$-channel case, eq.~\eqref{eq:scalar_lowT}. However, unlike $f(x)$ in eq.~\eqref{eq:scalar_lowT}, $f_t(x)$ decreases as $x\rightarrow0$, rather than approaching a constant; this reflects the suppressed probability of finding an initial scattering particle in the colder sector as the thermal abundance drops.  
We find the $t$-channel rate to be suppressed by a factor of six relative to the $s$-channel rate even when $x\rightarrow 1$.  In a similar manner one can also show that the $t$-channel contribution remains subdominant for other theories considered in appendix~\ref{appendix:s-channel}.  Thus our neglect of the $t$-channel  contribution to $\mathcal{C}_E$ is justified.

\bibliographystyle{JHEP}
\bibliography{references}

\providecommand{\href}[2]{#2}\begingroup\raggedright\begin{thebibliography}{10}

\bibitem{Ade:2015xua}
{\scshape Planck} collaboration, \emph{{Planck 2015 results. XIII. Cosmological
  parameters}},
  \href{https://doi.org/10.1051/0004-6361/201525830}{\emph{Astron. Astrophys.}
  {\bfseries 594} (2016) A13}
  [\href{https://arxiv.org/abs/1502.01589}{{\ttfamily 1502.01589}}].

\bibitem{Aghanim:2018eyx}
{\scshape Planck} collaboration, \emph{{Planck 2018 results. VI. Cosmological
  parameters}},  \href{https://arxiv.org/abs/1807.06209}{{\ttfamily
  1807.06209}}.

\bibitem{Escudero:2016gzx}
M.~Escudero, A.~Berlin, D.~Hooper and M.-X. Lin, \emph{{Toward (Finally!)
  Ruling Out Z and Higgs Mediated Dark Matter Models}},
  \href{https://doi.org/10.1088/1475-7516/2016/12/029}{\emph{JCAP} {\bfseries
  1612} (2016) 029} [\href{https://arxiv.org/abs/1609.09079}{{\ttfamily
  1609.09079}}].

\bibitem{Alexander:2016aln}
J.~Alexander et~al., \emph{{Dark Sectors 2016 Workshop: Community Report}},
  2016, \href{https://arxiv.org/abs/1608.08632}{{\ttfamily 1608.08632}},
  \href{https://inspirehep.net/record/1484628/files/arXiv:1608.08632.pdf}{https://inspirehep.net/record/1484628/files/arXiv:1608.08632.pdf}.

\bibitem{Carlson:1992fn}
E.~D. Carlson, M.~E. Machacek and L.~J. Hall, \emph{{Self-interacting dark
  matter}}, \href{https://doi.org/10.1086/171833}{\emph{Astrophys. J.}
  {\bfseries 398} (1992) 43}.

\bibitem{Feng:2008mu}
J.~L. Feng, H.~Tu and H.-B. Yu, \emph{{Thermal Relics in Hidden Sectors}},
  \href{https://doi.org/10.1088/1475-7516/2008/10/043}{\emph{JCAP} {\bfseries
  0810} (2008) 043} [\href{https://arxiv.org/abs/0808.2318}{{\ttfamily
  0808.2318}}].

\bibitem{Sigurdson:2009uz}
K.~Sigurdson, \emph{{Hidden Hot Dark Matter as Cold Dark Matter}},
  \href{https://arxiv.org/abs/0912.2346}{{\ttfamily 0912.2346}}.

\bibitem{Cheung:2010gj}
C.~Cheung, G.~Elor, L.~J. Hall and P.~Kumar, \emph{{Origins of Hidden Sector
  Dark Matter I: Cosmology}},
  \href{https://doi.org/10.1007/JHEP03(2011)042}{\emph{JHEP} {\bfseries 03}
  (2011) 042} [\href{https://arxiv.org/abs/1010.0022}{{\ttfamily 1010.0022}}].

\bibitem{Pappadopulo:2016pkp}
D.~Pappadopulo, J.~T. Ruderman and G.~Trevisan, \emph{{Dark matter freeze-out
  in a nonrelativistic sector}},
  \href{https://doi.org/10.1103/PhysRevD.94.035005}{\emph{Phys. Rev.}
  {\bfseries D94} (2016) 035005}
  [\href{https://arxiv.org/abs/1602.04219}{{\ttfamily 1602.04219}}].

\bibitem{Dror:2016rxc}
J.~A. Dror, E.~Kuflik and W.~H. Ng, \emph{{Codecaying Dark Matter}},
  \href{https://doi.org/10.1103/PhysRevLett.117.211801}{\emph{Phys. Rev. Lett.}
  {\bfseries 117} (2016) 211801}
  [\href{https://arxiv.org/abs/1607.03110}{{\ttfamily 1607.03110}}].

\bibitem{Berlin:2016gtr}
A.~Berlin, D.~Hooper and G.~Krnjaic, \emph{{Thermal Dark Matter From A Highly
  Decoupled Sector}},
  \href{https://doi.org/10.1103/PhysRevD.94.095019}{\emph{Phys. Rev.}
  {\bfseries D94} (2016) 095019}
  [\href{https://arxiv.org/abs/1609.02555}{{\ttfamily 1609.02555}}].

\bibitem{Dror:2017gjq}
J.~A. Dror, E.~Kuflik, B.~Melcher and S.~Watson, \emph{{Concentrated dark
  matter: Enhanced small-scale structure from codecaying dark matter}},
  \href{https://doi.org/10.1103/PhysRevD.97.063524}{\emph{Phys. Rev.}
  {\bfseries D97} (2018) 063524}
  [\href{https://arxiv.org/abs/1711.04773}{{\ttfamily 1711.04773}}].

\bibitem{Georg:2019jld}
J.~Georg, B.~Melcher and S.~Watson, \emph{{Primordial Black Holes and
  Co-Decaying Dark Matter}},
  \href{https://arxiv.org/abs/1902.04082}{{\ttfamily 1902.04082}}.

\bibitem{Faraggi:2000pv}
A.~E. Faraggi and M.~Pospelov, \emph{{Selfinteracting dark matter from the
  hidden heterotic string sector}},
  \href{https://doi.org/10.1016/S0927-6505(01)00121-9}{\emph{Astropart. Phys.}
  {\bfseries 16} (2002) 451}
  [\href{https://arxiv.org/abs/hep-ph/0008223}{{\ttfamily hep-ph/0008223}}].

\bibitem{Forestell:2016qhc}
L.~Forestell, D.~E. Morrissey and K.~Sigurdson, \emph{{Non-Abelian Dark Forces
  and the Relic Densities of Dark Glueballs}},
  \href{https://doi.org/10.1103/PhysRevD.95.015032}{\emph{Phys. Rev.}
  {\bfseries D95} (2017) 015032}
  [\href{https://arxiv.org/abs/1605.08048}{{\ttfamily 1605.08048}}].

\bibitem{Dienes:2016vei}
K.~R. Dienes, F.~Huang, S.~Su and B.~Thomas, \emph{{Dynamical Dark Matter from
  Strongly-Coupled Dark Sectors}},
  \href{https://doi.org/10.1103/PhysRevD.95.043526}{\emph{Phys. Rev.}
  {\bfseries D95} (2017) 043526}
  [\href{https://arxiv.org/abs/1610.04112}{{\ttfamily 1610.04112}}].

\bibitem{Forestell:2018dnu}
L.~Forestell and D.~E. Morrissey, \emph{{Infrared Effects of Ultraviolet
  Operators on Dark Matter Freeze-In}},
  \href{https://arxiv.org/abs/1811.08905}{{\ttfamily 1811.08905}}.

\bibitem{Kang:2019izi}
Z.~Kang, \emph{{Slightly Ultra-violet Freeze-in a Hidden Gluonic Sector}},
  \href{https://arxiv.org/abs/1901.10934}{{\ttfamily 1901.10934}}.

\bibitem{Blanco:2019eij}
C.~Blanco, M.~S. Delos, A.~L. Erickcek and D.~Hooper, \emph{{Annihilation
  Signatures of Hidden Sector Dark Matter Within Early-Forming Microhalos}},
  \href{https://arxiv.org/abs/1906.00010}{{\ttfamily 1906.00010}}.

\bibitem{Zhang:2015era}
Y.~Zhang, \emph{{Long-lived Light Mediator to Dark Matter and Primordial Small
  Scale Spectrum}},
  \href{https://doi.org/10.1088/1475-7516/2015/05/008}{\emph{JCAP} {\bfseries
  1505} (2015) 008} [\href{https://arxiv.org/abs/1502.06983}{{\ttfamily
  1502.06983}}].

\bibitem{Hodges:1993yb}
H.~M. Hodges, \emph{{Mirror baryons as the dark matter}},
  \href{https://doi.org/10.1103/PhysRevD.47.456}{\emph{Phys. Rev.} {\bfseries
  D47} (1993) 456}.

\bibitem{Berezhiani:1995am}
Z.~G. Berezhiani, A.~D. Dolgov and R.~N. Mohapatra, \emph{{Asymmetric
  inflationary reheating and the nature of mirror universe}},
  \href{https://doi.org/10.1016/0370-2693(96)00219-5}{\emph{Phys. Lett.}
  {\bfseries B375} (1996) 26}
  [\href{https://arxiv.org/abs/hep-ph/9511221}{{\ttfamily hep-ph/9511221}}].

\bibitem{Adshead:2016xxj}
P.~Adshead, Y.~Cui and J.~Shelton, \emph{{Chilly Dark Sectors and Asymmetric
  Reheating}}, \href{https://doi.org/10.1007/JHEP06(2016)016}{\emph{JHEP}
  {\bfseries 06} (2016) 016}
  [\href{https://arxiv.org/abs/1604.02458}{{\ttfamily 1604.02458}}].

\bibitem{Halverson:2019kna}
J.~Halverson, C.~Long, B.~Nelson and G.~Salinas, \emph{{On Axion Reheating in
  the String Landscape}},  \href{https://arxiv.org/abs/1903.04495}{{\ttfamily
  1903.04495}}.

\bibitem{Kolb:2003ke}
E.~W. Kolb, A.~Notari and A.~Riotto, \emph{{On the reheating stage after
  inflation}}, \href{https://doi.org/10.1103/PhysRevD.68.123505}{\emph{Phys.
  Rev.} {\bfseries D68} (2003) 123505}
  [\href{https://arxiv.org/abs/hep-ph/0307241}{{\ttfamily hep-ph/0307241}}].

\bibitem{Drewes:2013iaa}
M.~Drewes and J.~U. Kang, \emph{{The Kinematics of Cosmic Reheating}},
  \href{https://doi.org/10.1016/j.nuclphysb.2013.07.009,
  10.1016/j.nuclphysb.2014.09.008}{\emph{Nucl. Phys.} {\bfseries B875} (2013)
  315} [\href{https://arxiv.org/abs/1305.0267}{{\ttfamily 1305.0267}}].

\bibitem{Hardy:2017wkr}
E.~Hardy and J.~Unwin, \emph{{Symmetric and Asymmetric Reheating}},
  \href{https://doi.org/10.1007/JHEP09(2017)113}{\emph{JHEP} {\bfseries 09}
  (2017) 113} [\href{https://arxiv.org/abs/1703.07642}{{\ttfamily
  1703.07642}}].

\bibitem{Reece:2015lch}
M.~Reece and T.~Roxlo, \emph{{Nonthermal production of dark radiation and dark
  matter}}, \href{https://doi.org/10.1007/JHEP09(2016)096}{\emph{JHEP}
  {\bfseries 09} (2016) 096}
  [\href{https://arxiv.org/abs/1511.06768}{{\ttfamily 1511.06768}}].

\bibitem{Guth:1980zm}
A.~H. Guth, \emph{{The Inflationary Universe: A Possible Solution to the
  Horizon and Flatness Problems}},
  \href{https://doi.org/10.1103/PhysRevD.23.347}{\emph{Phys. Rev.} {\bfseries
  D23} (1981) 347}.

\bibitem{Linde:1981mu}
A.~D. Linde, \emph{{A New Inflationary Universe Scenario: A Possible Solution
  of the Horizon, Flatness, Homogeneity, Isotropy and Primordial Monopole
  Problems}}, \href{https://doi.org/10.1016/0370-2693(82)91219-9}{\emph{Phys.
  Lett.} {\bfseries 108B} (1982) 389}.

\bibitem{Albrecht:1982wi}
A.~Albrecht and P.~J. Steinhardt, \emph{{Cosmology for Grand Unified Theories
  with Radiatively Induced Symmetry Breaking}},
  \href{https://doi.org/10.1103/PhysRevLett.48.1220}{\emph{Phys. Rev. Lett.}
  {\bfseries 48} (1982) 1220}.

\bibitem{Martin:2013tda}
J.~Martin, C.~Ringeval and V.~Vennin, \emph{{Encyclopedia Inflationaris}},
  \href{https://doi.org/10.1016/j.dark.2014.01.003}{\emph{Phys. Dark Univ.}
  {\bfseries 5-6} (2014) 75} [\href{https://arxiv.org/abs/1303.3787}{{\ttfamily
  1303.3787}}].

\bibitem{Amin:2014eta}
M.~A. Amin, M.~P. Hertzberg, D.~I. Kaiser and J.~Karouby,
  \emph{{Nonperturbative Dynamics Of Reheating After Inflation: A Review}},
  \href{https://doi.org/10.1142/S0218271815300037}{\emph{Int. J. Mod. Phys.}
  {\bfseries D24} (2014) 1530003}
  [\href{https://arxiv.org/abs/1410.3808}{{\ttfamily 1410.3808}}].

\bibitem{Lozanov:2016hid}
K.~D. Lozanov and M.~A. Amin, \emph{{Equation of State and Duration to
  Radiation Domination after Inflation}},
  \href{https://doi.org/10.1103/PhysRevLett.119.061301}{\emph{Phys. Rev. Lett.}
  {\bfseries 119} (2017) 061301}
  [\href{https://arxiv.org/abs/1608.01213}{{\ttfamily 1608.01213}}].

\bibitem{Turner:1983he}
M.~S. Turner, \emph{{Coherent Scalar Field Oscillations in an Expanding
  Universe}}, \href{https://doi.org/10.1103/PhysRevD.28.1243}{\emph{Phys. Rev.}
  {\bfseries D28} (1983) 1243}.

\bibitem{Traschen:1990sw}
J.~H. Traschen and R.~H. Brandenberger, \emph{{Particle Production During
  Out-of-equilibrium Phase Transitions}},
  \href{https://doi.org/10.1103/PhysRevD.42.2491}{\emph{Phys. Rev.} {\bfseries
  D42} (1990) 2491}.

\bibitem{Kofman:1994rk}
L.~Kofman, A.~D. Linde and A.~A. Starobinsky, \emph{{Reheating after
  inflation}}, \href{https://doi.org/10.1103/PhysRevLett.73.3195}{\emph{Phys.
  Rev. Lett.} {\bfseries 73} (1994) 3195}
  [\href{https://arxiv.org/abs/hep-th/9405187}{{\ttfamily hep-th/9405187}}].

\bibitem{Emond:2018ybc}
W.~T. Emond, P.~Millington and P.~M. Saffin, \emph{{Boltzmann equations for
  preheating}},  \href{https://arxiv.org/abs/1807.11726}{{\ttfamily
  1807.11726}}.

\bibitem{Abbott:1982hn}
L.~F. Abbott, E.~Farhi and M.~B. Wise, \emph{{Particle Production in the New
  Inflationary Cosmology}},
  \href{https://doi.org/10.1016/0370-2693(82)90867-X}{\emph{Phys. Lett.}
  {\bfseries 117B} (1982) 29}.

\bibitem{Albrecht:1982mp}
A.~Albrecht, P.~J. Steinhardt, M.~S. Turner and F.~Wilczek, \emph{{Reheating an
  Inflationary Universe}},
  \href{https://doi.org/10.1103/PhysRevLett.48.1437}{\emph{Phys. Rev. Lett.}
  {\bfseries 48} (1982) 1437}.

\bibitem{Chung:1998rq}
D.~J.~H. Chung, E.~W. Kolb and A.~Riotto, \emph{{Production of massive
  particles during reheating}},
  \href{https://doi.org/10.1103/PhysRevD.60.063504}{\emph{Phys. Rev.}
  {\bfseries D60} (1999) 063504}
  [\href{https://arxiv.org/abs/hep-ph/9809453}{{\ttfamily hep-ph/9809453}}].

\bibitem{Adshead:2015pva}
P.~Adshead, J.~T. Giblin, T.~R. Scully and E.~I. Sfakianakis,
  \emph{{Gauge-preheating and the end of axion inflation}},
  \href{https://doi.org/10.1088/1475-7516/2015/12/034}{\emph{JCAP} {\bfseries
  1512} (2015) 034} [\href{https://arxiv.org/abs/1502.06506}{{\ttfamily
  1502.06506}}].

\bibitem{Adshead:2015kza}
P.~Adshead and E.~I. Sfakianakis, \emph{{Fermion production during and after
  axion inflation}},
  \href{https://doi.org/10.1088/1475-7516/2015/11/021}{\emph{JCAP} {\bfseries
  1511} (2015) 021} [\href{https://arxiv.org/abs/1508.00891}{{\ttfamily
  1508.00891}}].

\bibitem{Kane:2015jia}
G.~Kane, K.~Sinha and S.~Watson, \emph{{Cosmological Moduli and the
  Post-Inflationary Universe: A Critical Review}},
  \href{https://doi.org/10.1142/S0218271815300220}{\emph{Int. J. Mod. Phys.}
  {\bfseries D24} (2015) 1530022}
  [\href{https://arxiv.org/abs/1502.07746}{{\ttfamily 1502.07746}}].

\bibitem{Co:2015pka}
R.~T. Co, F.~D'Eramo, L.~J. Hall and D.~Pappadopulo, \emph{{Freeze-In Dark
  Matter with Displaced Signatures at Colliders}},
  \href{https://doi.org/10.1088/1475-7516/2015/12/024}{\emph{JCAP} {\bfseries
  1512} (2015) 024} [\href{https://arxiv.org/abs/1506.07532}{{\ttfamily
  1506.07532}}].

\bibitem{Krnjaic:2015mbs}
G.~Krnjaic, \emph{{Probing Light Thermal Dark-Matter With a Higgs Portal
  Mediator}}, \href{https://doi.org/10.1103/PhysRevD.94.073009}{\emph{Phys.
  Rev.} {\bfseries D94} (2016) 073009}
  [\href{https://arxiv.org/abs/1512.04119}{{\ttfamily 1512.04119}}].

\bibitem{Evans:2017kti}
J.~A. Evans, S.~Gori and J.~Shelton, \emph{{Looking for the WIMP Next Door}},
  \href{https://doi.org/10.1007/JHEP02(2018)100}{\emph{JHEP} {\bfseries 02}
  (2018) 100} [\href{https://arxiv.org/abs/1712.03974}{{\ttfamily
  1712.03974}}].

\bibitem{Arkani-Hamed:2016rle}
N.~Arkani-Hamed, T.~Cohen, R.~T. D'Agnolo, A.~Hook, H.~D. Kim and D.~Pinner,
  \emph{{Solving the Hierarchy Problem at Reheating with a Large Number of
  Degrees of Freedom}},
  \href{https://doi.org/10.1103/PhysRevLett.117.251801}{\emph{Phys. Rev. Lett.}
  {\bfseries 117} (2016) 251801}
  [\href{https://arxiv.org/abs/1607.06821}{{\ttfamily 1607.06821}}].

\bibitem{Chu:2011be}
X.~Chu, T.~Hambye and M.~H.~G. Tytgat, \emph{{The Four Basic Ways of Creating
  Dark Matter Through a Portal}},
  \href{https://doi.org/10.1088/1475-7516/2012/05/034}{\emph{JCAP} {\bfseries
  1205} (2012) 034} [\href{https://arxiv.org/abs/1112.0493}{{\ttfamily
  1112.0493}}].

\bibitem{egs}
J.~E. Evans, C.~Gaidau and J.~Shelton, ``{Leak-in Dark Matter and Hidden
  Sectors Below the Equilibration Floor}.''.

\bibitem{McDonald:2001vt}
J.~McDonald, \emph{{Thermally generated gauge singlet scalars as
  selfinteracting dark matter}},
  \href{https://doi.org/10.1103/PhysRevLett.88.091304}{\emph{Phys. Rev. Lett.}
  {\bfseries 88} (2002) 091304}
  [\href{https://arxiv.org/abs/hep-ph/0106249}{{\ttfamily hep-ph/0106249}}].

\bibitem{Hall:2009bx}
L.~J. Hall, K.~Jedamzik, J.~March-Russell and S.~M. West, \emph{{Freeze-In
  Production of FIMP Dark Matter}},
  \href{https://doi.org/10.1007/JHEP03(2010)080}{\emph{JHEP} {\bfseries 03}
  (2010) 080} [\href{https://arxiv.org/abs/0911.1120}{{\ttfamily 0911.1120}}].

\bibitem{Dufaux:2006ee}
J.~F. Dufaux, G.~N. Felder, L.~Kofman, M.~Peloso and D.~Podolsky,
  \emph{{Preheating with trilinear interactions: Tachyonic resonance}},
  \href{https://doi.org/10.1088/1475-7516/2006/07/006}{\emph{JCAP} {\bfseries
  0607} (2006) 006} [\href{https://arxiv.org/abs/hep-ph/0602144}{{\ttfamily
  hep-ph/0602144}}].

\bibitem{Birrell:2014uka}
J.~Birrell, C.-T. Yang and J.~Rafelski, \emph{{Relic Neutrino Freeze-out:
  Dependence on Natural Constants}},
  \href{https://doi.org/10.1016/j.nuclphysb.2014.11.020}{\emph{Nucl. Phys.}
  {\bfseries B890} (2014) 481}
  [\href{https://arxiv.org/abs/1406.1759}{{\ttfamily 1406.1759}}].

\end{thebibliography}\endgroup
\end{document}